\documentclass{vldb}
\usepackage{graphicx}
\usepackage{balance}  
\usepackage{listings}
\usepackage{paralist}
\usepackage[table]{xcolor}
\usepackage{xspace}
\usepackage{proof}
\usepackage{fancyvrb}
\usepackage{booktabs}
\usepackage{array}
\usepackage{enumitem}
\usepackage{url}
\usepackage{tikz}
\usepackage{forest}
\usepackage{subcaption}
\usepackage{dirtree}
\usepackage{graphicx}
\usepackage{balance}  
\usepackage{listings}
\usepackage{paralist}
\usepackage{xcolor}
\usepackage{xspace}
\usepackage{proof}
\usepackage{fancyvrb}
\usepackage{booktabs}
\usepackage{array}
\usepackage{enumitem}
\usepackage{url}
\usepackage{tikz}
\usepackage{forest}
\usepackage{subcaption}
\usepackage{adjustbox}
\usepackage{amssymb}

\usetikzlibrary{positioning,shadows,arrows,trees,shapes,fit,shapes.geometric,backgrounds,arrows.meta}
\usepackage{tikz-qtree}

\newcolumntype{L}{>{$}l<{$}}

\vldbTitle{Scalable Querying of Nested Data}
\vldbAuthors{Jaclyn Smith, Michael Benedikt, Milos Nikolic, and Amir Shaikhha}
\vldbDOI{https://doi.org/10.14778/xxxxxxx.xxxxxxx}
\vldbVolume{12}
\vldbNumber{xxx}
\vldbYear{2019}

\begin{document}

\title{Scalable Querying of Nested Data}


\numberofauthors{1}
\author{
   \alignauthor
   Jaclyn Smith\raisebox{4pt}{$\dagger$}, Michael Benedikt\raisebox{4pt}{$\dagger$}, Milos Nikolic\raisebox{4pt}{$\S$}, Amir Shaikhha\raisebox{4pt}{$\dagger$}  \\[3pt]
   \affaddr{\quad\;\raisebox{4pt}{$\dagger$}University of Oxford, UK \qquad\qquad\; \raisebox{4pt}{$\S$}University of Edinburgh, UK} \\
   \email{\raisebox{4pt}{$\dagger$}name.surname@cs.ox.ac.uk \quad \raisebox{4pt}{$\S$}name.surname@ed.ac.uk}
}

\lstdefinelanguage{NRC}{
  morekeywords={for, in, union, if, then, else, match, let},%
  sensitive,%
  morecomment=[l]//,%
  morecomment=[s]{/*}{*/},%
  morestring=[b]",%
  morestring=[b]',%
  showstringspaces=false,%
  breaklines=true,%
  mathescape=true,%
  showspaces=false,
  showtabs=false, 
  showstringspaces=false,
  breakatwhitespace=true,
  aboveskip=1pt,
  belowskip=1pt,
  lineskip=0pt,
  basicstyle=\small\ttfamily\color{white!15!black},
  keywordstyle=\small\ttfamily\bfseries\color{blue!80!black},%
  columns=fullflexible,
  escapeinside={(*@}{@*)}
}[keywords,comments,strings]%

\lstdefinelanguage{Scala}%
{morekeywords={abstract,%
  case,catch,char,class,%
  def,else,extends,final,finally,for,%
  if,import,implicit,%
  match,module,%
  new,null,%
  object,override,%
  package,private,protected,public,%
  for,public,return,super,%
  this,throw,trait,try,type,%
  val,var,%
  with,while,%
  yield%
  },%
  sensitive,%
  morecomment=[l]//,%
  morecomment=[s]{/*}{*/},%
  morestring=[b]",%
  morestring=[b]',%
  showstringspaces=false,%
    breaklines=true,%
  mathescape=true,%
  showspaces=false,
  showtabs=false,
  showstringspaces=false,
  breakatwhitespace=true,
  xleftmargin=2em,
  aboveskip=1pt,
  belowskip=1pt,
  lineskip=-0.2pt,
   basicstyle=\ttfamily,
  keywordstyle=\ttfamily\color{blue!80!black},
  columns=fullflexible,
}[keywords,comments,strings]%

\newcommand{\myeat}[1]{}
\newcommand{\punto}{$\hspace*{\fill}\Box$}
\newcommand{\NULL}{\texttt{NULL}}
\newcommand{\tab}{\;\;\;}
\newcommand\mydots{\makebox[1em][c]{.\hfil.\hfil.}}

\newcounter{magicrownumbers}
\newcommand\rownumber{\stepcounter{magicrownumbers}\arabic{magicrownumbers}}
\newcommand{\linenumber}{\makebox[2ex][r]{$\scriptscriptstyle\rownumber\;\;$}}

\newcommand{\michael}[1]{{\bf\color{green} Michael: #1}}
\newcommand{\jac}[1]{{\bf \color{violet} Jaclyn: #1}}
\newcommand{\milos}[1]{{\bf \color{red} Milos: #1}}

\newcommand{\bagtype}{\textit{Bag}\hspace{0.5mm}}
\newcommand{\labeltype}{\textit{Label}\xspace}
\newcommand{\stringtype}{\textit{string}}
\newcommand{\datetype}{\textit{date}}
\newcommand{\doubletype}{\textit{real}}
\newcommand{\inttype}{\textit{int}}
\newcommand{\booltype}{\textit{bool}}

\newcommand{\Part}{\texttt{\textup{Part}}\xspace}
\newcommand{\Lineitem}{\texttt{Lineitem}\xspace}
\newcommand{\COP}{\texttt{\textup{COP}}\xspace}
\newcommand{\cid}{\texttt{\textup{cid}}\xspace}
\newcommand{\cname}{\texttt{\textup{cname}}\xspace}
\newcommand{\cop}{\texttt{\textup{cop}}\xspace}
\newcommand{\corders}{\texttt{\textup{corders}}\xspace}
\newcommand{\odate}{\texttt{\textup{odate}}\xspace}
\newcommand{\oparts}{\texttt{\textup{oparts}}\xspace}
\newcommand{\pid}{\texttt{\textup{pid}}\xspace}
\newcommand{\pname}{\texttt{\textup{pname}}\xspace}
\newcommand{\pprice}{\texttt{\textup{price}}\xspace}
\newcommand{\lqty}{\texttt{\textup{qty}}\xspace}
\newcommand{\total}{\texttt{\textup{total}}\xspace}
\newcommand{\pkey}{\texttt{\textup{id}}\xspace}
\newcommand{\sumBy}{\texttt{\textup{sumBy}}}

\newcommand{\Order}{\texttt{Order}\xspace}
\newcommand{\Customer}{\texttt{Customer}\xspace}
\newcommand{\Region}{\texttt{Region}\xspace}
\newcommand{\rid}{\texttt{\textup{rid}}\xspace}
\newcommand{\rname}{\texttt{\textup{rname}}\xspace}
\newcommand{\rnations}{\texttt{\textup{rnations}}\xspace}
\newcommand{\Nation}{\texttt{Nation}\xspace}
\newcommand{\nid}{\texttt{\textup{nid}}\xspace}
\newcommand{\nname}{\texttt{\textup{nname}}\xspace}
\newcommand{\ncusts}{\texttt{\textup{ncusts}}\xspace}
\newcommand{\oid}{\texttt{\textup{oid}}\xspace}
\newcommand{\RNCOP}{\texttt{\textup{RNCOP}}\xspace}
\newcommand{\cogroup}{\texttt{\kw{COGROUP}}\xspace}

\newcommand{\nrcagg}{\kw{NRC}}
\newcommand{\nrcaggshr}{\nrcagg^{Lbl+\lambda}}
\newcommand{\nrcaggshrmat}{\nrcagg^{Lbl}}

\newcommand{\kwd}[1]{\texttt{\textup{#1}}\xspace}
\newcommand{\idfor}{\kwd{for}}
\newcommand{\idin}{\kwd{in}}
\newcommand{\idunion}{\kwd{union}}
\newcommand{\idif}{\kwd{if}}
\newcommand{\idthen}{\kwd{then}}
\newcommand{\idelse}{\kwd{else}}
\newcommand{\idlet}{\kwd{let}}
\newcommand{\idmatch}{\kwd{match}}

\newcommand{\ckwd}[1]{\kwd{\color{blue!80!black} #1}}
\newcommand{\cidfor}{\ckwd{\idfor}}
\newcommand{\cidin}{\ckwd{\idin}}
\newcommand{\cidunion}{\ckwd{\idunion}}
\newcommand{\cidif}{\ckwd{\idif}}
\newcommand{\cidthen}{\ckwd{\idthen}}
\newcommand{\cidelse}{\ckwd{\idelse}}
\newcommand{\cidlet}{\ckwd{\idlet}}
\newcommand{\cidmatch}{\ckwd{\idmatch}}

\newcommand{\getkw}{\kwd{get}}
\newcommand{\dedup}{\kwd{dedup}}
\newcommand{\groupby}{\kwd{groupBy}}
\newcommand{\sumby}{\kwd{sumBy}}
\newcommand{\lookup}{\kwd{Lookup}}
\newcommand{\dicttreeunion}{\kwd{DictTreeUnion}}
\newcommand{\newlabelof}{\kwd{NewLabel}}
\newcommand{\matlookup}{\kwd{MatLookup}}
\newcommand{\flatc}{\textup{F}}
\newcommand{\dictc}{\textup{D}}
\newcommand{\shredf}[1]{\mathcal{F}(#1)}
\newcommand{\shredd}[1]{\mathcal{D}(#1)}
\newcommand{\FUN}{\kwd{fun}}
\newcommand{\CHILD}{\kwd{child}}

\newcommand{\groupatt}{\kwd{group}}
\newcommand{\labelatt}{\kwd{label}}
\newcommand{\valueatt}{\kwd{value}}

\newcommand{\NetworkFull}{$\kw{BN}_1$\xspace}
\newcommand{\SeqOnto}{$\kw{BF}_3$\xspace}

\newcommand{\olabelatt}{\kwd{olabel}}
\newcommand{\clabelatt}{\kwd{clabel}}
\newcommand{\nlabelatt}{\kwd{nlabel}}
\newcommand{\rlabelatt}{\kwd{rlabel}}

\newcommand{\topexpr}{\kw{TopExpr}}
\newcommand{\topbag}{\kw{TopBag}}
\newcommand{\labeldomain}{\kw{LabDomain}}

\newcommand{\bagunion}{\uplus}

\newcommand{\vshred}{\kw{vshred}}
\newcommand{\vunshred}{\kw{vunshred}}
\newcommand{\labelcode}{\kw{idgen}}

\newcommand{\matdict}{\kw{MatDict}}
\newcommand{\shredalg}{\kw{ShredQ}}
\newcommand{\sysname}{\kw{BINDS}}
\newcommand{\abinterp}{\kw{AbInterp}}
\newcommand{\langeq}{==}
\newcommand{\join}{\bowtie}
\newcommand{\coloneq}{~:=~}
\newcommand{\ateq}{:=}
\newcommand{\leteq}{:=}
\newcommand{\assigneq}{\Leftarrow}
\newcommand{\sng}[1]{\{ #1 \}}

\newcommand{\linearize}{\kw{Linearize}\xspace}
\newcommand{\materialize}{\kw{Materialize}\xspace}
\newcommand{\translate}{\kw{Translate}\xspace}
\newcommand{\bagtomatdict}{\texttt{BagToDict}\xspace}
\newcommand{\agg}{\kw{AGG}\xspace}
\newcommand{\heavykeys}{\texttt{heavyKeys}\xspace}

\newcommand{\oplang}{\kw{OPL}}
\newcommand{\totalmult}{\kw{TotalMult}}
\newcommand{\topout}{\kw{OutTopSubq}}
\newtheorem{proposition}{Proposition}
\newcommand{\renaming}{\kw{Relabeling}}
\newcommand{\relabeling}{\renaming}
\newcommand{\mapunion}{\kw{MapUnion}}
\newcommand{\topdict}{\kw{TopDict}}
\newcommand{\fulldict}{\dict}
\newcommand{\inputlabels}{\kw{InputLabels}}
\newcommand{\nrcplus}{NRC^+}
\newcommand{\nrc}{\kw{NRC}}
\newcommand{\freevars}{\kw{FreeVars}}
\newcommand{\transformbag}{\kw{TransformQueryBag}}
\newcommand{\transformtuple}{\kw{TransformQueryTuple}}
\newcommand{\transformtupleelement}{\kw{TransformQueryTupleElement}}
\newcommand{\pretransformbag}{\kw{TransformQueryBagAux}}
\newcommand{\outputlabels}{\kw{OutputLabels}}
\newcommand{\inputdict}{\kw{InputDict}}
\newcommand{\extractfromlabel}{\kw{Extrctval}}
\newcommand{\extracttag}{\kw{Extrcttag}}
\newcommand{\extract}{\extractfromlabel}
\newcommand{\extractfromlabelall}{\kw{ExtractFromLabelAll}}
\newcommand{\simplenrc}{\kw{SimpleNRC}}
\newcommand{\simplenrclet}{\kw{SimpleNRC}^{\kw{Let}}}
\newcommand{\nextmaximal}{\kw{NextMaximal}}
\newcommand{\mult}{\kw{Mult}}
\newcommand{\maximalsubexpressions}{\kw{MaxSub}}
\newcommand{\parent}{\kw{Parent}}
\newcommand{\typeof}{\kw{TypeOf}}

\newcommand{\variants}{\kw{Variants}}
\newcommand{\gensingleton}{\kw{WeightedSing}}
\newcommand{\flatten}{\kw{Flatten}}
\newcommand{\indextype}{\kw{IndexType}}
\newcommand{\indicesof}{\kw{IndicesOf}}
\newcommand{\evalexpr}[2]{{\llbracket #1\rrbracket}_{#2}}
\newcommand{\shredobj}{\kw{Shredded}}
\newcommand{\shreddedvars}{\kw{ShreddedVars}}
\newcommand{\outcome}[3][]{\ensuremath{\llbracket#2\ifstrempty{#1}{}{\mid#1}\rrbracket_{#3}}}
\newcommand{\interpret}[2]{\outcome{#1}{#2}}
\newcommand{\IdFor}{\kw{For}}
\newcommand{\emptytype}{()}
\newcommand{\IdUnion}{\kw{Union}}
\newcommand{\IdIn}{\kw{In}}
\newcommand{\queryunion}{\uplus}
\newcommand{\tuple}[1]{{\langle#1\rangle}}
\newcommand{\pred}{pred}
\newcommand{\emptybag}{\{\}_{\bagtype(T)}}
\newcommand{\emptybagq}{\emptyset_{\bagtype(F)}}
\newcommand{\emptybaggen}[1]{\emptyset_{\bagtype(#1)}}
\newcommand{\emptysetq}{\emptyset_{\settype(T)}}
\def\lBrack{\lbrack\!\lbrack}
\def\rBrack{\rbrack\!\rbrack}
\newcommand{\Bracks}[1]{\lBrack#1\rBrack}
\newcommand{\<}{\langle}
\renewcommand{\>}{\rangle}

\newcommand{\datalogarrow}{ :=}
\newcommand{\aschema}{\mathcal{S}}
\newcommand{\powerset}[1]{\mathcal{P}(#1)}
\newcommand{\LFPP}[3]{\LFP_{#1,#2}^{#3}}
\newcommand{\LFPA}[2]{\LFP_{#1,#2}}
\newcommand{\GFPA}[2]{\GFP_{#1,#2}}
\newcommand{\nf}{\text{strict normal form}\xspace}
\newcommand{\logic}[1]{\textup{\small #1}\xspace}
\renewcommand{\vec}[1]{\boldsymbol{#1}}
\newcommand{\V}{\mathcal{V}}
\newcommand{\cA}{\mathcal{A}}
\newcommand{\cB}{\mathcal{B}}
\newcommand{\cC}{\mathcal{C}}
\newcommand{\cD}{\mathcal{D}}
\newcommand{\cE}{\mathcal{E}}
\newcommand{\cF}{\mathcal{F}}
\newcommand{\cG}{\mathcal{G}}
\newcommand{\cM}{\mathcal{M}}
\newcommand{\cN}{\mathcal{N}}
\newcommand{\cO}{\mathcal{O}}
\newcommand{\cS}{\mathcal{S}}
\newcommand{\cT}{\mathcal{T}}
\newcommand{\cU}{\mathcal{U}}
\newcommand{\cV}{\mathcal{V}}
\newcommand{\cW}{\mathcal{W}}
\newcommand{\cH}{\mathcal{H}}
\newcommand{\cI}{\mathcal{I}}
\newcommand{\cP}{\mathcal{P}}
\newcommand{\cQ}{\mathcal{Q}}
\newcommand{\A}{\mathbb{A}}
\newcommand{\B}{\mathbb{B}}
\newcommand{\C}{\mathbb{C}}
\newcommand{\D}{\mathbb{D}}
\newcommand{\U}{\mathbb{U}}
\newcommand{\N}{\mathbb{N}}
\newcommand{\Ninf}{\mathbb{N_\infty}}
\newcommand{\fA}{\mathfrak{A}}
\newcommand{\fB}{\mathfrak{B}}
\newcommand{\fU}{\mathfrak{U}}
\newcommand{\fM}{\mathfrak{M}}
\newcommand{\set}[1]{\left\{{#1}\right\}}
\newcommand{\sset}[1]{\{{#1}\}}
\newcommand{\size}[1]{\lvert #1 \rvert}
\newcommand{\width}[1]{\operatorname{width}( #1 )}
\newcommand{\restrict}[2]{ #1\restriction_{#2} }
\let\Lslash\L
\renewcommand{\L}{\mathrm{L}}
\newcommand{\R}{\mathrm{R}}
\newcommand{\interpolant}[1]{\theta_{#1}}
\newcommand{\free}[1]{\operatorname{free}(#1)}
\newcommand{\dom}[1]{\operatorname{dom}(#1)}
\newcommand{\gn}{\textup{GN}\xspace}
\newcommand{\gnfo}{\logic{GNF}}
\newcommand{\gso}{\logic{GSO}}
\newcommand{\mso}{\logic{MSO}}
\newcommand{\gnf}{\gnfo}
\newcommand{\gf}{\logic{GF}}
\newcommand{\lfp}{\logic{LFP}}
\newcommand{\unfo}{\logic{UNF}}
\newcommand{\fo}{\logic{FO}}
\newcommand{\unf}{\unfo}
\newcommand{\Lmu}{\textup{L}_{\mu}}
\newcommand{\gnfpup}{\logic{GNFP-UP}}
\newcommand{\unfpup}{\logic{UNFP-UP}}
\newcommand{\gnfpplus}{\gnfpup}
\newcommand{\unfpplus}{\unfpup}
\newcommand{\GFP}{\operatorname{\bf gfp}}
\newcommand{\gfp}{\logic{GFP}}
\newcommand{\LFP}{\operatorname{\bf lfp}}
\newcommand{\unravel}[1]{\cT(#1)}
\newcommand{\gnfp}{\logic{GNFP}}
\newcommand{\unfp}{\logic{UNFP}}
\newcommand{\unfpk}{\unfp^k}
\newcommand{\gnfk}{\gnf^k}
\newcommand{\unfk}{\unf^k}
\newcommand{\lfpl}{\logic{LFP}}
\newcommand{\complexity}[1]{\textsc{#1}}
\newcommand\ptime{\complexity{PTime}\xspace}
\newcommand\exptime{\complexity{ExpTime}\xspace}
\newcommand\nexptime{\complexity{NExpTime}\xspace}
\newcommand\twoexptime{\complexity{2-ExpTime}\xspace}
\newcommand{\twoexp}{\twoexptime}
\newcommand\np{\complexity{NP}\xspace}
\newcommand\conp{\complexity{co-NP}\xspace}
\newcommand{\dmodality}[1]{\langle #1 \rangle}
\newcommand{\bmodality}[1]{[ #1 ]}
\newcommand{\valuation}{\ell}
\newcommand{\codom}[1]{\operatorname{rng}(#1)}
\newcommand{\decode}[1]{\mathfrak{D}(#1)}
\newcommand{\tree}{\cT}
\newcommand{\cL}{\mathcal{L}}
\newcommand{\gnfpk}{\text{$\gnfp^k$}\xspace}
\renewcommand{\varphi}{\phi}
\newcommand{\kw}[1]{\textsc{#1}}
\newcommand{\atypesig}[3]{\kw{AT}_{#2,#3}(#1)}
\newcommand{\ExactLabel}[1]{\kw{ExactLabel}(#1)}
\newcommand{\GNLabel}[1]{\kw{GNLabel}(#1)}
\newcommand{\sigmag}{\sigma_g}
\newcommand{\sigmaprimeg}{\sigma'_g}
\newcommand{\gnl}{\gn^{l}}
\newcommand{\sigk}{\tilde{\sigma}_k}
\newcommand{\gnkinvar}{\logic{GN}^k}
\newcommand{\sgnkinvar}{\logic{BGN}^k}
\newcommand{\sunkinvar}{\logic{BUN}^k}
\newcommand{\gnlinvar}{\logic{GN}^l}
\newcommand{\gnninvar}{\logic{GN}^n}
\newcommand{\gnminvar}{\logic{GN}^m}
\newcommand{\ginvar}{\logic{G}}
\newcommand{\bunravelk}[2]{\cU_{\textup{BGN}^k[#2]}( #1 )}
\newcommand{\unravelk}[2]{\cU_{\textup{GN}^k[#2]}( #1 )}
\newcommand{\unravelm}[2]{\cU_{\textup{GN}^m[#2]}( #1 )}
\newcommand{\unravell}[2]{\cU_{\textup{GN}^l[#2]}( #1 )}
\newcommand{\sunravelk}[2]{\cU^{\textup{plump}}_{\textup{BGN}^k[#2]}( #1 )}
\newcommand{\gunravel}[2]{\cU_{\textup{G}[#2]}( #1 )}
\newcommand{\sigcode}[2]{\Sigma^{\text{code}}_{#1,#2}}
\newcommand{\dloglite}{\logic{DATALOG-LITE}}
\newcommand{\dlog}{\logic{DATALOG}}
\newcommand{\unravellingcode}{\cU_{\logictarget}(\fB)}
\newcommand{\lemmafwd}{\hyperref[lemma:forward-gso]{Lemma~Fwd}\xspace}
\newcommand{\lemmagfpbwd}[1]{\hyperref[lemma:backwards-gfp]{#1}\xspace}
\newcommand{\lemmagnfpbwd}[1]{\hyperref[lemma:backwards-gnfp]{#1}\xspace}
\newcommand{\logictarget}{\cL_1}
\newcommand{\sigtarget}{\sigma'}
\newcommand{\sigoriginal}{\sigma}
\newcommand{\Bekic}{Beki\v{c}\xspace}
\newcommand{\convertnf}[1]{\operatorname{convert}(#1)}
\newcommand{\indices}[1]{\textup{names}(#1)}
\newcommand{\dup}{\uparrow}
\newcommand{\dstay}{0}
\newcommand{\ddown}{\downarrow}
\newcommand{\Pri}{\mathrm{Pri}}
\newcommand{\Dir}{\mathrm{Dir}}
\newcommand{\gameonstart}[3]{\cG(#1, #2, #3)}
\newcommand{\gameon}[2]{\cG(#1, #2)}
\newcommand{\Sigmap}{\Sigma_p}
\newcommand{\Sigmaa}{\Sigma_a}
\newcommand{\QE}{Q_{E}}
\newcommand{\QA}{Q_{A}}
\newcommand{\simgn}[1]{\sim_{#1}}
\newcommand{\subsetgn}[1]{\subseteq_{#1}}
\newcommand{\remove}[2]{\text{remove}_{#2}(#1)}
\newcommand{\cJ}{\mathcal{J}}
\newcommand{\namesk}{U_k}
\newcommand{\atype}[2]{\kw{AT}_{#2}(#1)}
\newcommand{\propsguardedneg}[1]{\kw{GNProps}(#1)}
\newcommand{\propsneg}[1]{\kw{NProps}(#1)}
\newcommand{\bagextend}[1]{\kw{BagLabels}(#1)}
\newcommand{\GuardedLabels}{\kw{GLabels}}
\newcommand{\GuardedLabelsNeighbor}[2]{\kw{GLabels}(#1,#2)}
\newcommand{\IntNodeLabels}{\kw{InterfaceLabels}}
\newcommand{\BagNodeLabels}{\kw{BagLabels}}
\newcommand{\EdgeLabels}{\kw{Edges}}
\newcommand{\EdgeLabelsRestrict}[1]{\kw{Edges}(#1)}
\newcommand{\NodeLabels}{\kw{NodeLabels}}
\newcommand{\tforward}[1]{ {#1}^{\rightarrow} }
\newcommand{\tbackward}[1]{ {#1}^{\leftarrow} }
\newcommand{\struct}[1]{\mathfrak{D}(#1)}
\newcommand{\readymode}{\text{select}}
\newcommand{\waitmode}{\text{move}}
\newcommand{\views}{\kw{Views}}
\newcommand{\view}{I}
\newcommand{\dneighbor}{\updownarrow}
\newcommand{\strategy}{\zeta}
\newcommand{\correct}[1]{\textup{correct}(#1)}
\newcommand{\correctr}[1]{\textup{correct}_{r}(#1)}
\newcommand{\Fpart}[1]{\textup{match}(#1)}
\newcommand{\id}{\textup{id}}
\newcommand{\consistencyform}{\phi_{\textup{consistent}}}
\newcommand{\elem}[1]{\operatorname{elem}(#1)}
\newcommand{\guardg}[2]{\operatorname{guard}^{\sigmag}_{#1}(#2)}
\newcommand{\Guarded}[1]{\kw{Gdd}(#1)}
\newcommand{\GuardedDom}[1]{\guardedg_{\dom{#1}}} 
\newcommand{\GuardedRng}[1]{\guardedg_{\codom{#1}}} 
\newcommand{\GuardDom}[2]{\operatorname{guard-dom}(#1,#2)} 
\newcommand{\GuardRng}[2]{\operatorname{guard-rng}(#1,#2)}
\newcommand{\phiL}{\phi_{\L}}
\newcommand{\phiR}{\phi_{\R}}
\newcommand{\cform}[2]{\text{consistent}_{#1,#2}}
\newcommand{\sigR}{\sigma_{\R}}
\newcommand{\fG}{\mathfrak{G}}
\newcommand{\gnnf}{\text{\textup{GN}-normal form}\xspace}
\newcommand{\myparagraph}[1]{{\bf #1.}}
\newcommand{\true}{\kw{True}}
\newcommand{\false}{\kw{False}}

\newcommand{\mattranstwo}[2]{\mathcal{M}\lBrack#1\rBrack_{#2}}
\newcommand{\mattransbasic}[2]{\mathcal{M}\lBrack#1\rBrack_{#2}}
\newcommand{\mattrans}[3]{\mattranstwo{#1}{#2,#3}}

\newcommand{\Flat}{\kw{Flat}\xspace}
\newcommand{\Fplus}{$\kw{Flat}^+$\xspace}
\newcommand{\Fpplus}{\kw{Standard} } 
\newcommand{\bioendtoend}{\kw{E2E}}
\newcommand{\Stepi}{$\kw{Step}_1$\xspace}
\newcommand{\Stepii}{$\kw{Step}_2$\xspace}
\newcommand{\Stepiii}{$\kw{Step}_3$\xspace}
\newcommand{\Stepiv}{$\kw{Step}_4$\xspace}
\newcommand{\Stepv}{$\kw{Step}_5$\xspace}
\newcommand{\Ci}{$\kw{C}_1$\xspace}
\newcommand{\Cii}{$\kw{C}_2$\xspace}
\newcommand{\Ciii}{$\kw{C}_3$\xspace}
\newcommand{\OccurFull}{$\kw{BN}_2$\xspace}
\newcommand{\CNVFull}{$\kw{BF}_2$\xspace}
\newcommand{\ExprFull}{$\kw{BF}_1$\xspace}
\newcommand{\Network}{$\kw{BN}_1$\xspace}

\newcommand{\superunshred}{\kw{U}}
\newcommand{\Shred}{\kw{Shred}\xspace}
\newcommand{\Unshred}{\kw{Unshred}\xspace}
\newcommand{\Shrdom}{$\Shred^{\kw{Dom}}$\xspace}
\newcommand{\Fskew}{$\Fpplus_{\kw{skew}}$\xspace}
\newcommand{\Sskew}{$\Shred_{\kw{skew}}$\xspace}
\newcommand{\Uskew}{$\Shred_{\kw{skew}}^{+ \superunshred}$\xspace}
\newcommand{\Sdskew}{$\Shred^{\kw{Dom}}_{\kw{skew}}$\xspace}

\DeclareRobustCommand{\ojoin}{\rule[0.10ex]{.3em}{.4pt}\llap{\rule[1.40ex]{.3em}{.4pt}}}
\newcommand{\leftouterjoin}{\mathrel{\ojoin\mkern-6.5mu\Join}}
\newcommand{\rightouterjoin}{\mathrel{\Join\mkern-6.5mu\ojoin}}
\DeclareRobustCommand{\ounnest}{\rule[2.0ex]{.3em}{.4pt}\llap{\rule[1.0ex]{.3em}{.4pt}}}
\newcommand\textequal{%
 \rule[.3ex]{3pt}{0.4pt}\llap{\rule[.7ex]{3pt}{0.4pt}}}
\newcommand{\outerunnest}{\mathrel{\textequal\!\mu}}

\newtheorem{example}{Example}

\newcommand{\revcolor}{black}
\newcommand{\revision}[1]{{\color{\revcolor} #1}}
\newcommand{\revisionomit}[1]{} 

\newcommand{\Occur}{\texttt{Occurrences}\xspace}
\newcommand{\OccurGroup}{\texttt{OccurGrouped}\xspace}
\newcommand{\OccurJoin}{\texttt{OccurCNVJoin}\xspace}
\newcommand{\OccurAgg}{\texttt{OccurCNVAgg}\xspace}
\newcommand{\CopyNum}{\texttt{CopyNumber}\xspace}
\newcommand{\Pathway}{\texttt{Pathway}\xspace}
\newcommand{\GeneExpr}{\texttt{GeneExpression}\xspace}
\newcommand{\Biospec}{\texttt{Samples}\xspace}
\newcommand{\SOImpact}{\texttt{SOImpact}\xspace}
\newcommand{\MNetwork}{\texttt{MappedNetwork}\xspace}
\newcommand{\FNetwork}{\texttt{FNetwork}\xspace}
\newcommand{\SNetwork}{\texttt{SampleNetwork}\xspace}
\newcommand{\Biomart}{\texttt{Biomart}\xspace}
\newcommand{\Hybrid}{\texttt{HybridMatrix}\xspace}
\newcommand{\Effect}{\texttt{EffectMatrix}\xspace}
\newcommand{\Connect}{\texttt{ConnectMatrix}\xspace}
\newcommand{\Connects}{\texttt{Connectivity}\xspace}
\newcommand{\aid}{\texttt{\textup{aid}}\xspace}
\newcommand{\aliquot}{\texttt{\textup{aliquot}}\xspace}
\newcommand{\sample}{\texttt{\textup{sample}}\xspace}
\newcommand{\gid}{\texttt{\textup{gid}}\xspace}
\newcommand{\gene}{\texttt{\textup{gene}}\xspace}

\newcommand{\protein}{\texttt{\textup{protein}}\xspace}
\newcommand{\impact}{\texttt{\textup{impact}}\xspace}
\newcommand{\conseq}{\texttt{\textup{conseq}}\xspace}
\newcommand{\cnum}{\texttt{\textup{cnum}}\xspace}

\newcommand{\trans}{\texttt{\textup{transcripts}}\xspace}
\newcommand{\cands}{\texttt{\textup{candidates}}\xspace}
\newcommand{\nodes}{\texttt{\textup{nodes}}\xspace}
\newcommand{\conseqs}{\texttt{\textup{consequences}}\xspace}
\newcommand{\hyscores}{\texttt{\textup{scores}}\xspace}
\newcommand{\hscore}{\texttt{\textup{score}}\xspace}
\newcommand{\sift}{\texttt{\textup{sift}}\xspace}
\newcommand{\siftpred}{\texttt{\textup{siftpredict}}\xspace}
\newcommand{\poly}{\texttt{\textup{poly}}\xspace}
\newcommand{\polypred}{\texttt{\textup{polypredict}}\xspace}
\newcommand{\proj}{\texttt{\textup{project}}\xspace}
\newcommand{\nodep}{\texttt{\textup{nodeProtein}}\xspace}
\newcommand{\edges}{\texttt{\textup{edges}}\xspace}
\newcommand{\edgep}{\texttt{\textup{edgeProtein}}\xspace}
\newcommand{\score}{\texttt{\textup{score}}\xspace}
\newcommand{\genes}{\texttt{\textup{genes}}\xspace}
\newcommand{\mutations}{\texttt{\textup{mutations}}\xspace}
\newcommand{\dist}{\texttt{\textup{distance}}\xspace}
\newcommand{\case}{\texttt{\textup{case}}\xspace}
\newcommand{\contig}{\texttt{\textup{contig}}\xspace}
\newcommand{\mstart}{\texttt{\textup{start}}\xspace}
\newcommand{\mend}{\texttt{\textup{end}}\xspace}
\newcommand{\mref}{\texttt{\textup{reference}}\xspace}
\newcommand{\alt}{\texttt{\textup{alternate}}\xspace}
\newcommand{\mutid}{\texttt{\textup{mutationId}}\xspace}
\newcommand{\uid}{\texttt{\textup{uid}}\xspace}
\newcommand{\gibbs}{\texttt{\textup{gibbs}}\xspace}
\newcommand{\extern}{\texttt{\textup{external}}\xspace}
\newcommand{\interm}{\texttt{\textup{E}}\xspace}
\newcommand{\fpkm}{\texttt{\textup{fpkm}}\xspace}
\newcommand{\pathway}{\texttt{\textup{pathway}}\xspace}
\newcommand{\pathways}{\texttt{\textup{pathways}}\xspace}
\newcommand{\burden}{\texttt{\textup{burden}}\xspace}

\maketitle

\begin{abstract}
While large-scale distributed
data processing platforms have become
an attractive target for query processing, these systems are problematic
for applications that deal with nested collections.
Programmers are forced either to perform non-trivial translations 
of collection programs or to employ automated flattening procedures, 
both of which lead to performance problems. These challenges 
only worsen for nested collections with skewed cardinalities,  
where both handcrafted rewriting and automated flattening
are unable to enforce load balancing across partitions.

In this work, we propose a framework that translates a 
program manipulating nested collections  
into a set of semantically equivalent 
shredded queries that can be efficiently evaluated.
The framework employs a combination of
query compilation techniques, an efficient data representation for nested collections, and automated
skew-handling. We provide an extensive experimental evaluation,
demonstrating significant improvements provided by the framework in diverse scenarios for nested collection programs.
\end{abstract}

\section{Introduction}\label{sec:introduction}

Large-scale, distributed data processing platforms such as Spark~\cite{sparkpaper}, Flink~\cite{nephele}, and Hadoop~\cite{mapreduce} have become indispensable tools for 
modern data analysis.
Their wide adoption stems from powerful functional-style APIs that 
allow programmers to express complex analytical tasks while abstracting distributed resources and data parallelism.
These systems  use an underlying data model that allows for data to be described as a collection of tuples 
whose values may themselves be collections. 
This {\em nested data} representation arises naturally in many domains, such as web, biomedicine, and business intelligence.
The widespread use of nested data also contributed to the rise in NoSQL databases~\cite{cassandra,couchdb,mongodb}, 
where the nested data model is central.
Thus, it comes as no surprise that nested data accounts for most large-scale, structured data processing at major web companies~\cite{dremel}.

Despite natively supporting nested data, existing systems fail to harness the full potential of processing nested collections at scale. 
One implementation difficulty is the discrepancy between the tuple-at-a-time processing of local programs and the bulk processing used in distributed settings. 
Though maintaining similar APIs, local programs that use Scala collections often require non-trivial and error-prone translations to distributed Spark programs.
Beyond programming difficulties, nested data processing requires 
scaling in the presence of large or skewed inner collections. 
We next elaborate these difficulties by means of an example. 

\begin{example} \label{ex:run}
\em
Consider a variant of the TPC-H database containing a flat relation $\Part$ with information about parts and a nested relation $\COP$ with information about customers, their orders, and the parts purchased in the orders.
The relation $\COP$ stores customer names, order dates, part IDs and purchased quantities per order. The type of $\COP$ is: 

{\small
\vspace{-6pt}
\begin{align*}
\bagtype\,(\<\, & \cname:\stringtype,\, \corders:\bagtype\,(\<\, \odate: \datetype, \\[-2pt]
& \hspace{1.8cm}\oparts: \bagtype\,(\<\, \pid: \inttype,\, \lqty: \doubletype \,\>) \,\>) \,\>).
\end{align*}
\vspace{-6pt}
}

Consider now a high-level collection program
that returns for each customer and for each of their orders, the total amount spent per part name. This requires joining $\COP$ with $\Part$ on $\pid$ to get the name and price of each part and then summing up the costs per part name. We can express this using nested relational calculus~\cite{complexobj,limsoonthesis} as:

\begin{lstlisting}[language=NRC,escapechar=?]
  $\!\cidfor ~ cop ~ \cidin ~ \COP ~ \cidunion \hspace{2cm} {\color{gray}}$
  ?\vrule? $ \{\<\, \cname \ateq cop.\cname,$ 
  ?\vrule? $\phantom{\{\<\,} \corders \ateq$ 
  ?\vrule? $\tab\tab \cidfor ~ co ~ \cidin ~ cop.\corders ~ \cidunion$
  ?\vrule\tab\tab\;\;\vrule? $\{\<\, \odate \ateq co.\odate,$ 
  ?\!\!$Q$\tab\tab\hspace{0.45mm}\vrule? $\phantom{\{\<\,} \oparts \ateq \sumBy_{\hspace{0.15mm}\pname}^{\hspace{0.15mm}\total}($
  ?\vrule\tab\tab\;\;\vrule \tab\tab\tab\tab\tab\tab\tab\tab \vrule? $\cidfor ~ op ~ \cidin ~ co.\oparts ~ \cidunion$
  ?\vrule\tab$Q_{\corders}$ \tab\tab\tab\tab\hspace{0.89mm} \vrule? $\tab\cidfor ~ p ~ \cidin ~ \Part ~ \cidunion$
  ?\vrule\tab\tab\;\;\vrule \tab\tab\tab\tab\tab\tab\tab\tab \vrule? $\tab\tab\cidif ~ op.\pid == p.\pid ~ \cidthen$
  ?\vrule\tab\tab\;\;\vrule \tab\tab\tab\tab\tab\tab $Q_{\oparts}$? $\tab\,\{\<\, \pname \ateq p.\pname,$
  ?\vrule\tab\tab\;\;\vrule \tab\tab\tab\tab\tab\tab\tab\tab \vrule? $\tab\tab\tab\tab\total \ateq op.\lqty * p.\pprice \,\>\})\>\}\>\}$
\end{lstlisting}

\vspace{9pt}
The program navigates to $\oparts$ in $\COP$, joins each bag with $\Part$, computes the amount spent per part, and aggregates using 
$\sumby$ which sums up the total amount spent for each distinct part name.
The output type is:

{\small
\vspace{-6pt}
\begin{align*}
&\bagtype\,(\<\, \cname:\stringtype,\, \corders:\bagtype\,(\<\, \odate: \datetype, \\[-2pt]
& \hspace{1.2cm} \oparts: \bagtype\,(\<\, \pname: \stringtype,\, \total: \doubletype \,\>)\, \>)\, \>).\phantom{\quad\textrm{$\Box$}}
\end{align*}

}

A key motivation for our work is \emph{analytics pipelines}, which typically 
consist of a sequence of transformations with the output of one transformation 
contributing to the input of the next, as above. 
Although the final output consumed by an end user  or
application may be flat, the intermediate
nested outputs may still be important in themselves: either because
the intermediate  outputs are used in multiple follow-up transformations, or because
the pipeline is expanded and modified as the data is  explored. Our interest is thus
in \emph{programs consisting of sequence of queries, where both inputs and outputs may be nested}.


We next discuss the challenges associated with processing such examples
 using distributed frameworks.

{\bf Challenge 1: Programming mismatch.}
    The translation of collection programs from local to distributed settings is not always straightforward. 
    Distributed processing systems natively distribute collections only at the granularity of top-level tuples
    and provide no direct support for using nested loops over different distributed collections.
    To address these limitations, programmers resort to techniques that are either prohibitively expensive or error-prone.
    In our example, a common error is to try to iterate over $\Part$, which is a collection distributed across worker nodes, 
    from within the map function over $\COP$ that navigates to $\oparts$, which are collections local to one worker node. 
    Since $\Part$ is distributed across worker nodes, it cannot be referenced inside another distributed collection.
    Faced with this impossibility, programmers could either replicate $\Part$ to each worker node, which can be too expensive, or rewrite 
    to use an explicit join operator, which requires flattening $\COP$ to bring $\pid$ values needed for the join to the top level.
    A natural way to flatten  $\COP$ is by pairing each $\cname$ with every tuple from $\corders$ and subsequently with every tuple from $\oparts$; however,  
    the resulting $(\cname, \odate, \pid, \lqty)$ tuples encode incomplete information -- e.g., omit customers with no orders -- which can eventually cause an incorrect result.
    Manually rewriting such a computation while preserving its correctness is non-trivial.

{\bf Challenge 2: Input and output data representation.}
    Default top-level partitioning of nested data can lead to poor scalability, 
    particularly with few
    top-level tuples and/or large inner collections. The number of top-level 
    tuples limits the level of parallelism by requiring all data on lower levels to persist 
    on the same node, which is detrimental regardless of the size of these inner collections. 
    Conversely, large inner collections can overload worker nodes and increase data transfer costs, 
    even with larger numbers of top-level tuples.
    In our example, a small number of customers in $\COP$ may lead to poor cluster utilization because few worker nodes process the inner 
collections of $\COP$.

    To cope with limited parallelism, an alternative is to flatten nested data and redistribute processing over more worker nodes. 
    In addition to the programming challenges, unnesting collections also leads to data duplication and consequently redundancy in computation.
    Wide flattened tuples increase memory pressure and the amount of data shuffled among the worker nodes. Alleviating problems such as disk spillage and load imbalance requires rewriting of programs on an ad hoc basis, without a principled approach.

Representation  of nested \emph{outputs} from transformations is likewise an issue.
Flat representations of these outputs can result in significant inefficency in  space usage,
since  parts of the input can appear as subobjects in many places within an output.
Exploration of succinctness and efficiency for different representations of nested  output has not, to our
knowledge,  been the topic of significant study.

{\bf Challenge 3: Data skew.}
    Last but not least, data skew can significantly degrade performance of large-scale data processing, even when dealing with flat data only.
    Skew can lead to load imbalance where some workers perform significantly more work than others, prolonging run times on platforms with synchronous execution such as Spark, Flink, and Hadoop.  
    The presence of nested data only exacerbates the problem of load imbalance:
    inner collections may have skewed cardinalities -- e.g., very few customers can have very many orders -- 
    and the standard top-level partitioning places each inner collection entirely on a single worker node.
    Moreover, programmers must ensure that inner collections
    do not grow large enough to cause disk spill or crash worker nodes due to insufficient memory.
\punto
\end{example}

Prior work has addressed some of these challenges.
High-level scripting languages such as Pig Latin~\cite{piglatinpaper} and Scope~\cite{scope} ease the programming of distributed data processing systems.
Apache Hive~\cite{hive}, F1~\cite{f1google,f1query}, Dremel~\cite{dremel}, and BigQuery~\cite{bigquery} provide SQL-like languages with extensions for querying nested structured data at scale. 
Emma~\cite{emma} and DIQL~\cite{fegaras2018compile}  
support for-comprehensions and respectively SQL-like syntax via DSLs deeply-embedded in Scala,
and target several distributed processing frameworks.
But alleviating the performance problems caused by skew and manually flattening nested collections remains an open problem.

{\bf Our approach.}
To address these challenges and achieve scalable processing of nested data, 
we advocate an approach that relies on three key aspects:

\myparagraph{Query Compilation} 
When designing programs over nested data, programmers often first conceptualize desired transformations using high-level languages such as nested relational calculus~\cite{complexobj,limsoonthesis}, monad comprehension calculus~\cite{fegarasmaier}, or the collection constructs of functional programming languages (e.g., the Scala collections API).
These languages lend themselves naturally to centralized execution, thus making them the weapon of choice for rapid prototyping and testing 
 over small datasets.

To relieve programmers from manually rewriting program for scalable execution (Challenge 1), 
we propose using a compilation framework that automatically restructures programs   to distributed settings. 
In this process, the framework applies optimizations that are often overlooked by programmers such as column pruning, selection pushdown, and pushing nesting and aggregation operators past joins. These optimizations may
\revisionomit{, in particular,} significantly cut communication overheads. 


\myparagraph{Query and Data Shredding} 
To achieve parallelism beyond top-level records (Challenge 2), we argue for processing over {\em shredded} data.
In this processing model, nested collections are encoded as flat collections and
nested collection queries are translated into relational queries.
Early work on nested data \cite{vandenbussche, limsoonthesis} proposed a flattening
approach in which a nested collection is converted into a single collection containing
both top-level attributes, lower-level attributes, and identifiers encoding inner collections.
We refer to these as \emph{flattened representations} below.
We will follow later work which considers a normalized representation in which inner collections
are stored separately. Regardless of whether flattened or shredded representations are used,
it is necessary to rewrite queries over nested data to run over the encoded structure.
For example \cite{cheneyshred} proposes a transformation -- denoted there
as \emph{query shredding} -- which converts  nested relational queries to SQL queries
that can be run on top of
commercial relational processor. Closer to our work is \cite{pods16}  which proposes
a shredded data representation and query transformation, applying it
to  incremental evaluation~\cite{pods16}. Both of these works
focus on centralized setting.

The motivation for using shredding in distributed settings is twofold.
First, a shredded representation of query inputs can enable full parallel 
processing of large nested collections and their even distribution among worker nodes.
The second motivation concerns query outputs.
In the setting
of complex programs consisting of a sequence of nested-to-nested data transformations,
shredding eliminates the need for the reconstruction of intermediate nested results. Compared
to flattening, the intermediate outputs are much more succinct. In fact, shredded
outputs can even be succinct compared to a native representation, due to the ability to share
identifiers for inner bags among distinct outer bags that contain them; we refer to 
these benefits as the \emph{succinct representation} used in shredding.
The shredded 
representation  
also provides more opportunities for reduction of data through aggregation, since aggregates
can often be applied locally at a given level of the data.
When an aggregation or selection operation on the source object
can be translated into a similar operation running on only one component of a shredded representation,
then we say that the operator is \emph{localized}.

In our example, the shredded form of $\COP$ consists of one flat top-level collection containing labels in place of inner collections and two dictionaries, one for each level of nesting, associating labels with flat collections. 
The top-level collection and dictionaries are distributed among worker nodes. 
We transform the example program to compute over the shredded form of $\COP$ and produce shredded output.
The first two output levels are those from the shredded input.
Computing the last output level requires joining one dictionary of $\COP$ with $\Part$, followed by the sum aggregate; both operations can be done in a distributed fashion.
In either a flattened or a native representation, the aggregation would be performed
over a much larger object,
thus resulting in a much larger amount of shuffled data. 
The shredded output can serve as input to another constituent query in a pipeline.
If required  downstream, the nested output can be restored by joining together the computed flat collections.

\myparagraph{Skew-Resilient Query Processing} To handle data skew (Challenge 3), we propose using different evaluation plans for processing skewed and non-skewed portions of data. 
We transparently rewrite evaluation plans to avoid partitioning collections on values that appear frequently in the data, thus consequently avoid overloading few worker nodes with significantly more data. 
Our processing strategy deals with data skew appearing at the top level of distributed collections, 
but when coupled with the shredding transformation, it can also effectively handle data skew present in inner collections.


In summary,  we make the following contributions:

\begin{compactitem}
    \item We propose a compilation framework that transforms declarative programs over nested collections into plans that are executable in distributed environments. 
    \item The framework provides transformations that enable compute over shredded relations;
    this includes an extension of the symbolic shredding approach, originating in \cite{pods16}, to a larger query language, along with a \emph{materialization phase} which better suites 
    programs with nested output to be compiled to parallel programs.
    \item We introduce techniques that adapt generated plans to the presence of skew. 
    \item We develop a micro-benchmark based on TPC-H and also a benchmark based on a real-word biomedical dataset,
suitable for evaluating nesting collection programs. Using these we show that our framework can scale
    as nesting and skew increase, even when alternative methods are unable to complete at all.
\end{compactitem}
    

\section{Background}\label{sec:background}

We describe here our source language, which is based on nested relational calculus~\cite{complexobj,limsoonthesis}, and our query plan language, based on the intermediate object algebra of \cite{fegarasmaier}.

\subsection{Nested Relational Calculus (NRC)}\label{sec:src_lang}

We support a rich collection programming language $\nrcagg$ as our source language.
Programs are sequences of assignments of variables to expressions,
where the expression language  extends the standard nested relational
calculus with primitives for aggregation and deduplication.
Figure~\ref{fig:nrcagg_syn} gives the syntax of NRC. 
We work with a standard nested data model with typed data items. 
The $\nrcagg$ types are built up from the basic scalar types (integer, string, etc.), tuple type $\<a_1: T_1 \ldots a_n:T_n\>$, and bag type $\bagtype(T)$. 
For ease of presentation, we restrict the bag content $T$ to be either a tuple type or a scalar type, as shown in Figure~\ref{fig:nrcagg_syn}. 
Set types are modeled as bags with multiplicity one. 
We refer to a bag of tuples whose each attribute has a scalar type as a \emph{flat bag}.

\begin{figure}[t]
\begin{align*}
\textit{P} & ::= (\textit{var} \assigneq \textit{e})^* \\
\textit{e} &::= \; \emptybagq  \mid \{ \textit{e}\,\} \mid \getkw(\textit{e}\,)\\
& \quad\; \mid c \mid \textit{var} \mid \textit{e}.a \mid \< a_1 \ateq \textit{e}, \mydots, a_n \ateq \textit{e} \,\> \\
& \quad\; \mid \cidfor~ \textit{var}~ \cidin ~ \textit{e} ~
       \cidunion~ \textit{e}  
 \mid  \textit{e}  ~ \uplus ~ \textit{e}  \\
& \quad\; \mid \cidlet  ~ \textit{var} := \textit{e} ~ \cidin ~ \textit{e} 
\mid \textit{e} ~ PrimOp ~ \textit{e} \\
& \quad\; \mid \cidif   ~ \textit{cond} ~ \cidthen ~ \textit{e}
 \mid \dedup(\textit{e}\,) \\
& \quad\; \mid \groupby_{\textit{key}}(\textit{e}\,) \mid \sumby_{\textit{key}}^{\textit{value}}(\textit{e}\,) \\
\textit{cond} & ::= \textit{e} ~ RelOp ~ \textit{e} \mid \neg \textit{cond} \mid \textit{cond} ~ BoolOp ~ \textit{cond} \\
T & ::= S \mid C \\
C & ::= \bagtype(F) &\hspace{-1.5cm}\textit{ -- Collection Type} \\
F & ::= \<a_1: T, \ldots, a_n:T\> \mid S &\hspace{-1.5cm}\textit{ -- Flat Type}\\
S & ::= \inttype \mid \doubletype \mid \stringtype \mid \booltype \mid \datetype &\hspace{-1.5cm}\textit{ -- Scalar Type}
\end{align*}
\vspace{-6pt}
\caption{Syntax of $\nrcagg$.}\label{fig:nrcagg_syn}
\vspace{-6pt}
\end{figure}

\myeat{
\begin{small}
\begin{figure}[t]
\centering
$\infer{\emptybaggen{F}: \bagtype(F)}{}$\tab\tab
$\infer{\{ \textit{e} \,\}: \bagtype(F)}{\textit{e}: F}$\tab\tab
$\infer{\getkw(\textit{e}): F}{\textit{e}: \bagtype(F)}$
\\\vspace{0.15cm}
$\infer{\Gamma \vdash \textit{var}: T}{\textit{var}: T \in \Gamma}$ \tab
$\infer{\<a_1 \ateq \textit{e}_1, \mydots, a_n \ateq \textit{e}_n \>: \<a_1 : T_1, \mydots, a_n: T_n \>}{\textit{e}_1: T_1 \tab \mydots \tab \textit{e}_n: T_n}$ 
\\\vspace{0.15cm}
$\infer{\textit{e}.a_i: T_i}{\textit{e}: \<a_1 : T_1, \mydots, a_n: T_n \>}$ \tab\tab
$\infer{\textit{e}_1 \uplus \textit{e}_2: \bagtype(F)}{\textit{e}_1: \bagtype(F) \tab \textit{e}_2: \bagtype(F)}$
\\\vspace{0.15cm}
$\infer{\Gamma \vdash \idfor~ \textit{var}~ \idin ~ \textit{e}_1 ~
       \idunion~ \textit{e}_2: \bagtype(F_2)}{\Gamma \vdash \textit{e}_1: \bagtype(F_1) \tab \Gamma, \textit{var}: F_1 \vdash \textit{e}_2: \bagtype(F_2)}$
\\\vspace{0.15cm}
$\infer{\textit{e}_1 ~ RelOp ~ \textit{e}_2: \booltype}{\textit{e}_1: S  \tab \textit{e}_2: S}$ \tab\tab
$\infer{\textit{c}_1 ~ BoolOp ~ \textit{c}_2: \booltype}{\textit{c}_1: \booltype  \tab \textit{c}_2: \booltype}$
\\\vspace{0.15cm}
$\infer{\idif  ~ \textit{c} ~ \idthen ~ \textit{e}_1 ~\idelse ~ \textit{e}_2: T}{\textit{c}: \booltype \tab \textit{e}_1: T \tab \textit{e}_2: T}$ \tab\tab
$\infer{\dedup(\textit{e}\,): \bagtype(F)}{\textit{e}: \bagtype(F) \tab \textit{e} \text{ is flat bag}}$
\\\vspace{0.15cm}
$\infer{\Gamma \vdash \idlet  ~ \textit{var} := \textit{e}_1 ~ \idin ~ \textit{e}_2 : T_2}{\Gamma \vdash \textit{e}_1: T_1 \tab \Gamma, \textit{var}: T_1 \vdash \textit{e}_2: T_2}$ \tab\tab
$\infer{\textit{e}_1 ~ PrimOp ~ \textit{e}_2: S}{\textit{e}_1: S  \tab \textit{e}_2: S}$
\\\vspace{0.15cm}
$\infer{\groupby_{\textit{key}}(\textit{e}\,): \bagtype(\< \textit{key}: T_1, \groupatt: \bagtype(\<\textit{value} : T_2\>)\>)}{\textit{e}: \bagtype(\<\textit{key} : T_1, \textit{value} : T_2\>)}$
\\\vspace{0.15cm}
$\infer{\sumby_{\textit{key}}^{\textit{value}}(\textit{e}\,) : \bagtype(\<\textit{key}: T_1, \textit{value}: T_2\>)}{\textit{e}: \bagtype(\<\textit{key} : T_1, \textit{value} : T_2\>)}$
\\\vspace{0.15cm}
\caption{Type system of $\nrcagg$.}
\label{fig:nrcagg_type}
\end{figure}
\end{small}
}

$RelOp$ is a comparison operator on scalars (e.g., $\langeq$, $\leq$), $PrimOp$ is a primitive function
on scalars (e.g., $+$, $*$), and $BoolOp$ is a boolean operator (e.g., $\&\&$, $||$). 
A simple term can be a constant, a variable, or an expression {\em e.a}. 
Variables can  be free, e.g. representing input objects, or 
can be  introduced in \idfor or \idlet constructs. 
$\sng{\textit{e}\,}$ takes the expression \textit{e} and returns a singleton bag. 
Conversely, $\getkw(\textit{e}\,)$ takes a singleton bag and returns
its only element; if $e$ is  empty or has more than one element, $\getkw$ returns a default value.
$\emptybaggen{F}$ returns an empty bag. 

The if-then-else constructor for expressions of bag type can be expressed using the if-then and union constructors.

$\dedup(\textit{e}\,)$ takes the bag \textit{e} and returns a bag with the same elements, but with
all multiplicities changed to one.
We impose a restriction that will be useful in our
query shredding methods (Section \ref{sec:pipeline_shredding}): the input to $\dedup$ must be a flat bag.

$\groupby_{key}(\textit{e}\,)$ groups the tuples of bag $\textit{e}$ by  a collection of attributes $\textit{key}$ (in the figure
we assume a single attribute for simplicity) and for each distinct values of $\textit{key}$, produces a bag named $\textsc{group}$ containing the tuples from $\textit{e}$ with the $\textit{key}$ value, 
projected on the remaining  attributes.

$\sumby_{\textit{key}}^{\textit{value}}(\textit{e}\,)$ groups the tuples of bag $\textit{e}$ by the values of their
 $\textit{key}$ attributes and for each distinct value of $\textit{key}$, sums up the  attributes $\textit{value}$ of the tuples with the $\textit{key}$ value.

In both $\groupby$ and $\sumby$ operators, we restrict the grouping attributes $\textit{key}$ to be flat.

\subsection{Plan Language}\label{sec:plan_language}

Our framework also makes use of 
a language with algebraic operators, variants of those described in~\cite{fegarasmaier}.

Our plan language includes selection $\sigma$, projection $\pi$, join $\join$, and left outer join $\leftouterjoin$, as known from relational algebra. 

The unnest operator $\mu^{a}$ takes a nested bag with top-level bag-valued attribute $a$ and pairs each tuple of the outer bag with each tuple of $a$, while projecting away $a$.
The outer-unnest operator $\outerunnest^{a}$ is a variant of $\mu^{a}$ that in addition
extends each tuple from the outer bag with a unique ID and
pairs such tuples with {\NULL} values when $a$ is the empty bag.

The nest operator ${\Gamma^{\textit{agg}}\,}_{key}^{value}$ defines a key-based reduce, parameterized by the aggregate function $\textit{agg}$, which could be addition ($+$) or bag union ($\bagunion$). 
This operator also casts $\NULL$ values introduced by the outer operators into $0$ for the $+$ aggregate and the empty bag for the $\bagunion$ aggregate.

\myeat{

\begin{figure}[t]
\begin{center}
\begin{tabular}{ll@{}}
\toprule
Plan Operator & Name \\[3pt]
\midrule
$\sigma_{p}$ &
Selection \\[3pt]
$\pi_{\,a_1,\ldots,a_k}$ &
Projection \\[3pt]
$\join_{a}$ &
Equijoin \\[3pt]
$\leftouterjoin_{a}$ &
Left outer join \\[3pt]
$\mu^{a}$ &
Unnest \\[3pt]
$\outerunnest^{a}$ &
Outer-unnest \\[3pt]
${\Gamma^{\bagunion,\,}}^{value}_{key}$  &
Nest\\[3pt]
${\Gamma^{+\,}}^{value}_{key}$ &
Sum aggregate\\[3pt]
\bottomrule
\end{tabular}
\end{center}
\caption{Plan language operators.}\label{fig:plan_operators}
\end{figure}
}

\section{Compilation Framework}\label{sec:pipeline_basic}

\begin{figure}[t]
 \center
 \includegraphics[width=0.74\linewidth]{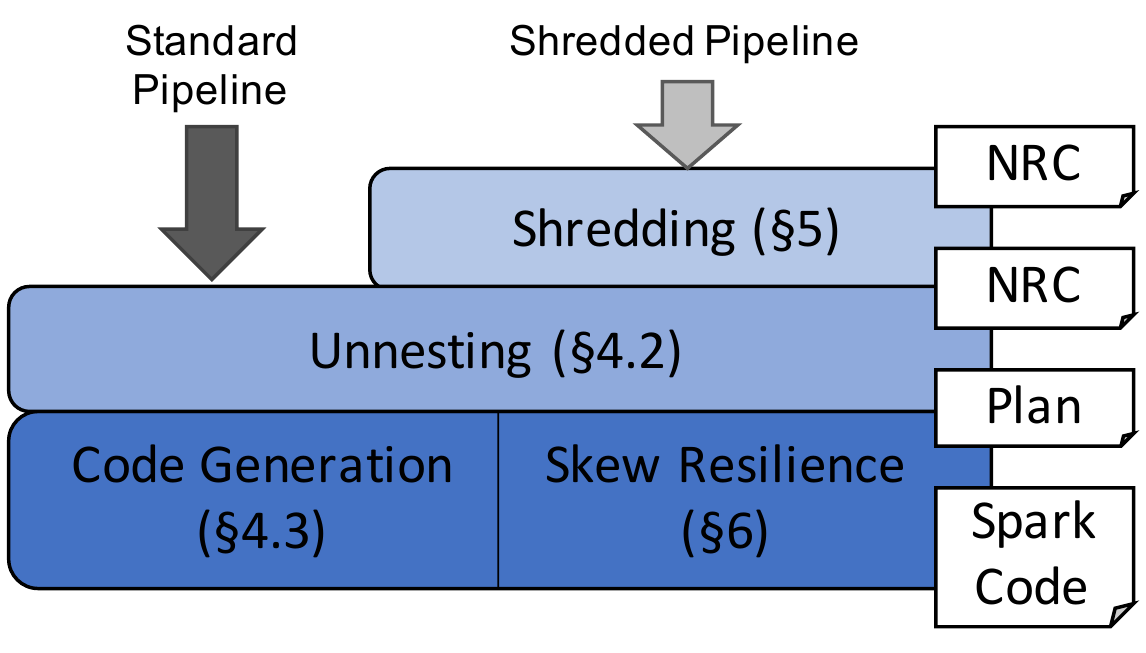}
 \caption{System architecture.}
 \label{fig:overview}
 \vspace{-12pt}
\end{figure}

Our framework transforms high-level programs into programs optimized for distributed execution. 
It can generate code operating directly over nested data,
as well as code operating over shredded data.
Figure~\ref{fig:overview} shows our architecture.

This section focuses on the standard compilation method, which
uses a variant of the unnesting algorithm~\cite{fegarasmaier} to transform an 
input NRC program into an algebraic plan expressed in our
plan language. 
A plan serves as input to the code generator producing executable code for a chosen target platform. For this paper, we consider Apache Spark as the target platform; other platforms such as Apache Flink, Cascading, and Apache Pig could also be supported. 

The shredded compilation adds on a shredding phase that transforms an NRC program into a shredded NRC program that operates over flat collections with shredding-specific labels, dictionaries, and dictionary lookup operations. 
Section~\ref{sec:pipeline_shredding} describes the shredded data representation and the shredding algorithm. 
The shredded program then passes through the unnesting and code generation phases, as in the standard method.

Both compilation routes can produce skew-resilient code during code generation. 
Our skew-resilient processing first uses a lightweight sampling mechanism to detect skew in data. Then, for each operator affected by data skew (e.g., joins), our framework generates different execution strategies that process non-skewed and skewed parts of the data. 
Section~\ref{sec:skew} describes our skew-resilient processing technique.

We describe here our standard compilation route for producing distributed 
programs from high-level queries.  
It relies
on existing techniques from unnesting \cite{fegarasmaier} and 
code-generation for distributed processing platforms
(e.g., \cite{paxquery}).  
It generates code that automatically inserts IDs and processes NULL values 
-- techniques that play a role in the manually-crafted solutions
discussed in the introduction --  while
adding optimizations and guaranteeing correctness. Further, 
this route forms the basis for our shredded variant and skew-resilient processing module.

\subsection{Unnesting} \label{subsec:unnest}

The unnesting stage translates each constituent query in an $\nrcagg$ program into a query plan consisting of the operators of the plan language, following the prior work~\cite{fegarasmaier}, 
to facilitate code generation for batch processing frameworks.

The unnesting algorithm starts from the outermost level of a query and recursively builds up a subplan $subplan^{\textit{e}}$ for each nested expression $e$. 
The algorithm detects joins written as nested loops with equality conditions and
also translates each $\idfor$ loop over a bag-typed attribute $x.a$ to an unnest operator $\mu^a$. 

The algorithm enters a new nesting level when a tuple expression contains a bag expression.
When at a non-root level, the algorithm generates the outer counterparts of joins and unnests. 
At each level, the algorithm maintains a set $\mathcal{G}$ of attributes used as a prefix 
in the key parameter of any generated nest operator at that level.
The set $\mathcal{G}$ is initially empty. Before entering a new nesting level, the algorithm expands $\mathcal{G}$ to include the unique ID attribute and other output attributes from the current level.
The $\sumby_{\textit{key}}^{\textit{value}}(\textit{e}\,)$ operator
translates to 
${\Gamma^{+}\,}_{\mathcal{G}, \textit{key}\,}^{\textit{value}\,}(subplan^{\textit{e}})$, while 
$\groupby_{\textit{key}}(\textit{e}\,)$ translates 
to ${\Gamma^{\bagunion}\,}_{\mathcal{G}, \textit{key}\,}^{\textit{value}}(subplan^{\textit{e}})$, where $value$ represents the remaining non-key attributes in $e$. These $\Gamma$ operators connect directly to the subplan built up from unnesting $e$.


The example below details the plans produced by the unnesting algorithm on the running example. We will see that while unnesting addresses
the programming mismatch, it highlights the  challenges introduced by flattening mentioned in the introduction.

\begin{figure}[t]
\begin{minipage}{\textwidth}
\begin{minipage}{0.63\textwidth}
 \includegraphics[width=0.43\textwidth]{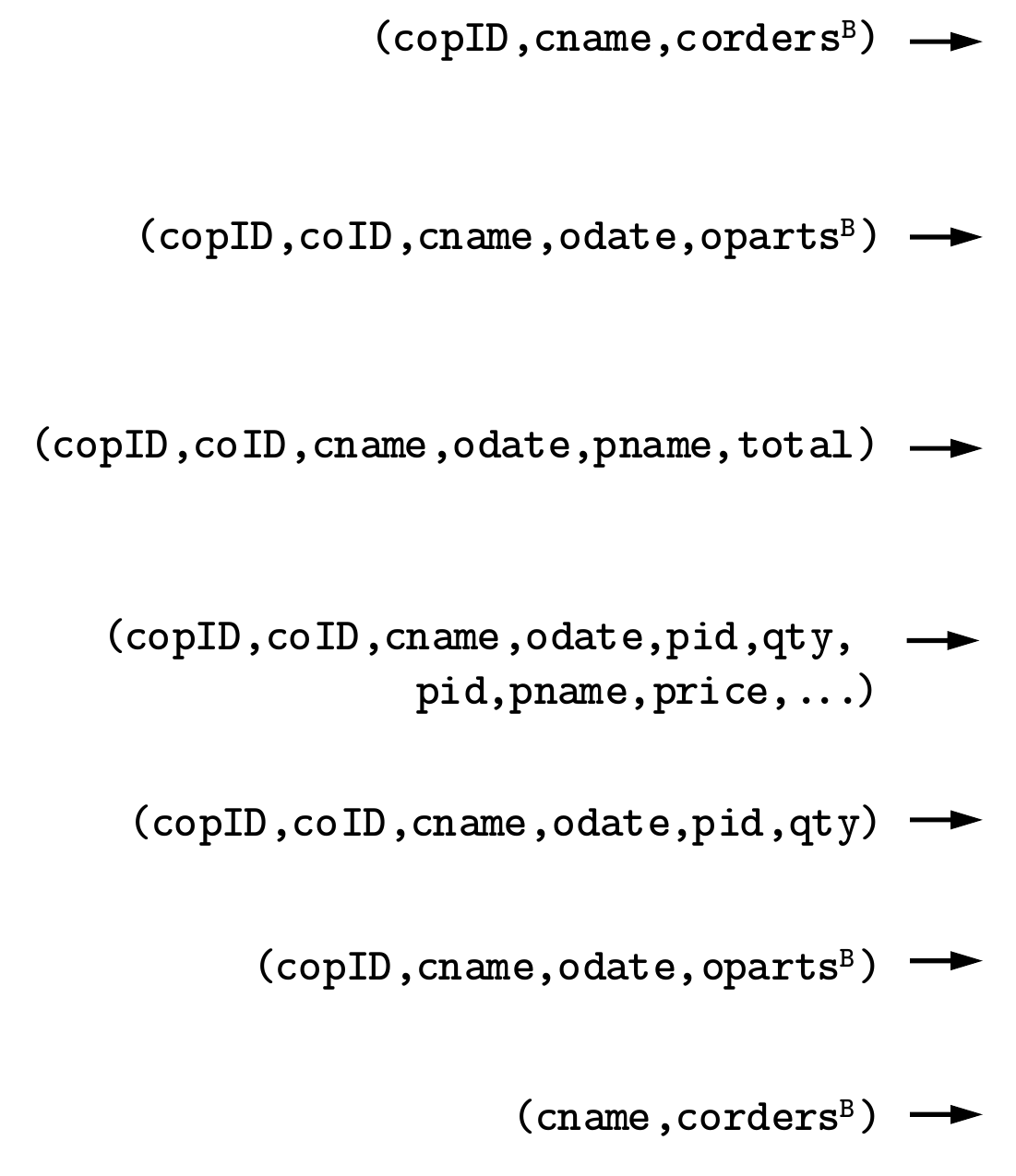}
 \vspace{3pt}
\end{minipage}
\hspace{-6.5cm}
\begin{minipage}{0.21\textwidth}
\begin{forest}
for tree={
  s sep=10mm,
  l sep = 0.75em, 
  l = 0,
},
[$\pi_{\cname, \corders}$
  [${\mbox{\large$\Gamma^{\bagunion\,}$}}^{\odate, \oparts}_{\texttt{copID}, \cname}$
    [${\mbox{\large$\Gamma^{\bagunion\,}$}}^{\pname, \total}_{\texttt{copID}, \texttt{coID}, \cname, \odate}$
      [${\mbox{\large$\Gamma^{+\,}$}}^{\lqty*\pprice}_{\texttt{copID}, \texttt{coID}, \cname, \odate, \pname}$
        [ $\leftouterjoin_{\pid}$
          [$\outerunnest^{\oparts}$
            [$\outerunnest^{\corders}$
              [$\COP$]]]
          [$\Part$]
          ]]]]]
\end{forest}
\end{minipage}
\end{minipage}
\vspace{0.5em}
\caption{A query plan for the running example with output types.
Bag types are annotated with ${}^B$.} \label{fig:unnest_plan}
\end{figure}

\begin{example} \label{ex:run_unnest}
\em
Figure~\ref{fig:unnest_plan} shows a query plan output by an unnesting
algorithm for the running example, along with the output type of each operator. For readability, we omit renaming operators from the query plan. 

Starting at the bottom left,
the plan performs an outer-unnest on  $cop.\corders$.
To facilitate the upcoming nest operation at $\corders$, 
the outer-unnest associates a unique ID to each tuple $cop$.
The set $\mathcal{G}$ of grouping attributes at this point is $\{\texttt{copID}, \cname \}$. 

The plan continues with an 
outer-unnest for $co.\oparts$, extending each tuple $co$ with a unique ID. 
The set of grouping attributes is $\{\texttt{copID}, \texttt{coID}, \cname, \odate \}$. 
At this point the input is fully flattened.
The next operation is an outer join between the plan built up to this point and the $\Part$ relation. 
This comprises the subplan corresponding to the input for the $\sumBy$ expression in the source query.

The $\sumBy$ expression is translated to a sum aggregate $\Gamma^{+}$ with the key consisting of the grouping attributes from 
the first two levels and $p.\pname$ from the key attribute of the $\sumBy$. 
The second-level subplan continues with a nest operation  $\Gamma^{\bagunion}$
created from entering the second level at $co.\oparts$ with grouping attributes $\{\texttt{copID}, \texttt{coID}, \cname, \odate \}$,
followed by another nest operation with grouping attributes $\{\texttt{copID}, \cname\}$ and a projection on $(\cname,\corders)$.

The unnesting algorithm may produce suboptimal query plans.
We can optimize the plan from Figure~\ref{fig:unnest_plan} by projecting away unused attributes from the $\Part$ relation. 
We can push the sum aggregate $\Gamma^{+}$ past the join to compute partial sums of $\lqty$ values over the 
output of $\outerunnest^{\oparts}$, grouped by $\{\texttt{copID}, \texttt{coID}, \cname, \odate, \pid\}$. However, a similar local aggregation over $\Part$ brings no benefit since $\pid$ is the primary key of $\Part$.
\punto
\end{example}

\myeat{
The example below details the unnesting algorithm on the running example. We will see that while unnesting addresses
the programming mismatch, it highlights the  challenges introduced by flattening mentioned in the introduction.

\begin{example} \label{ex:run_unnest}
\em
Figure~\ref{fig:unnest_plan} shows a query plan for the running example and the output type of each operator. For readability, we omit renaming operators from the query plan. 

The unnesting algorithm starts at the first $\idfor$ expression,
entering a new level at the $\corders$ attribute.
The algorithm is at the first non-root level and returns an outer-unnest at the $\idfor$ expression associated to $cop.\corders$.
To facilitate the upcoming nest operation at $\corders$, 
the outer-unnest associates a unique ID to each tuple $cop$.
The set $\mathcal{G}$ of grouping attributes at this point is $\{\texttt{copID}, \cname \}$.

The algorithm then enters the second non-root level and returns an 
outer-unnest for $co.\oparts$, extending each tuple $co$ with a unique ID. 
The set of grouping attributes is $\{\texttt{copID}, \texttt{coID}, \cname, \odate \}$. 
At this point the input is fully flattened.
The next $\idfor$ expression iterates over $\Part$ and defines an equality condition on $\pid$. 
Since the algorithm is at a non-root level,
this pattern corresponds to an outer join between the plan built up to this point and the $\Part$ relation. 
The algorithm has now produced a subplan for the $\sumBy$ expression. 

The $\sumBy$ expression is translated to a sum aggregate $\Gamma^{+}$ with the key consisting of the grouping attributes from 
the first two levels and $p.\pname$ from the key attribute of the $\sumBy$. 
The second-level subplan is finalized with a nest operation 
created from entering the second level at $co.\oparts$ with grouping attributes $\{\texttt{copID}, \texttt{coID}, \cname, \odate \}$. 
The first-level subplan is finalized with a nest operation
created from entering the first level at $\corders$ with grouping attributes $\{\texttt{copID}, \cname\}$. 
The algorithm finalizes with a projection on $(\cname,\corders)$.

The unnesting algorithm may produce suboptimal query plans.
The  plan from Figure~\ref{fig:unnest_plan} can be optimized by projecting away unused attributes from the $\Part$ relation. 
The sum aggregate $\Gamma^{+}$ can be pushed past the join to compute partial sums of $\lqty$ values over the 
output of $\outerunnest^{\oparts}$, grouped by $\{\texttt{copID}, \texttt{coID}, \cname, \odate, \pid\}$. However, a similar local aggregation over $\Part$ brings no benefit, since $\pid$ is the primary key of $\Part$.
\punto
\end{example}
}

\subsection{Code Generation}\label{sec:codegen}
Code generation works bottom-up over a plan produced by 
the unnesting stage, translating an NRC program 
into a parallel collection program. 
For this paper, we describe our compilation in terms of the Spark collection API~\cite{sparkpaper}.


We model input/output bags as Spark \emph{Datasets}, which are strongly-typed, immutable collections of objects; by default, collections are distributed at the granularity of top-level tuples. 
We translate operators of the plan language into operations over Datasets,
taking into account that $\Gamma^{\bagunion}$ and $\Gamma^{+}$ also cast $\NULL$ values to the empty bag and respectively 0; implementation details can be found online~\cite{ourgit}.

The design choice to use Datasets was based on an initial set of experiments
(Appendix \ref{sec:rddvsds}), 
which revealed that an alternative encoding -- using 
RDDs of case classes -- incurs much higher memory and processing overheads. 
Datasets use encoder objects to operate directly on the compact Tungsten binary format, 
thus significantly reducing memory footprint and serialization cost. 
Datasets also allow users to explicitly state which attributes are used in each operation,
providing valuable meta-information to the Spark optimizer. 


Operators effect the partitioning 
guarantee (``partitioner'' in  Spark) of a Dataset. 
All partitioners are key-based and 
ensure all values associated with the same 
key are assigned to the same location. 
Partitioning guarantees affect the amount 
of data movement across partitions (shuffling), which 
can significantly impact the execution cost.
Inputs that have not been altered by an operator have 
no partitioning guarantee. 
An operator inherits the 
partitioner from its input 
and can either 
preserve, drop, or redefine it
for the output. 
Join and nest operators are the only operators 
that directly alter partitioning since 
they require moving values associated with keys to the same partition.
These operators control all partitioning in the standard compilation route.

\subsection{Optimization} \label{subsec:opt}
The generated plan is subject to standard database optimizations; this includes
pushing down selection, projection, and nest operators. Aggregations are pushed down when
the key is known to be unique, based on schema information for inputs. When an 
aggregation is pushed down, the aggregation is only executed locally at each partition. 
Additional optimizations are applied during code generation;
for example, 
a join followed by a nest operation is directly 
translated into a cogroup, a Spark primitive 
that leverages 
logical grouping to avoid separate grouping and join 
operations. This  is beneficial when
building up nested objects with large input bags. 
These optimizations collectively eliminate unnecessary data as 
early as possible and thus reduce data transfer during distributed execution.

\myeat{
The generated plan is subject to standard database optimizations. This includes
pushing down selection, projection, and nesting operators.  Aggregations are pushed down when
the aggregation key is  known to be unique. We assume schema information that provides such information
on inputs, and we perform some simple propagation of this ``object key'' information to subqueries: e.g., uniqueness
of a top-level attribute
is preserved under selection operation and grouping on other attributes. 
The optimized plan is traversed during code 
generation to apply the aforementioned cogroup optimization.
Together these optimizations eliminate unnecessary data as early as possible and thus reduce data transfer during distributed execution.
}

\section{Shredded compilation}\label{sec:pipeline_shredding}

On top of the standard compilation route we build a shredded variant. 
This shredded compilation route provides
a novel approach that begins with an extension of
symbolic query shredding from \cite{pods16}, applies a materialization
phase that makes the resulting queries appropriate for bulk implementation, optimizes
the resulting expressions, and then applies a custom translation into bulk operations. 


Our \emph{shredded representation} encodes a nested bag $b$ whose type $T$ includes bag-valued attributes
 $a_1, \dots, a_k$ by a \emph{flat} bag $b^{\flatc}$ of type $T^\flatc$ where each $a_i$ has a special
type \labeltype and an identifier of a lower-level bag.
The representation of $b$ also includes
a \emph{dictionary tree} $b^\dictc$ of type $T^\dictc$ capturing the 
association between labels and flat bags at each level.

The type $T^\flatc$, always a flat bag type,  and the type $T^\dictc$, always a tuple type, are defined recursively.
For a tuple type $T= \< a_1: T_1, \ldots, a_n:T_n \>$,
$T^\flatc$ is formed by replacing each attribute $a_i$ of bag type with a corresponding attribute of type $\labeltype$.
$T^\dictc$ includes attributes $a_i^\FUN$ and $a_i^\CHILD$ for each bag-valued $a_i$, where
$a_i^\FUN$ denotes the \emph{dictionary} for $a_i$ of type $\labeltype \rightarrow \bagtype(T_i^\flatc)$,
while $a_i^\CHILD$ denotes the dictionary tree of type $T_i^\dictc$.
Similarly, for a bag type $T = \bagtype(T_1)$, we have $T^\flatc = \bagtype(T^\flatc_1)$ and $T^\dictc = T_1^\dictc$;
for scalar types, we have $T^\flatc=T$ and $T^\dictc=\<\>$.

\begin{example} \label{ex:runshreddedinput}
\em
Recall the type of $\COP$ from Example~\ref{ex:run}:
{\small
\begin{align*}
\bagtype\,(\<\, & \cname:\stringtype,\, \corders:\bagtype\,(\<\, \odate: \datetype, \\[-2pt]
& \hspace{1.8cm}\oparts: \bagtype\,(\<\, \pid: \inttype,\, \lqty: \doubletype \,\>) \,\>) \,\>).
\end{align*}
}
The shredded representation of $\COP$ consists of a top-level flat bag ${\COP}^{\flatc}$ of type 
$\bagtype(\<\, \cname: \stringtype, \,\corders: \labeltype \,\>)$
and a dictionary tree ${\COP}^{\dictc}$ of tuple type 
{\small
\begin{align*}
\<\,&\corders^\FUN: \labeltype \rightarrow \bagtype(\<\, \odate: \datetype, \oparts: \labeltype \,\>),\\
&\corders^\CHILD: \bagtype(\<\,\\
&\tab\tab \oparts^\FUN: \labeltype \rightarrow \bagtype(\<\, \pid: \stringtype, \lqty: \doubletype \,\>),\\
&\tab\tab \oparts^\CHILD: \bagtype(\<\, \,\>) \quad\>)\quad\>
\end{align*}
}
The ${\COP}^{\dictc}$ dictionary tree encodes 
the $\corders^\FUN$ dictionary for $\corders$ labels,
the $\oparts^\FUN$ dictionary for $\oparts$ labels,
and the nesting structure via the child attributes.
Since the type system prevents nesting tuples inside tuples, 
each child dictionary tree is wrapped in a singleton bag.
\punto
\end{example}

\vspace{-5pt}
Note that the shredded representation allows opportunities for sharing in the output,
since a given label identifying a nested object may be referenced many times. Appendix \ref{sec:succintexp} provides a microexperiment that quantifies the benefits of a succinct representation in real-world datasets.

We can convert nested objects to their shredded representations and vice versa.
A \emph{value shredding} function takes as input a nested object $o$ and returns
a top-level bag $o^\flatc$ and a dictionary tree $o^\dictc$. This function associates a unique label to each lower-level bag.
A \emph{value unshredding} function performs the opposite conversion \cite{pods16}.
High-level definitions of the shredding and unshredding of a value of type $T$ are straightforward, defined by induction on $T$.

\subsection{Shredded NRC}

Our goal in \emph{query shredding} is to convert a source $\nrc$ program to a 
program capable of working over shredded representations of input and output.

\myparagraph{Shredding approaches}
Past implementations of shredding focused on producing a collection of 
stand-alone subqueries, e.g. in SQL \cite{cheneyshred}, which are well-suited 
for the parallel execution strategy of relational query engines. 
Collectively, these subqueries make up the shredded query,
with one component query for the top-level bag and one component
for  each dictionary. 
To ensure each query can be executed independently, 
the component query for a dictionary $D$ will need to independently
materialize the label domain for $D$; this can result in duplicated
computation of different components as well as repetitive iteration 
steps, which could be problematic for queries with several levels of nesting.
We refer to this as \emph{monolithic shredding} below.

We adopt a \emph{sequential approach}, which is well-designed for the sequential execution strategy of distributed processing systems. 
Our method produces a sequence of queries, one for each level of the output,
with the query  for a given dictionary  $D$ depending on the output of its parent.
The advantage of this approach is that the labels of $D$ will  already
be materialized  in the parent object, which avoids the duplication of computation and 
repeated iteration artifacts of monolithic shredding. 

We now begin the detailed discussion of our shredding approach, starting
with the target languages involved.

As an intermediate stage in shredding, we will make use of
a \emph{symbolic} representation for output dictionaries,  defined as partial functions from labels to bags.
We will use an intermediate query language, denoted $\nrcaggshr$, which extends $\nrc$
with a new atomic type for labels
and a function type for dictionaries.
The grammar of $\nrcaggshr$ is:

\hspace{-0.3cm}
\begin{minipage}{\linewidth}
\begin{align*}
\textit{e} &::= \textit{[Similar to Figure~\ref{fig:nrcagg_syn}]} \\
& \mid \newlabelof(\textit{var}, \mydots) \mid \cidmatch ~ \textit{e} =  \newlabelof(\textit{var}, \mydots) ~ \cidthen ~ \textit{e}\\
& \mid \lookup(\textit{e},\textit{e}) \mid \matlookup(\textit{e},\textit{e}) \mid \lambda \textit{var}.\textit{e} \\
& \mid \textit{e} ~ \dicttreeunion ~ \textit{e}\\
T & ::= \textit{[Similar to Figure~\ref{fig:nrcagg_syn}]}\\
& \mid \labeltype \hspace{5.1cm}\textit{ -- Label Type} \\
& \mid \labeltype \rightarrow \bagtype(F) \hspace{2.7cm}\textit{ -- Dictionary Type}
\end{align*}
\end{minipage}

\vspace{7pt}
The $\newlabelof(x_1, \mydots, x_n)$ construct creates a new label 
encapsulating the values of variables $x_1, \mydots, x_n$ of flat types. 
To deconstruct labels we have a ``label matching construct'':

\vspace{5pt}
\begin{lstlisting}[language=NRC]
          $\cidmatch ~ l = \newlabelof(\vec{x}) ~ \cidthen ~ F(\vec{x}, \vec y)$
\end{lstlisting}
\vspace{5pt}
where $F$ is a bag expression, and $\vec x$ and $\vec y$ are tuples of variables.
Formally, 
given any label $l$ and a binding for $\vec y$,
we find the unique $\vec x$ such that $l=\newlabelof(\vec x)$ and evaluate $F$.
If there is no such $\vec x$, $F$ returns the empty bag.
In this expression, $\vec x$ becomes bound although it is free in $F$.

We have the standard $\lambda$ abstraction restricted to label parameters:
if $e$ is an expression of type $T$ and $l$ is a label variable, then we can form $\lambda l.e$
of type $\labeltype \rightarrow T$.
We also have the standard function application:
if $\textit{e}_1$ is an expression of type $\labeltype \rightarrow T$ and $\textit{e}_2$ an expression of type \labeltype, then we can form
$\lookup(\textit{e}_1,\textit{e}_2)$ of type $T$.
$\matlookup(\textit{e}_1,\textit{e}_2)$ corresponds to a lookup of the label returned by $e_2$ within the bag of label-bags pairs returned by $e_1$.
Finally, we have a variation of the binary union for expressions representing dictionary trees, denoted as $\dicttreeunion$.

\subsection{Query Shredding Algorithm}


A query shredding algorithm takes as input an expression $e$ of type $T$ and produces
an expression $e^\flatc$ of type $T^\flatc$ and a dictionary tree $e^\dictc$ of tuple type $T^\dictc$. 
The algorithm transforms the evaluation problem for $e$ to evaluation problems
for $e^\flatc$ and $e^\dictc$:
When the output of $e$ on an input $i$ is $o$,
then the outputs of $e^\flatc$ and $e^\dictc$ on the shredded representation of $i$ 
will be $o^\flatc$ and $o^\dictc$, respectively.

Our shredding transformation consists of two phases. 
We proceed first with a \emph{symbolic query shredding} phase, producing
succinct expressions manipulating intermediate and output dictionaries defined via
$\lambda$ expressions. 
This phase adapts the work from~\cite{pods16} to our source language.
The key to the succinctness of the representation is that the dictionaries
are not presented explicitly as a list of label/bag pairs, but rather
are \emph{over-approximated} by $\lambda$-terms that provide  general recipes
for computing a bag from an arbitrary label. 

In the second phase, the shredding transformation is followed by a 
\emph{materialization transformation}
that produces a sequence of expressions that is free of $\lambda$ abstractions.

\myeat{
\michael{restore this for archive?}
We list some invariants of our symbolic shredding algorithm.
First, if we have an expression $E$ with free variables $\mathcal{X}$, then the output $E^\flatc$
has free variables contained in
the  \emph{shredded variables} corresponding to
 $\mathcal{X}$. For each variable $x$ of type $T$, we have
a variable $x^\flatc$  whose type depends on that of $x$ as described above.
\milos{should come when presenting the algo as this is the base case}
We also have variables $x^\dictc_\sigma$ where $\sigma$
is a dictionary index of $T$, where the type of the variable is 
$\labeltype \rightarrow \bagtype(T^\flatc)$.
Informally,
$x^\dictc_\sigma$ represents the lookup function associated
with dictionary index $\sigma$.
For example, if $R$ is a variable of type $T=\bagtype(\texttt{a}: \inttype, \texttt{b}: \bagtype(\texttt{c}:\inttype, \texttt{d}:\inttype))$
then the shredded dictionary variables for $R$ are
$R^\flatc$ of type $\bagtype(\texttt{a}: \inttype, \texttt{b}: \labeltype)$, and
 $R^\dictc_{a}$, with type  $\labeltype \rightarrow \bagtype(\texttt{c}:\inttype, \texttt{d}:\inttype)$.
The latter gives a function for dereferencing nested bags of $R$.
\milos{notation: R or b}

Although the  expressions $E^\dictc_\sigma$ producing dictionaries are parameterized by free variables,
we want to avoid having a different dictionary for every variable binding.
Thus, our algorithm will ensure that the expressions  $E^\dictc_\sigma$ will have no free variables of \emph{tuple} type but
will only have variables of dictionary type corresponding to the free variables of $E$. 
\milos{how does the algorithm (not yet discussed) ensures that?}
Further we will guarantee that:
\emph{any free variables of $E^\dictc$ or $E^\flatc$ in the shredding
 corresponds to free variables of $E$}.
 \milos{already said above?}
}

Figure~\ref{fig:shredalg} shows the shredding transformation in terms of recursive functions $\mathcal{F}$ and $\mathcal{D}$.
Given a source expression $e$, the invocation $\shredf{e}$ returns the expression $e^\flatc$ computing the flat version of the output, while $\shredd{e}$ returns the dictionary tree $e^\dictc$ corresponding to $e^\flatc$.

The base cases are those of constants and variables:
for a scalar $c$, we have $\shredf{c}=c$ and $\shredd{c}$ is an empty dictionary tree (line 1);
for a variable $x$ of type $T$, the two functions return $x^\flatc$ and $x^\dictc$ whose types $T^\flatc$ and $T^\dictc$ depend on $T$ as described above (line 2).

The interesting cases are those of tuple construction and tuple projection.
For each bag-valued attribute $a_i$ in a tuple constructor (line 3),
$\mathcal{F}$ replaces the bag expression $e_i$ with a new label encapsulating the (flat) free variables of $e_i$.
The dictionary tree of the tuple constructor includes two attributes:
$a_i^\FUN$, which represents the mapping from such a label to the flat variant $e_i^\flatc$ of $e_i$; 
and $a_i^\CHILD$, which represents the dictionary tree for $e_i^\flatc$.
To conform with the type system,
each child dictionary tree is wrapped as a singleton bag as in Example \ref{ex:runshreddedinput}.
For each scalar-valued attribute $a_j$ in a tuple constructor (line 4), 
$\mathcal{F}$ recurs on the scalar expression $e_j$ to produce $e_j^\flatc$.
Since $e_j$ is already flat, $e_j^\dictc$ is empty; thus, we omit it from the dictionary tree of the tuple constructor. 
When accessing a bag-valued attribute $a$ in a tuple expression $e$ (line 5),
$\mathcal{F}$ returns a $\lookup$ construct computing the flat bag of $e.a$, based on the dictionary $\shredd{e}.a^\FUN$ and label $\shredf{e}.a$ formed during tuple construction. The returned child dictionary tree serves to dereference any label-valued attributes in the corresponding flat bag.
When $e.a$ is a scalar expression (line 6), $\mathcal{F}$ recursively computes $e^\flatc.a$, while $\mathcal{D}$ returns an empty dictionary tree.

For the binary union, the same label-valued attribute may correspond to labels that depend on two different sets of free variables. To encapsulate two dictionary trees, the function $\mathcal{D}$ constructs a $\dicttreeunion$ of tuple type (line 11).
For the remaining constructs, the functions $\mathcal{F}$ and $\mathcal{D}$ proceed recursively,
maintaining the invariant that for any source expression $e$, 
the free variables of $e^\flatc$ or $e^\dictc$ in the shredded representation correspond to the free variables of $e$.

\begin{figure*}  
\centering\small
{
\setcounter{magicrownumbers}{0}
\renewcommand{\arraystretch}{1.2}
\begin{tabular}{@{}L@{\;}LLL@{}}
\toprule
& \multicolumn{1}{l}{Pattern of $e$} & \multicolumn{1}{l}{$\shredf{e}$} & \multicolumn{1}{l}{$\shredd{e}$}\\
\midrule
\linenumber & c & c & \<\> \\[3pt]


\linenumber & \textit{var} & \textit{var\,}^\flatc & \textit{var\,}^\dictc\\[3pt]

\linenumber &
\<\mydots\, a_i \ateq {e_i}, \mydots\quad e_i \mbox{ of bag type} & 
 \< \mydots\, a_i \ateq \newlabelof((x^\flatc)_{x\in\textit{\freevars}(e_i)}), \mydots &
\begin{tabular}{@{}l@{}}
$\<\mydots\, a_i^\FUN \ateq \lambda l. ~ \idmatch ~ l= \newlabelof(\mydots) ~ \idthen ~ \shredf{e_i},$\\
$\phantom{\<\mydots\,}a_i^\CHILD \ateq \sng{\shredd{e_i}}, \mydots$
\end{tabular}
\\[3pt]

\linenumber &
\phantom{\<}\mydots\, a_j \ateq {e_j}, \mydots\quad  e_j \mbox{ of scalar type} \> & 
\phantom{\<}\mydots\, a_j \ateq \shredf{e_j}, \mydots\> &
\phantom{\>}\mydots \> \quad\quad \mbox{\color{gray}     // omit empty dictionary tree for $a_j$}\\[3pt]

\linenumber & 
e.a \qquad \mbox{ of bag type} & 
\lookup(\shredd{e}.a^{\FUN}, \shredf{e}.a) &
\getkw(\shredd{e}.a^{\CHILD}) \\[3pt]

\linenumber & e.a \qquad \mbox{ of scalar type} & \shredf{e}.a & \<\>  \\[3pt]

\linenumber & \sng{e} & \sng{\shredf{e}} & \shredd{e} \\[3pt]

\linenumber & 
\cidfor ~ var ~ \cidin ~ {e}_1 ~ \cidunion ~ \textit{e}_2 & 
\begin{tabular}{@{}l@{}}
$\cidlet ~ var^\dictc \leteq \shredd{e_1} ~ \cidin$ \\
$\tab \cidfor ~ var^\flatc ~ \cidin ~ \shredf{{e}_1} ~ \cidunion ~ \shredf{\textit{e}_2}$
\end{tabular} &
\cidlet ~ var^\dictc \leteq \shredd{e_1} ~ \cidin ~ \shredd{e_2}
\\[9pt]

\linenumber &
\cidlet ~ var \leteq ~ {e}_1 ~ \cidin ~ \textit{e}_2 & 
\begin{tabular}{@{}l@{}}
$\cidlet ~ var^\dictc \leteq \shredd{e_1} ~ \cidin$ \\
$\tab \cidlet ~ var^\flatc ~ \idin ~ \shredf{{e}_1} ~ \cidin ~ \shredf{\textit{e}_2}$
\end{tabular} &
\cidlet ~ var^\dictc \leteq \shredd{e_1} ~ \cidin ~ \shredd{e_2}
\\[9pt]

\linenumber &
\cidif ~ cond ~ \cidthen ~ e &
\cidif ~ \shredf{cond} ~ \cidthen ~ \shredf{e} &
\shredd{e} \\[3pt]

\linenumber &
e_1 \uplus e_2 &
\shredf{e_1} \uplus \shredf{e_2} &
\shredd{e_1} ~ \dicttreeunion ~ \shredd{e_2}\\[3pt]

\linenumber & op(e) 
& op(\shredf{e}) & \shredd{e} \\[3pt]

\linenumber & 
op(e_1, e_2) 
& op(\shredf{e_1}, \shredf{e_2}) & \<\> \\[3pt]





\bottomrule
\end{tabular}
}
\vspace{3pt}
\caption{Query shredding algorithm.}\label{fig:shredalg}
\vspace{-6pt}
\end{figure*}


\myeat{
\begin{figure*}  
\begin{center}
{
\renewcommand{\arraystretch}{1.3}
\begin{tabular}{L L L}
\toprule
\multicolumn{1}{l}{$\textit{e}$} & \multicolumn{1}{l}{\color{red}$\shredf{e}$} & \multicolumn{1}{l}{\color{red}$\shredd{\sigma}{e}$}\\
\midrule
{\color{red} x} &
  x^\flatc & 
  x^\dict_{\sigma}\\[3pt]
  x.a &
\begin{cases}
  \lookup(x^\dict_{\color{red}a}, x^\flatc.a) & \mbox{ if } x.a \mbox{ of bag type } \\ 
x^\flatc.a & \mbox{ if } x.a \mbox{ of scalar type } 
\end{cases}&
x^\dict_{\color{red}a \sigma} \\[3pt]
\<a_i \ateq {\color{red}x_i}\> &
\begin{cases}
  \<a_i \ateq \newlabelof_{x_i}(x_i^\flatc)\>
   & \mbox{ if } x_i \mbox{ of bag type } \\
  \<a_i \ateq x_i^\flatc\> & \mbox{ if } x_i \mbox{ of scalar type } 
\end{cases}
  & 
  \begin{cases} 
  \lambda l. ~ \idmatch ~ l= \newlabelof_{x_i}(\color{red}x_i^\flatc) &\mbox{ if } \sigma=a_i \\ \tab \idthen ~ x_i^\flatc  \\
\shredd{\sigma'}{x_i} &\mbox{ if } \sigma=a_i \sigma'
\end{cases} \\[3pt]
\sng{x}&
  \sng{x^\flatc} & 
  x^\dict_{\sigma} \\[3pt]
    \shortstack[l]{$\idfor ~ x ~ \idin ~ X$\\$\tab\idunion ~ \textit{e}_1$}
  &
  \shortstack[l]{$\idfor  ~ x^\flatc ~ \idin ~ X^\flatc ~  \idunion$\\ 
  $\tab\; \idlet ~   \bigwedge_{\sigma'}  x^\dict_{\sigma'} \leteq X^\dict_{\sigma'} ~ \idin ~
  \shredf{\textit{e}_1}$} &
  \idlet ~   \bigwedge_{\sigma'}  x^\dict_{\sigma'} \leteq X^\dict_{\sigma'} ~ \idin ~
  \shredd{\sigma}{\textit{e}_1}\\[3pt]
  \shortstack[l]{$\idlet ~ x \leteq \textit{e}_1$ \\ $\tab \idin ~ \textit{e}_2$} &
  \shortstack[l]{$\idlet ~   \bigwedge_{\sigma'}  x^\dict_{\sigma'} \leteq \shredd{\sigma'}{\textit{e}_1} ~ \idin$\\ 
  $\tab\; \idlet ~ x^\flatc:=\shredf{\textit{e}_1} ~ \idin ~ 
  \shredf{\textit{e}_2}$} &
  \idlet ~   \bigwedge_{\sigma'}  x^\dict_{\sigma'} \leteq \shredd{\sigma'}{\textit{e}_1} ~ \idin ~ \shredd{\sigma}{\textit{e}_2}\\[3pt]
  \idif ~ x ~ \idthen ~ \textit{e}_1 &
  \idif ~ x^\flatc ~ \idthen ~ \shredf{\textit{e}_1} &
  \shredd{\sigma}{\textit{e}_1} \\[3pt]
  \textit{x}_1 ~ \uplus ~ \textit{x}_2&
  \textit{x}^\flatc_1 ~ \uplus ~ \textit{x}^\flatc_2 &
  \shredd{\sigma}{\textit{x}_1} ~ \dictunion ~ \shredd{\sigma}{\textit{x}_2}\\[3pt]
  \dedup(x)&
  \dedup(x^\flatc) & 
  x^\dict_{\sigma} \\[3pt]
  \groupby_{key}(x) & \groupby_{key}(x^\flatc) & \shredd{\sigma}{x}\\[3pt]
  \sumby_{key}^{value}(x) & \sumby_{key}^{value}(x^\flatc) & \shredd{\sigma}{x}\\[3pt]
\bottomrule
\end{tabular}
}
\end{center}
\caption{Query shredding algorithm.}\label{fig:shredalg}
\end{figure*}
}

\begin{example} \label{ex:shredalg}
\em
We exhibit the shredding algorithm from Figure~\ref{fig:shredalg} on the query $Q$ from Example~\ref{ex:run}.
The algorithm produces a query $Q^\flatc$ computing the top-level flat bag and a query $Q^\dictc$ computing the dictionary tree for $Q^\flatc$.
We match $Q$ to the $\idfor$ construct in $\mathcal{F}$ (line 8) and recur to derive $Q^\flatc$:

\vspace{3pt}
\begin{lstlisting}[language=NRC]
   $\cidfor ~ cop^\flatc ~ \cidin ~ {\COP}^\flatc ~ \cidunion$ 
     $\{\,\<\,\cname \ateq cop^\flatc.\cname, \corders \ateq \newlabelof(cop^\flatc)\,\>\,\}$
\end{lstlisting}
\vspace{3pt}

Recall that $\COP^\flatc$ is a flat bag, so $Q^\flatc$ indeed computes a bag of flat tuples.
We drop here the unused $\idlet$ binding to $cop^\dictc$.
We derive $Q^\dictc$ after matching $\corders \ateq Q_{\corders}$ 
in the top-level tuple constructor of $Q$ (line 3):

\vspace{3pt}
\begin{lstlisting}[language=NRC]
  $\cidlet ~ cop^\dictc \leteq {\COP}^{\dictc} ~ \cidin$
  $\<\corders^{\FUN} \ateq \lambda l.\;\cidmatch \; l=\newlabelof(cop^\flatc) \; \cidthen$
     $\cidfor \; co^\flatc \; \cidin ~ \lookup(cop^{\dictc}.\corders,cop^\flatc.\corders) ~\cidunion$
      $\;\{\,\<\,\odate \ateq co^\flatc.\odate,\, \oparts \ateq \newlabelof(co^\flatc ) \,\>\, \}$,
  $\corders^{\CHILD} \ateq \{ Q_{\corders}^\dictc \}\,\>$
\end{lstlisting}
\vspace{3pt}

We omit the unused $\idlet$ binding to $co^\dictc$ in the expression computing the dictionary $\corders^{\FUN}$.
The query producing the lowest-level dictionary tree $Q_{\corders}^\dictc$ is: 

\vspace{3pt}
\begin{lstlisting}[language=NRC]
  $\cidlet ~ co^\dictc \leteq \getkw(cop^{\dictc}.\corders^{\CHILD}) ~ \cidin$
  $\<\,\oparts^{\FUN} \ateq \lambda l.\,\cidmatch ~ l=\newlabelof(co^\flatc) ~ \cidthen$
     $\sumby_{\pname}^{\total}($
      $\cidfor \; op^\flatc \; \cidin \; \lookup(co^{\dictc}.\oparts, co^\flatc.\oparts) \;  \cidunion$
       $\cidfor \; p^\flatc \; \cidin  \; \Part^\flatc \; \cidunion $
        $\cidif \; (p^\flatc.\pid \langeq op^\flatc.\pid) \; \cidthen $
         $\{ \,\<\, \pname \ateq p^\flatc.\pname,\,\total \ateq op^\flatc.\lqty*p^\flatc.\pprice\,\>\, \}),$
  $\,\oparts^{\CHILD} \ateq \{ \<\> \,\}\,\>\hspace{5cm}\Box$
\end{lstlisting}
\end{example}

Our implementation~\cite{ourgit} refines the shredding algorithm 
to retain only the relevant attributes of free variables in a $\newlabelof$, 
thus contributing to a succinct representation. 
For instance, the $\corders$ labels in $Q^\flatc$ need only capture the 
free variables referenced in $cop^\flatc.\corders$ and not the
$cop^\flatc.\cname$ values. 


\subsection{Materialization Algorithm}

The symbolic dictionaries produced by the symbolic query shredding algorithm are useful
in keeping the inductively-formed expressions succinct. 
The second phase of the shredding process, which we refer to as \emph{materialization}, 
eliminates $\lambda$ terms in favor of expressions that produce an explicit representation of the shredded 
output. 
In the case of a query with flat output, materialization will be a variant of well-known
normalization and conservativity algorithms \cite{conservativity,fegarasmaier}.
In the case of nested output, materialization will be the key distinguishing
feature of the sequential  approach from monolithic shredding.
Intuitively, the materialized expression for a given dictionary  
$D$  will make use of the label domain already materialized
for its parent, producing a join of these
labels that takes into account the $\lambda$ term associated with $D$ in symbolic shredding.

Our materialization phase produces a sequence of assignments, given a shredded expression and its dictionary tree.
For each symbolic dictionary $Q^\dictc_{\mathit{index}}$ in the dictionary tree,
the transformation creates one assignment $\texttt{\matdict}_{\mathit{index}} \Leftarrow e$,
where the expression $e$ computes a bag of tuples representing the materialized form of $Q^\dictc_{\mathit{index}}$.
Each assignment can depend on the output of the prior ones.
The strategy works downward on the dictionary tree, keeping track of the assigned variable for each symbolic dictionary. 
Prior to producing each assignment, 
the transformation rewrites the expression $e$ to replace any symbolic dictionary by its assigned variable and  
any $\lookup$ on a symbolic dictionary by a $\matlookup$ on its materialized variant.

Materialization needs to resolve the domain of a symbolic dictionary. 
In our baseline materialization, the expression $e$ computing $\texttt{\matdict}_{\mathit{index}}$ takes as input a \emph{label domain}, the set of labels produced in the parent assignment.
The expression $e$ then simply iterates over the label domain and 
evaluates the symbolic dictionary $Q^\dictc_{\mathit{index}}$ for each label. 

The materialization algorithm from Figure~\ref{fig:matalg_extended}
produces a sequence of assignments given a shredded expression, its dictionary tree, and a variable to represent the top-level expression.
The \textsc{Materialize} procedure first replaces the (input) symbolic dictionaries in $e^\flatc$ by their materialized counterparts (line 1) and then assigns this rewritten expression to the provided variable (line 2). 
Prior to traversing the dictionary tree, its dictionaries are simplified by inlining each $\idlet$ binding produced by the shredding algorithm (line 3).

The \textsc{MaterializeDict} procedure performs a depth-first traversal of the dictionary tree. 
For each label-valued attribute $a$, we first emit an assignment computing the set of labels produced in the parent assignment (lines 3-4). 
We then rewrite the dictionary $a^\FUN$ to replace all references to symbolic dictionaries, each of them guaranteed to have a matching assignment since the traversal is top-down.
We finally produce an assignment computing a bag of label-value tuples representing the materialized form of $a^\FUN$ (lines 6-8), before recurring to the child dictionary tree (line 9).

A technicality is that labels within a label domain can encode values of multiple
types, while the expressions above assume that we can match labels coming from each domain
against a single known tag $F$, always returning the same type.
To solve this we form separate label domains for each tag $\tau$
of a label $\newlabelof_\tau$ term. This assures that in each domain, the labels depend upon tuples of the same type.

The \textsc{ReplaceSymbolicDicts} function first recursively inlines all let bindings in the given expression and then performs the following actions:
1) replaces any $\lookup$ over an input symbolic dictionary with a $\matlookup$ over the corresponding materialized dictionary obtained using the $\mathit{resolver}$ function;
2) $\beta$-reduces any $\lookup$ over an intermediate symbolic dictionary (lambda), returning its body expression with the given label inlined; and
3) replaces any $\lookup$ over a $\dicttreeunion$ with a union of two recursively resolved lookups.

\begin{example} \label{ex:materialize}
\em
We showcase the \textsc{Materialize} procedure on the shredded queries $Q^\flatc$ and $Q^\dictc$ from Example~\ref{ex:shredalg}.  
Let \texttt{$\kw{MatCOP}$}, \texttt{$\kw{MatCOP}_{\corders}$}, and \texttt{$\kw{MatCOP}_{\corders\_\oparts}$}
denote the materializations of the top-level input bag $\COP^\flatc$ and 
the two symbolic dictionaries from $\COP^\dictc$, respectively. 
The procedure replaces $\COP^F$ by \texttt{$\kw{MatCOP}$} in $Q^\flatc$ (line 1), followed by emitting the assignment to the provided variable \texttt{$\topbag$}:

\vspace{3pt}
\begin{lstlisting}[language=NRC]
  $\topbag \assigneq \cidfor ~ cop^\flatc ~ \cidin ~ {\kw{MatCOP}} ~ \cidunion$
             $\{\< \cname \ateq cop^\flatc.\cname, \corders \ateq \newlabelof(cop^\flatc)\}$
\end{lstlisting}
\vspace{3pt}

\textsc{MaterializeDict} 
produces an assignment computing the set of $\corders$ labels from the parent \texttt{\topbag} (lines 3-4):

\vspace{3pt}
\begin{lstlisting}[language=NRC]
  $\labeldomain_{\corders} \assigneq$
     $\dedup(\cidfor \; x \; \cidin \; \topbag \; \cidunion \; \{\<  \labelatt \ateq \; x.\corders \> \})$
\end{lstlisting}
\vspace{3pt}

The function uses the labels from \texttt{$\labeldomain_{\corders}$} to materialize the $\corders^\FUN$ dictionary (lines 6-8):

\vspace{3pt}
\begin{lstlisting}[language=NRC]
$\matdict_{\corders} \assigneq$
  $\cidfor ~ l ~ \cidin ~ \labeldomain_{\corders} ~ \cidunion$
  $\{\,\<\, \labelatt \ateq$ l.$\labelatt$,
     $\valueatt \ateq \cidmatch ~ l.\labelatt=\newlabelof(cop^\flatc) ~ \cidthen$
        $\cidfor ~  co^\flatc ~ \cidin ~ \matlookup\big({\kw{MatCOP}_{\corders}}, cop^\flatc.\corders\big)$
        $\cidunion ~ \{\<\odate \ateq co^\flatc.\odate, \oparts \ateq \newlabelof(co^\flatc) \> \}$
\end{lstlisting}
\vspace{3pt}
The query computing \valueatt corresponds to the body of the $\corders^\FUN$ dictionary, 
with the $\lookup$ and its symbolic dictionary replaced by their materialized counterparts (line 5).  

The function finally recurs on the dictionary tree for $\oparts$, deriving the label domain for $\oparts^\FUN$ from \texttt{$\matdict_{\corders}$}.

\vspace{10pt}

\begin{lstlisting}[language=NRC]
 $\labeldomain_{\corders\_\oparts} \assigneq$
  $\dedup\big(\cidfor \;  x  \; \cidin \; \matdict_{\corders} \; \cidunion \; \{\<\, \labelatt \; \ateq$ x.$\oparts \,\>\}\big)$
 $\matdict_{\corders\_\oparts} \assigneq$
   $\cidfor ~ l ~ \cidin ~ \labeldomain_{\corders\_\oparts} ~ \cidunion$
    $\{\,\<\, \labelatt \ateq$l.$\labelatt$, 
      $\,\valueatt \ateq \cidmatch ~ l.\labelatt=\newlabelof(co^\flatc) ~ \cidthen$
       $\sumby_{\pname}^{\total}($
         $\cidfor ~op^\flatc~ \cidin ~ \matlookup\big({\kw{MatCOP}_{\corders\_\oparts}}, co^\flatc.\oparts\big)$
         $\cidunion~\cidfor ~ p^\flatc ~ \cidin  ~ \Part^\flatc ~ \cidunion$ 
                  $\cidif ~ (p^\flatc.\pid \langeq op^\flatc.\pid) ~ \cidthen$
                    $\{ \,\<\, \pname := p^\flatc.\pname, $
                      $\;\total := op^\flatc.\lqty*p^\flatc.\pprice \,\>\,\}\,)\,\>\,\}\hspace{0.7cm}\Box$
\end{lstlisting}
\end{example}

\myeat{
\begin{example} \label{ex:materialize}
\em
We exhibit the materialization algorithm on the
query from Example \ref{ex:run}. As with symbolic shredding, we include
here some simple post-processing, including trivial inlining and dropping unused $\idlet$ bindings.
The materialization sequence first computes the flat bag,
and stores the labels occurring in it:

\vspace{2pt}
\begin{lstlisting}[language=NRC]
$\topbag \assigneq$
  $\cidfor ~ cop ~ \cidin ~ {\COP}^\flatc ~ \cidunion$
    $\{\< \cname \ateq cop.\cname,$
      $\corders \ateq \newlabelof_F(\< \corders \ateq cop.\corders\>)\}$
$\labeldomain_{\corders} \assigneq$
  $\dedup(\cidfor \; cop \; \cidin \; \topbag \; \cidunion \; \{\<  \labelatt \ateq \; cop.\corders \> \})$
\end{lstlisting}
\vspace{2pt}

We then materialize the first-level dictionary, 
$Q^\dictc_{\corders}$, using
the labels from $\labeldomain_{\corders}$. Then, we materialize the labels occurring in the output of this
dictionary.

\vspace{2pt}
\begin{lstlisting}[language=NRC]
$\matdict_{\corders} \assigneq$
  $\cidfor ~ l ~ \cidin ~ \labeldomain_{\corders} ~ \cidunion$
    $\{\< \labelatt \ateq$ l.$\labelatt$,
      $\valueatt \ateq \cidmatch ~ l.\labelatt=\newlabelof_F(t) ~ \cidthen$
        $\cidfor ~  o ~ \cidin ~ \lookup\big({\COP}^{\dictc}_{\corders}, t.\corders\big) ~ \cidunion$
          $\{\<\odate \ateq o.\odate,$
            $\oparts \ateq \newlabelof_G(\<\oparts \ateq o.\oparts \>) \> \}$
  $\labeldomain_{\oparts} \assigneq$
    $\dedup\big(\cidfor \;  v  \; \cidin \; \matdict_{\corders} \; \cidunion $
            $\{\< \labelatt \; \ateq$ v.$\oparts \> \}\big)$
\end{lstlisting}
\vspace{2pt}

Finally, we use $\labeldomain_{oparts}$ to materialize the lower-level dictionary $Q^\dictc_{\corders\_\oparts}$:

\vspace{2pt}
\begin{lstlisting}[language=NRC]
$\matdict_{\corders\_\oparts} \assigneq$
  $\cidfor ~ l ~ \cidin ~ \labeldomain_{\oparts} ~ \cidunion$
    $\{\< \labelatt \ateq$l.$\labelatt$, 
      $\valueatt \ateq \cidmatch ~ l.\labelatt=\newlabelof_F(t) ~ \cidthen$
         $\sumby_{\pname}^{\total}($
           $\cidfor ~op~ \cidin ~ \lookup\big({\COP}^\dictc_{\oparts}, t.\oparts\big) ~ \cidunion$
             $\cidfor ~ p ~ \cidin  ~ \Part^\flatc ~ \cidunion$ 
               $\cidif ~ (p.\pid \langeq op.\pid) ~ \cidthen$
                 $\{ \< \pname := p.\pname, $
                   $\total := op.\lqty*p.\pprice\> \}$
           $)\>\}\hspace{6cm}\Box$
\end{lstlisting}
\end{example}
}


\subsection{Domain Elimination} \label{subsec:domainelim}

In many cases the materialization algorithm produces label domains that 
are redundant.  This represents one major shortcoming of sequential shredding:
in some cases making use of  labels materialized in the parent is unnecessary.

In order to detect such situations, we optimize our materialization procedure using \emph{domain elimination} rules.

The first rule recognizes that a child symbolic dictionary is an expression of the form:

\vspace{2pt}
\begin{lstlisting}[language=NRC]
    $\lambda l.\,\cidmatch ~ l=\newlabelof(x) ~ \cidthen$
       $\cidfor ~  y ~ \cidin ~  \lookup(D, x.a) ~ \cidunion ~ e$
\end{lstlisting}
%
where the only used attribute of $x$ is $a$ of label type.
We can  skip computing the domain of labels for this dictionary 
and compute the label-value pairs directly from the materialized dictionary \texttt{\kw{MatD}} corresponding to $D$:
\vspace{2pt}
\begin{lstlisting}[language=NRC]
    $\cidfor ~  z  ~ \cidin ~ \texttt{\kw{MatD}} ~ \cidunion$
      $\{\,\<\, \labelatt \ateq z.\labelatt, $
         $\valueatt \ateq \cidlet ~ x \ateq \<\, a \ateq z.\labelatt \,\> ~ \cidin$
                  $\,\cidfor ~ y ~ \cidin ~ z.\valueatt ~ \cidunion ~ e \,\,\>\,\}$
\end{lstlisting}
\vspace{2pt}
We benefit from this rule when the size of the label domain from the parent assignment is comparable to the size of $\texttt{\kw{MatD}}$.

\begin{example} \label{ex:domainelim}
\em
Returning to Example~\ref{ex:materialize}, 
applying this rule avoids producing \texttt{$\labeldomain_{\corders}$} as an intermediate result.
The materialization of $\corders^\FUN$ now corresponds to:
\vspace{30pt}
\begin{lstlisting}[language=NRC]
$\matdict_{\corders} \assigneq$
  $\cidfor ~ z  ~ \cidin ~  {\kw{MatCOP}_{\corders}} ~ \cidunion$
   $\{\,\<\,\labelatt \ateq z.\labelatt, $
      $\valueatt \ateq \cidfor ~ co^\flatc ~ \cidin ~ z.\valueatt ~ \cidunion$
               $\{\<\, \odate \ateq co^\flatc.\odate,\,\oparts \ateq \newlabelof(co^\flatc) \,\>\}$               
\end{lstlisting}
We can extend this rule to match a $\sumby$ construct around the $\idfor$ construct,
which would then also enable computing \texttt{$\matdict_{\corders\_\oparts}$} directly from \texttt{${\kw{MatCOP}_{\corders\_\oparts}}$}.
\punto
\end{example}

The second rule for domain elimination 
recognizes when a label $l$ encodes a non-label attribute $b$ filtering a bag: 
\vspace{2pt}
\begin{lstlisting}[language=NRC]
    $\lambda l.\, \cidmatch ~ l=\newlabelof(x) ~ \cidthen$
       $\cidfor ~  y ~ \cidin ~ Y ~ \cidunion$ $\idif ~  (y.a \langeq x.b) ~  \cidthen ~  e$
\end{lstlisting}
\vspace{2pt}
If no other attribute of $x$ appears in $e$,
we can produce the label-value pairs from $Y$ using the value of $y.a$:
\vspace{2pt}
\begin{lstlisting}[language=NRC]
  $\groupby_{\labelatt}\big(\cidfor ~  y  ~ \cidin ~ Y ~ \cidunion$
                 $\cidlet ~ x \ateq \<b \ateq y.a\> ~ \cidin$               
                 $\{\,\<\,  \labelatt \ateq \newlabelof(x), \valueatt \ateq e \,\>\,\}\big)$
\end{lstlisting}
\vspace{2pt}
This rule transforms the variable $x$ from free to bound,
allowing computing the materialized dictionary from $Y$ only.

\subsection{Partial Shredding}

Shredding can easily be tuned towards either partially-shredded inputs or partially-shredded outputs.
\begin{example} \label{ex:runningpartial}
\em
The shredded output in Example~4 in the body of the paper 
was produced by three queries, 
the first producing the top-level bag, 
the second producing the dictionary for $\corders$, 
and the third producing a dictionary for $\corders~\oparts$.
If there is little or no data skew in $\oparts$, 
we could decide not to shred $\corders$. 
The query producing the top-level bag would be unchanged,
but we modify the dictionary for $\corders$ so that it maps each label to an entire 
nested object.

\vspace{3pt}
\begin{lstlisting}[language=NRC]
  $\cidlet ~ cop^\dictc \leteq {\COP}^{\dictc} ~ \cidin$
  $\<\corders^{\FUN} \ateq \lambda l.\;\cidmatch \; l=\newlabelof(cop^\flatc) \; \cidthen$
     $\cidfor \; co^\flatc \; \cidin ~ \lookup(cop^{\dictc}.\corders,cop^\flatc.\corders) ~\cidunion$
      $\;\{\,\<\,\odate \ateq co^\flatc.\odate,\, \oparts \ateq 
            \sumby_{\pname}^{\total}($
                $\cidfor ~ op ~ \cidin ~ co^\flatc.\oparts ~ \cidunion$
                  $\cidfor ~ p^\flatc ~ \cidin ~ \Part^\flatc ~ \cidunion$
                    $\cidif ~ p^\flatc.\pid \langeq op.\pid ~ \cidthen$
                      $\{ \< \pname \ateq p^\flatc.\pname,$
                        $\total \ateq op.\lqty * p^\flatc.\pprice\> \}$
              $) \> \}$,
  $\corders^{\CHILD} \ateq \{ \<\> \,\}\,\>\hspace{5cm}\Box$
\end{lstlisting}
\vspace{3pt}

The modification over the fully-shredded version is to replace
a call to the $\newlabelof$ function at \oparts with the inlining of the $\oparts$ subquery.
\punto
\end{example}

\subsection{Extensions for Shredded Compilation}\label{sec:pipeline_shredding_ext}
Given that the output of materialization is an NRC program 
with expressions containing only a subset of $\nrcaggshr$, 
there are only a few extensions required 
to support compilation for  shredded data.
Dictionaries are simply considered to be a 
special instance of a bag, keyed with values of a label type
 and supporting lookup operations. 
A $\matlookup$ is translated directly 
into an outer join in the plan language, providing opportunities 
for pushed aggregation and merging with nest operators to 
form cogroups in the code generated for value unshredding. 

The shredded representation 
allows nested operations 
in the input query to be translated to operations working only 
on the dictionaries relevant to that level,  
these are the \textit{localized operations} first discussed 
in the introduction.
For example, the $\sumBy$ in the running example will operate 
directly over the \oparts dictionary. This produces a plan, 
similar to a subset of the plan in Figure~\ref{fig:unnest_plan}, 
that has isolated the join and aggregate operation to the 
lowest level dictionary. This avoids flattening the input via 
the series of unnest operators and avoids the programming mismatch 
also described in Section~\ref{sec:introduction}. 

The label-bag representation of 
dictionaries can lead to 
overheads when we want to join with a bag value.  Thus, a relational dictionary 
representation of label-value pairs can be useful for implementation. 
To provide flexibility in dictionary representation, 
we introduce an operator \bagtomatdict, which casts
a bag to a dictionary while delaying explicit representation until 
the relevant stage of compilation.

For our current code generation, dictionaries are represented the 
same as bags, Dataset[T] where T contains a label
column (\labelatt) as key.
Top-level bags that have not been altered by 
an operator have no partitioning 
guarantee and are distributed by the same default strategy 
as bags in the standard compilation route.
Dictionaries have a label-based partitioning guarantee, 
which is the key-based partitioning guarantee 
described in Section \ref{sec:codegen} where all values 
associated to the same label reside on the same partition.

\begin{figure}
\small
{
\renewcommand{\arraystretch}{1.3}

\setcounter{magicrownumbers}{0}
\begin{tabular}{@{}l@{}l@{\;}}
\toprule
\multicolumn{2}{l}{\textsc{Materialize}(expression $e^\flatc$, expression $e^\dictc$, var \texttt{\kw{Top}}, funct $\mathit{resolver}$)} \\
\midrule
\linenumber & $e_1^\flatc = \textsc{ReplaceSymbolicDicts}(e^\flatc, \mathit{resolver})$ {\color{gray}//by materialized dicts} \\
\linenumber & $\textsc{Emit}(\texttt{\kw{Top}} \assigneq e_1^\flatc)$ \\
\linenumber & $e_1^\dictc = \textsc{Normalize}(e^\dictc)$ {\color{gray}// recursively inline \idlet bindings} \\
\linenumber & $\textsc{MaterializeDict}(e_1^\dictc, \texttt{\kw{Top}}, \mathit{resolver})$ \\
\midrule
\end{tabular}
\\[6pt]
\setcounter{magicrownumbers}{0}
\begin{tabular}{@{}l@{}l}
\midrule
\multicolumn{2}{l}{\textsc{MaterializeDict}(expression $e^\dictc$, var $\texttt{\kw{Parent}}_{\mathit{index}}$, funct $\mathit{resolver}$)} \\
\midrule
\linenumber & $\mbox{\bf switch } e^\dictc$: \\
\linenumber & $\tab \mbox{\bf case } \< a_1 \ateq e_1, \mydots, a_n \ateq e_n \> \Rightarrow$ \\
\linenumber & $\tab\tab\mbox{\bf foreach } a^\FUN \in \{a_1, \mydots, a_n\}$\\
\linenumber & $\tab\tab\tab \textsc{Emit}\big(\texttt{\labeldomain}_{\mathit{index\_a}} \assigneq $\\
\linenumber & $\tab\tab\tab\tab\tab \dedup(\cidfor ~ x ~ \cidin ~ \texttt{\kw{Parent}}_{\mathit{index}} ~  \cidunion ~ \{\,\<\, \labelatt \ateq ~ x.a \,\>\, \})\big)$\\
\linenumber & $\tab\tab\tab fun = \textsc{ReplaceSymbolicDicts}(e^\dictc.a^\FUN, \mathit{resolver})$\\
\linenumber & $\tab\tab\tab \textsc{Emit}\big(\texttt{\matdict}_{\mathit{index\_a}} \assigneq $\\
\linenumber & $\tab\tab\tab\tab\tab \cidfor ~ l ~ \cidin ~ \texttt{\labeldomain}_{\mathit{index\_a}} ~  \cidunion$ \\
\linenumber & $\tab\tab\tab\tab\tab\tab\;\; \{\,\<\, \labelatt \ateq l.\labelatt,\, \valueatt \ateq fun(l.\labelatt) \,\>\, \}\big)$\\
\linenumber & $\tab\tab\tab \mathit{resolver} = \textsc{AddMapping}(\mathit{resolver}, e^\dictc.a^\FUN, \texttt{\matdict}_{\mathit{index\_a}})$ \\
\linenumber & $\tab\tab\tab \textsc{MaterializeDict}(\getkw(e^\dictc.a^{\CHILD}), \texttt{\matdict}_{\mathit{index\_a}}, \mathit{resolver})$\\
\linenumber & $\tab \mbox{\bf case } \dicttreeunion(e_1^\dictc, e_2^\dictc) \Rightarrow$ \\
\linenumber & $\tab\tab \textsc{MaterializeDict}(e_1^\dictc, \texttt{\kw{Parent}}_{\mathit{index}}, \mathit{resolver})$\\
\linenumber & $\tab\tab \textsc{MaterializeDict}(e_2^\dictc, \texttt{\kw{Parent}}_{\mathit{index}}, \mathit{resolver})$\\
\bottomrule
\end{tabular}
}
\vspace{3pt}
\caption{Materialization algorithms.}\label{fig:matalg_extended}
\vspace{-12pt}
\end{figure}

\section{Skew-Resilient Processing}\label{sec:skew}
We provide a novel approach to skew handling, which adapts a generated 
plan based on statistics of the data at runtime and supports shredding.
The framework up to this point has not addressed skew-related bottlenecks. 
Skew is a consequence of key-based partitioning where
all values with the same key are sent to the same partition. 
If the collected values of a key ($key_1$) are much larger 
than the collected values of another key ($key_2$), then the partition 
housing $key_1$ values will take longer to compute than others. 
We refer to such keys as \textit{heavy}, and all others 
as \textit{light}.
In the worst case, 
the partition is completely saturated and the evaluation 
is heavily burdened by reading and writing to disk. 
We mitigate the effects of skew by 
implementing skew-aware plan operators that automate the 
skew-handling process. 

The identification of heavy keys is central to
automating the skew-handling process. 
We use a sampling
procedure to determine heavy keys in a distributed bag, considering a 
key \textit{heavy} 
when at least a certain threshold
of the sampled tuples (10\%) in a partition are associated with that key; 
in our experiments, we use the threshold of 2.5\%.
Based on the heavy keys, the input bag is split into 
a light component and a heavy component. We refer to these 
three components (light bag, heavy bag, and heavy keys)
as a \textit{skew-triple}.

A skew-aware operator is a plan operator that 
accepts and returns a skew-triple. 
If the set of heavy keys is not known, then
the light and heavy components are unioned and a 
heavy key set is generated. This set 
of heavy keys remains associated to that 
skew-triple until the operator does something 
to alter the key, such as a join operation.

In essence, the skew-aware plan 
consists of two plans, 
one that works on the light component (\textit{light plan}), 
and one that works on 
the heavy component (\textit{heavy plan}).  
The \textit{light plan} follows the 
standard implementation for all operators. The light plan 
ensures that values corresponding to light keys reside in 
the same partition. 
The \textit{heavy plan} works on 
the heavy input following an alternative implementation. 
The heavy plan ensures that the 
values associated to heavy keys are distributed across 
partitions. This ability to preserve the partitioning 
guarantee of light keys and ensure the distribution 
of large collections associated to heavy keys enables
\textit{skew-resilient} processing.

Figure~\ref{fig:skewops} 
provides the implementation of the main skew-aware 
operations.
The skew-aware operators assume that the set of heavy keys is small. 
The threshold used to compute heavy keys puts an upper bound on their number; 
for example, the threshold of $2.5\%$ means there can be at most $\frac{100}{2.5}=40$ 
distinct heavy keys in the sampled tuples per partition.
This small domain allows for lightweight broadcasting of heavy keys and, 
in the case of the skew-aware join operations, the  
heavy values of the smaller relation. Broadcast operations 
maintain skew-resilience by ensuring the heavy key values 
in the larger relation remain at their current location and 
are not consolidated to the same node.

All nest operations merge the light and heavy components and follow the standard implementation, 
returning a skew-triple with an empty heavy component and a null set 
of heavy keys. Aggregation $\Gamma^{+}$ mitigates 
skew-effects by default by reducing the values of all keys. 
A grouping operation $\Gamma^{\bagunion}$  
cannot avoid skew-related bottlenecks. More importantly, 
skew-handling for nested output types would be detrimental to our design -- 
attempting to solve a problem that the shredded representation already 
handles gracefully.

The encoding of dictionaries as flat bags means that their distribution is skew-resilient by default.
In the shredded variant, input dictionaries come as skew triples, with known sets of heavy labels.
The skew-aware evaluation 
of a dictionary involves casting of a bag 
with the skew-aware \bagtomatdict operation. This maintains 
the skew-resilience of dictionaries by repartitioning 
only light labels, and leaving the heavy labels in their 
current location, as shown in Figure~\ref{fig:skewops}.

\begin{figure}[t]
\centering
\begin{tabular}{ll@{}}
\toprule
Plan Operator & Definition \\
\midrule
$X \join_{f(x) = g(y)} Y$ &
\begin{lstlisting}[language=Scala,basicstyle=\small\ttfamily] 
val (X_L, X_H, hk) = X.heavyKeys(f)
val Y_L = Y.filter(y => !hk(g(y)))
val Y_H = Y.filter(y => hk(g(y)))

val light = X_L.join(Y_L, f === g)
val heavy = X_H.join(
  Y_H.hint("broadcast"), f === g)

(light, heavy, hk)
\end{lstlisting} 
\vspace{3pt}
\\
\midrule
${\Gamma^{agg}\,}_{key}^{value\,} X$  &
\begin{lstlisting}[language=Scala,basicstyle=\small\ttfamily] 
val unioned = X_L.union(X_H)

// proceed with light plan
val light = ... 

(light, sc.emptyDataset, null)
\end{lstlisting} 
\vspace{3pt}
\\
\midrule
\bagtomatdict$X$ & 
\begin{lstlisting}[language=Scala,basicstyle=\small\ttfamily] 
val (X_L, X_H, hk) = 
  X.heavyKeys(x => x.label)

val light = X_L.repartition(x => x.label)
val heavy = X_H

(light, heavy, hk)
\end{lstlisting} 
\vspace{3pt}
\\
\bottomrule
\end{tabular}
\vspace{3pt}
\caption{Skew-aware implementation for the plan language operators using Spark Datasets.}\label{fig:skewops}
\end{figure}

\begin{figure*}[t]  
  \begin{subfigure}{0.49\textwidth}
    \centering   
    \includegraphics[width=\linewidth]{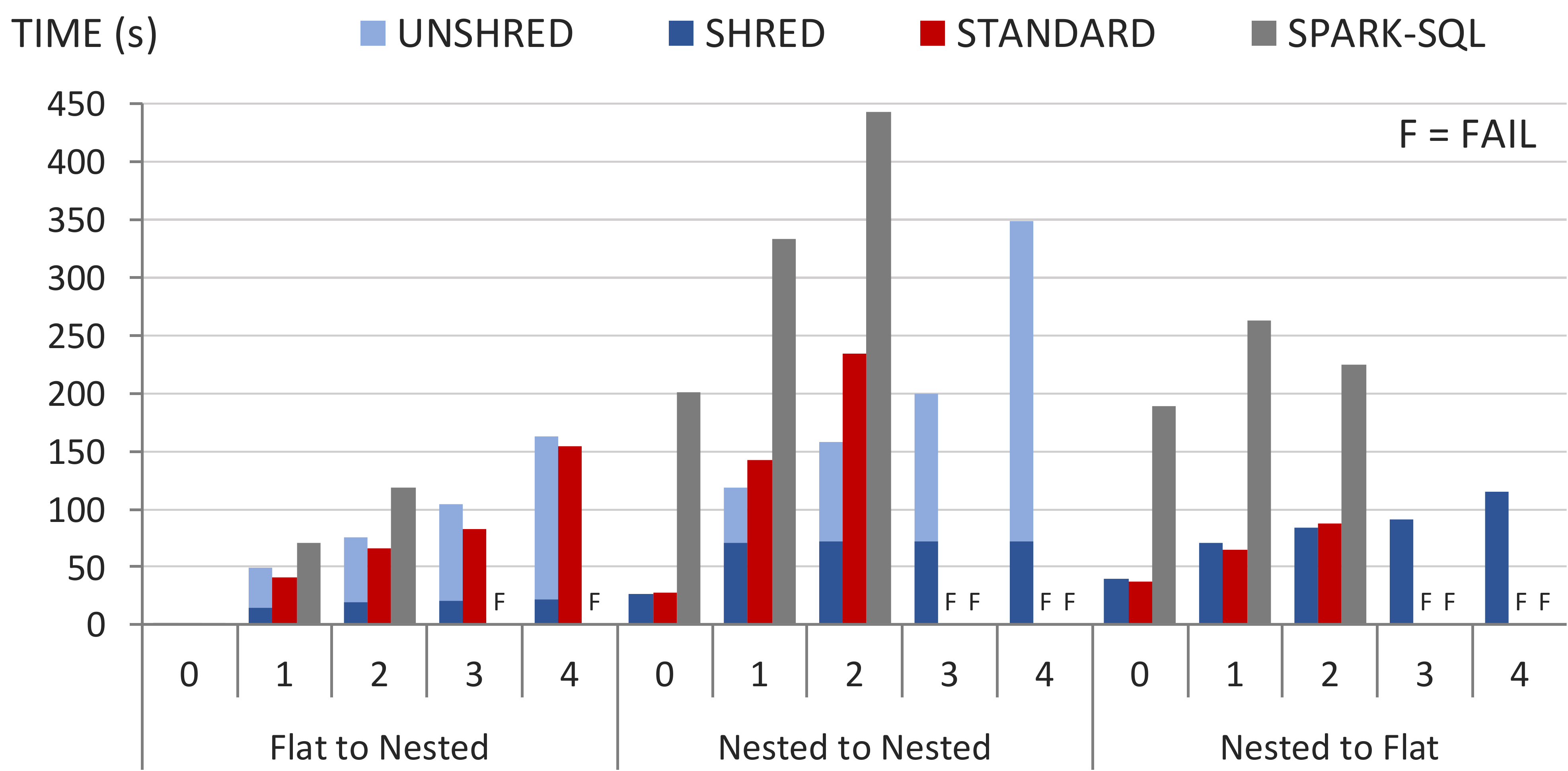}
    \caption{Narrow schema}
    \label{fig:exp_thin}
  \end{subfigure}
  \quad
  \begin{subfigure}{0.49\textwidth}
    \centering   
    \includegraphics[width=\linewidth]{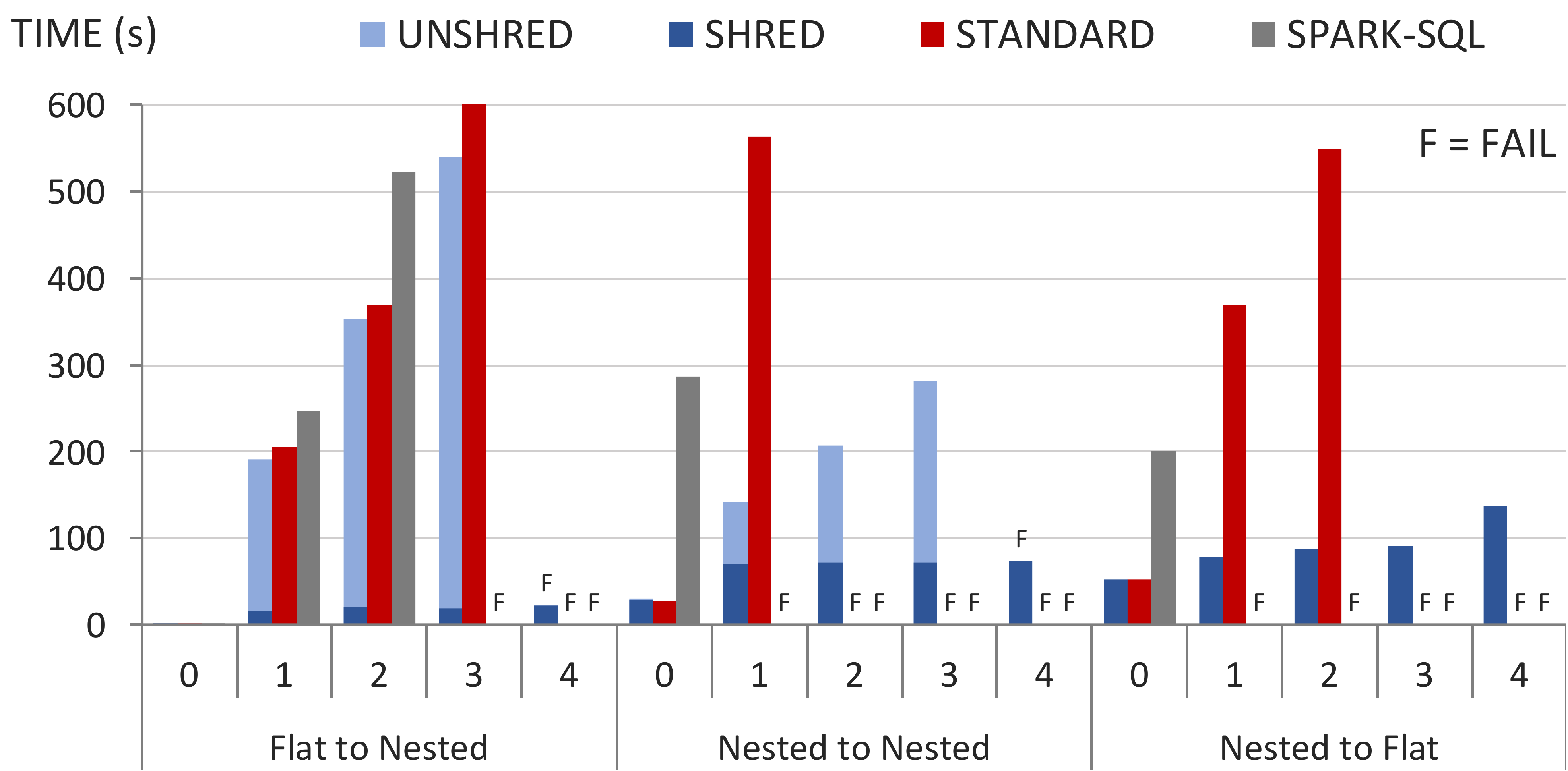}
    \caption{Wide schema}
    \label{fig:exp_wide}
  \end{subfigure}
  \vspace{5pt}
  \caption{Performance comparison of the narrow and wide benchmarked TPC-H queries with varying levels of nesting (0-4).}
  \label{fig:exp1}
  \vspace{-5pt}
\end{figure*}

\myeat{
\begin{figure*}[t]  
  \begin{subfigure}{0.33\textwidth}
    \centering   
    \includegraphics[width=\linewidth]{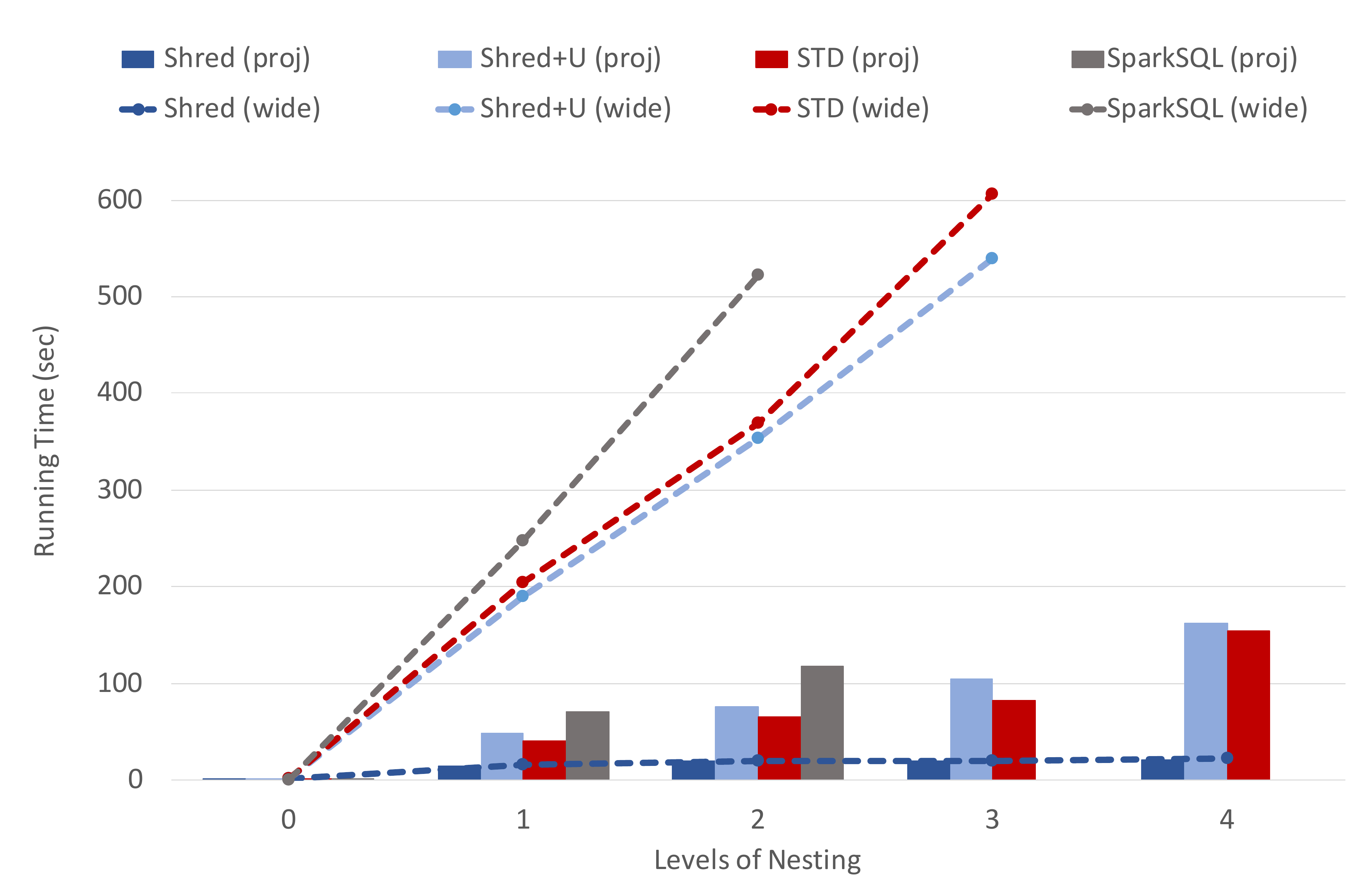}
    \caption{Flat to Nested}
    \label{fig:exp1}
  \end{subfigure}
  \;
  \begin{subfigure}{0.33\textwidth}
    \includegraphics[width=\linewidth]{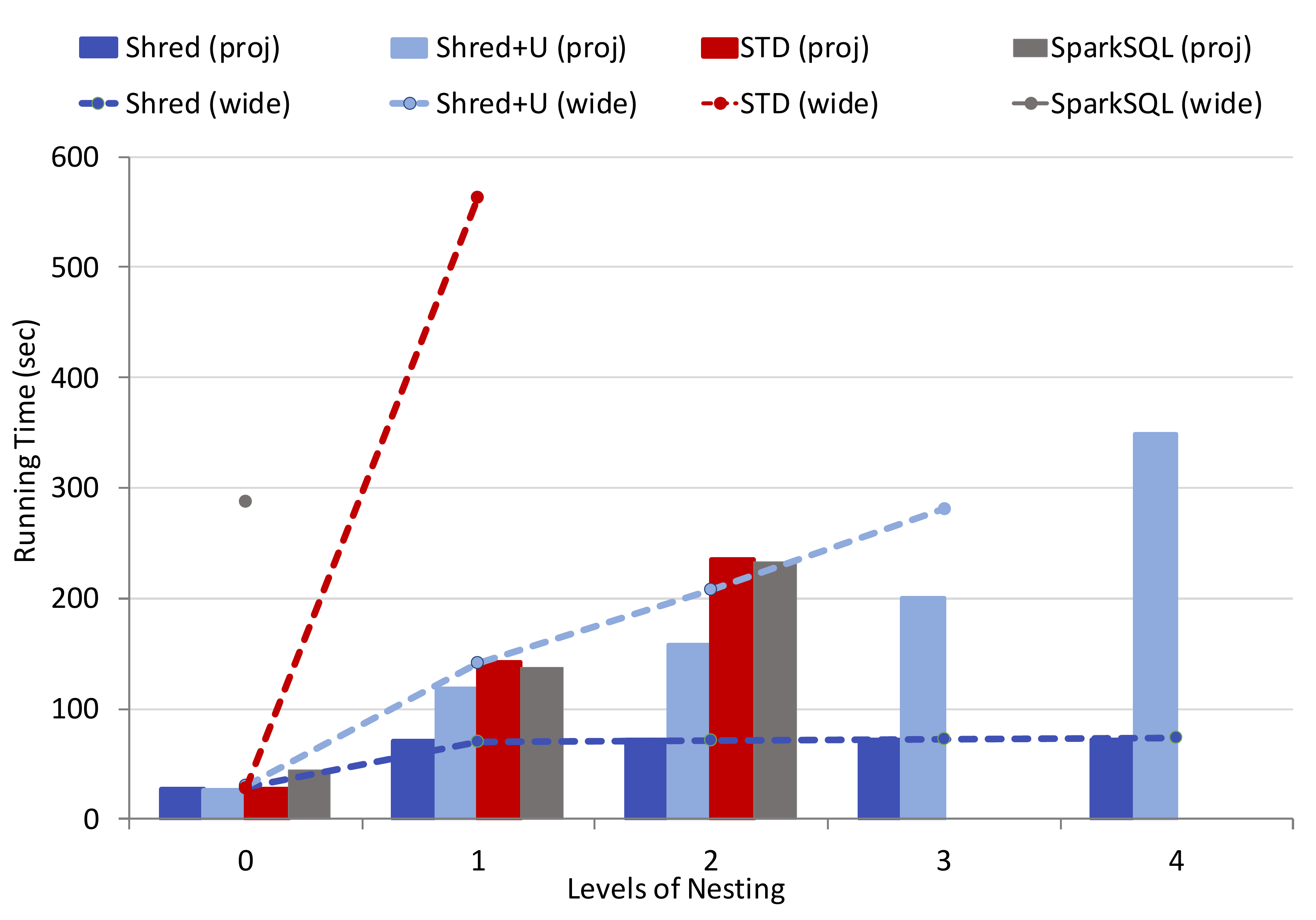}
    \caption{Nested to Nested}
    \label{fig:exp2}
  \end{subfigure}
  \;
  \begin{subfigure}{0.33\textwidth}
    \includegraphics[width=\linewidth]{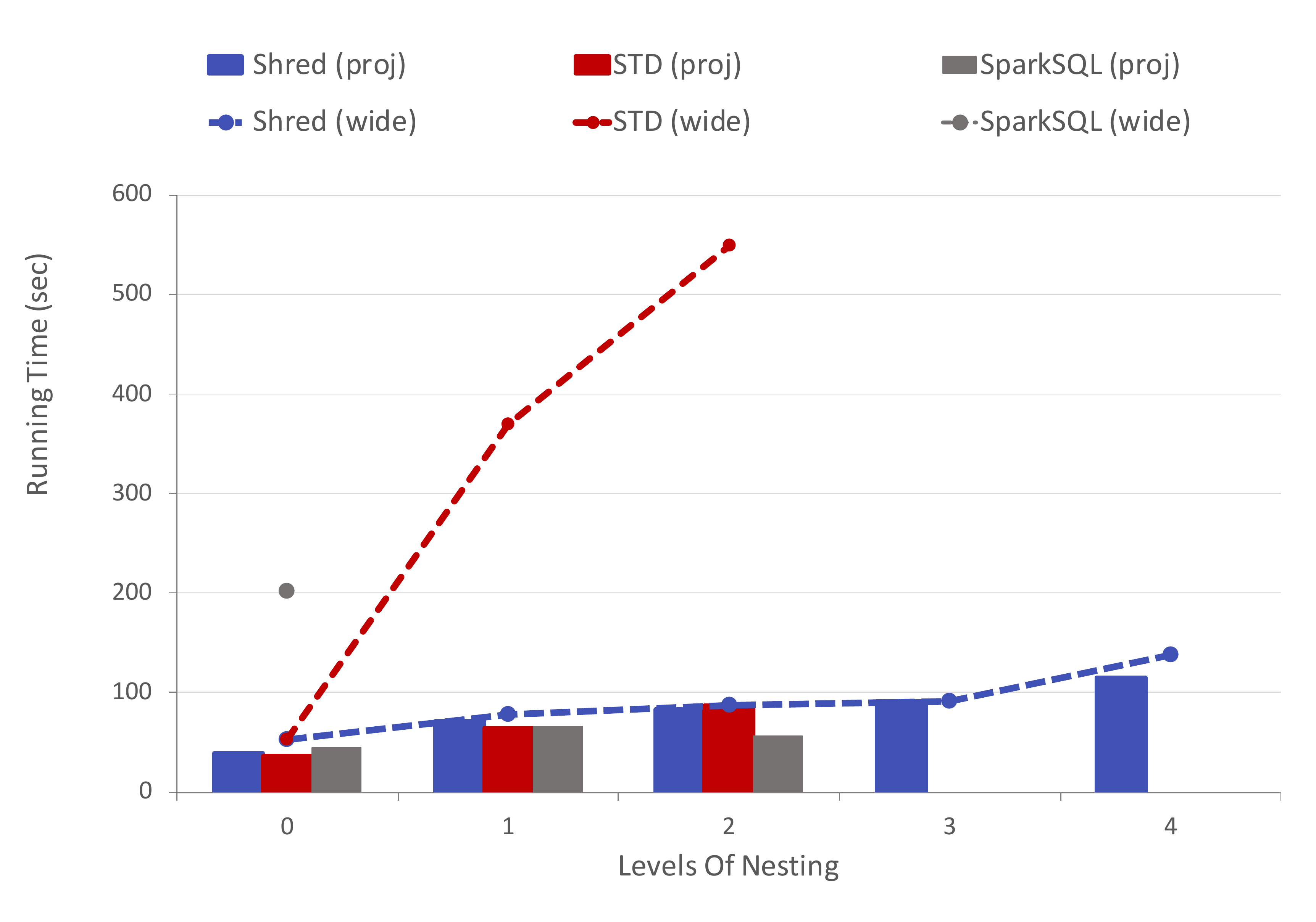}
    \caption{Nested to Flat}
    \label{fig:exp5}
  \end{subfigure}
  \vspace{5pt}
  \caption{Performance comparison of the benchmarked queries.}
\end{figure*}
}

\section{Experiments}\label{sec:experiments}
We evaluate the performance of the standard and 
shredded variants of our framework and compare against 
existing systems that 
support nested data processing for both 
generated and real-world datasets. 
One goal of the experiments is to evaluate how the 
succinct dictionary representation of shredding 
compares to the flattening methods. 
The following experiments look at how the strategies scale 
with increased levels of nesting, number of top-level tuples, 
and the size of inner collections.
We explore how aggregations can reduce 
dictionaries and downstream dictionary operations 
for nested and flat output types.
Finally, to highlight our ability to handle skewed data, 
we evaluate the performance of our framework
for increasing amounts of skew.

Our experimental results can be summarized as follows:
\begin{compactitem} 
\item For flat-to-nested queries, the shredded variant adds no overhead 
when producing nested results and shows up to 25x gain when producing shredded results.
\item For nested input, the shredded compilation route has a 16x
improvement for shallow nesting and scales to 
deeper levels which  flattening methods cannot handle.
\item The shredded data representation provides more opportunities for optimizations in nested-to-flat 
queries with aggregation, providing a 6x improvement over the flattening methods.
\item 
The skew-aware shredded compilation route outperforms flattening methods with 15x
improvement in runtime and 33x reduction in shuffling for moderate skew, as well as graceful handling of increasing amounts of skew \emph{while the flattening methods, skew-aware and skew-unaware, are unable to complete at all}.
\end{compactitem}
Full details of the experimental evaluation can be found in the Appendix. 
Framework and experimental implementation can be found in our github 
repository~\cite{ourgit}.

\myparagraph{Experimental environment} 
Experiments were run on a Spark 2.4.2 cluster (Scala 2.12, Hadoop 2.7) 
with five workers, each with 20 cores and 320G memory. 
Each experiment was run as a Spark application with 25 executors, 
4 cores and 64G memory per executor, 32G memory allocated to the driver, 
and 1000 partitions used for shuffling data. 
Broadcast operations 
are deferred to Spark, which broadcasts anything under 10MB.
Total reported runtime starts after caching all inputs. We 
provide summary information on shuffling cost -- full details of 
which can be found in the github repository.
All missing values correspond to
a run that crashed due to memory saturation of a node.

\myparagraph{Benchmarks} We create two NRC query benchmarks; 
a micro-benchmark based on the standard TPC-H 
schema and the second based on biomedical datasets
from the International Cancer Genome Consortium (ICGC) \cite{icgc}. 

The TPC-H benchmark contains a suite of flat-to-nested, 
nested-to-nested, and nested-to-flat queries -- all with  
$0$ to $4$ levels of nesting and organized such that the 
number of top-level tuples decrease as the level of nesting 
increases. All queries start with the Lineitem table at level 0,
then group across Orders, Customer, Nation, 
then  Region, as the level increases.
Each query has a \emph{wide} variant where we keep all the attributes,
and a \emph{narrow} variant which follows the grouping with a projection at each level.
For scale factor $100$, this organization gives query results with $600$ million, 
$150$ million, $15$ million, $25$, and $5$ top-level tuples. 
\emph{Flat-to-nested} queries perform the iterative grouping above to the relational inputs,
returning a nested output.  At the lowest
level we keep the partkey and quantity of a Lineitem. At the higher levels the narrow variant keeps only a single attribute,
e.g., order\_date for Orders, customer\_name for Customer, etc.
The \emph{nested-to-nested} queries take the materialized 
result of the flat-to-nested queries as input and perform a join with $\Part$ at
the lowest level, followed by $\sumby^{\lqty* \pprice}_{\pname}$, 
as in Example~\ref{ex:run}. 
The nested-to-nested queries thus produce the same hierarchy as
the flat-to-nested queries. 
Finally, the \emph{nested-to-flat} queries follow
the same construction as the nested-to-nested queries, but
apply $\sumby^{\lqty * \pprice}_{name}$ at top-level, where $name$ is one of the top-level
attributes; this returns a flat collection persisting only 
attributes from the outermost level.
We use the skewed TPC-H data generator~\cite{tpch} to 
generate 100GB datasets with a Zipfian distribution; 
skew factor 4 is the greatest amount of skew, with a few heavy keys 
occurring at a high frequency. Non-skewed data is generated with 
skew factor 0, generating keys at a normal distribution as in the 
standard TPC-H generator.

The biomedical benchmark includes an end-to-end
pipeline $\bioendtoend$, based on \cite{weicancer}, consisting of $5$ steps. 
Matching one of the major motivations of this work, results of the
intermediate steps produce  output that has significant nesting.

The inputs include a two-level nested relation
\OccurFull (280GB) \cite{icgc,vep}, a one-level nested relation \Network (4GB) \cite{string}, 
and five relational inputs -- the most notable of which are 
\ExprFull (23G), \CNVFull (34GB), and \SeqOnto (5KB) \cite{icgc, seqonto}.
The first two steps of $\bioendtoend$ are the most expensive.
\Stepi flattens the whole of 
\OccurFull, while performing a nested join on each level
(\CNVFull at level 1 and \SeqOnto at level 2),
aggregating the result, and finally grouping
to produce nested output. \Stepii joins and 
aggregates \Network on the first-level of the output 
of \Stepi. Information on the additional steps 
and additional queries of the biomedical benchmark 
can be found in Appendix \ref{sec:biobench}.




\myeat{
The benchmark also includes three queries reflecting web-based 
exploratory analysis that occurs through clinical user interfaces 
\cite{openmrs, i2b2}. The queries are nested-to-nested,  
representing a scenario where the clinician wants 
to explore nested data results, and
each query applying an additional operation on the 
next. \Ci groups $DN_2$ to return a three-level 
nested output. \Cii  
joins $DF_3$ at level 1 of $DN_2$ then groups 
as in \Ci. \Ciii proceeds the same way and aggregates the result 
of the join prior to grouping. 
}


\myparagraph{Evaluation strategies and competitors} 
We compare the standard compilation route 
(Section~\ref{sec:pipeline_basic}) 
to the shredded compilation route (Section~\ref{sec:pipeline_shredding}).
On each of our benchmarks, we have also compared to a wide array of external competitors:
an implementation via encoding in SparkSQL \cite{sparksql};
Citus, a distributed version of Postgres \cite{citus}; MongoDB \cite{mongodb}, and the recently-developed nested relational
engine DIQL \cite{diql}.
Since we found that SparkSQL outperformed
the other competitors, we include only SparkSQL in the body of the paper, deferring discussion of the others to the github repository.
SparkSQL does not support explode (i.e., UNNEST) operations in the SELECT clause, requiring 
the operator to be kept with the source relation which forces flattening 
for queries that take nested input. 
The SparkSQL queries were manually 
written based on this restriction. 
\Fpplus denotes the standard compilation route and
produces results that match the type of the input query.  \Fpplus also serves
as  a means to explore 
the general functionality of flattening methods. 
\Shred denotes the shredded compilation route with domain elimination, leaving its output
in shredded form. \Unshred represents the time required to 
unshred the materialized dictionaries, 
returning results that match the type of the input query. \Unshred 
is often considered in combination with \Shred to denote total 
runtime of the shredded compilation route when the output type is nested.
\Shred assumes a downstream operation can consume materialized 
dictionaries.
The above strategies do not implement skew-handling; 
we use \Fskew, \Sskew, and \Uskew \ to denote the 
skew-handling variations. 
When executing queries, we use the
optimal plan for a given strategy, which at a
minimum includes pushing 
projections and domain-elimination (Section \ref{subsec:domainelim}) 
any further optimizations are described within the context of each 
experiment.

\myparagraph{Flat-to-nested queries for non-skewed data}
The TPC-H flat-to-nested queries are used
to explore building nested structures from flat inputs 
when the data is not skewed. 
Figure~\ref{fig:exp1} shows the running times of 
SparkSQL, \Fpplus, \Shred, and \Unshred \ for 
all levels of nesting. 

\Shred runs to completion for all levels, remaining 
constant after the first level and exhibiting nearly
identical runtimes and max data shuffle for narrow and wide. 
\Unshred and \Fpplus have comparable runtimes overall and 
a max data shuffle that is 20x that of \Shred. 
For deeper levels of nesting,
these methods fail to store the local nested collections 
when tuples are wide.
The use of cogroup in these 
methods (Section \ref{sec:codegen}) 
provides performance advantages for shallow
nesting in comparison to SparkSQL, which is unable to 
scale for the small number of top-level tuples that 
occur at deeper levels of nesting even for narrow tuples.   
\Shred shows 6x improvement for narrow queries and 26x for wide queries.
The flat-to-nested case introduces a worst case for the shredded compilation route, 
where the shredded query is executed only to group everything up 
in the unshredding phase. Even in this case, 
the shredded compilation route adds no additional overhead in comparison 
to flattening.

\myparagraph{Nested-to-nested queries for non-skewed data}
The TPC-H nested-to-nested queries are used
to explore the effects of aggregation on non-skewed data. 
We use the wide version of the flat-to-nested queries as 
input to evaluate the impact of projections on nested input.
Figure~\ref{fig:exp1} shows the runtimes of SparkSQL, \Fpplus, \Shred, 
and \Unshred for increasing levels of nesting.
\myeat{Point zero shows the time needed to join \Lineitem and \Part 
and perform the sum aggregate. 
The runtime remains the
same across all strategies regardless of projections, 
showing that the shredded compilation route brings no overhead 
when performing SQL-like aggregate queries over flat relations.} 
SparkSQL exhibits consistently worse performance while maintaining 
the same total amount of shuffled memory as \Fpplus. 
This behavior starts for 0 levels of nesting, suggesting
overheads related to pushing projections that may be compounded 
when combined with the explode operator at deeper levels of 
nesting. SparkSQL does not survive even for one-level of nesting 
for wide tuples. 
The rest of the methods diverge with respect to runtime 
at one level of nesting, with \Shred 
exhibiting the best performance overall. 
\myeat{With no nesting, the aggregation groups by \pname thus upper bounding 
the size of the result to the size of $\Part$.
With one level of nesting, the grouping key 
is now (\texttt{orderkey}, \pname) and
returns a much larger result. This aggregation 
is the most expensive operation of 
\Shred, which has constant runtime for 
all levels of nesting.}

For wide queries \Fpplus is burdened by the 
top-level attributes that must be included in the
$\sumby^{\lqty*\pprice}_{\pname}$ aggregate
operation, and only finishes for one level of nesting.
In the narrow case, \Fpplus survives the aggregation but 
fails for the small number of top-level tuples 
in the deeper levels of nesting. 
\Unshred shows a linear drop in performance 
as the number of top-level tuples decrease from 150 million to 25 tuples, 
but at a much lower rate compared to the flattening methods. 
With 5 top-level tuples, \Unshred suffers from a poor distribution 
strategy that must maintain the wide tuples in nested local collections,
and crashes due to memory saturation. 
Overall, \Shred and \Unshred
show up to a 8x performance advantage for shallow 
levels of nesting and exhibit 3x less total shuffle.
Further, in comparison to the previous results without aggregation, 
the localized 
aggregation produces up to a 3x performance gain 
in unshredding. 
This experiment highlights how the succinct representation 
of dictionaries is a key factor 
in achieving scalability. 
The use of this succinct representation 
enables localized aggregation (Section \ref{sec:pipeline_shredding_ext})
which avoids data duplication, reduces the size of dictionaries, 
and lowers unshredding costs to support
processing at deeper levels of nesting even when 
the output to be returned is nested.

\myparagraph{Nested-to-flat with non-skewed inputs}
We use nested-to-flat queries to explore effects of aggregation 
for queries that navigate over multiple levels of the input; 
such queries are challenging for shredding-based approaches due to 
an increase in the number of joins in the query plan. 
Figure~\ref{fig:exp1} shows the runtimes for SparkSQL, \Fpplus and \Shred, where
the unshredding cost for flat outputs is zero.  Similar to 
the nested-to-nested results, \Fpplus is
unable to run for the small number of top-level tuples associated with 
deeper levels of nesting for both the narrow and wide case. 
SparkSQL has the worst performance overall, 
and cannot complete for even one level of nesting when 
tuples are wide.

\Shred outperforms \Fpplus with a 6x runtime advantage,
over a 2x total shuffle
advantage for wide queries, and runs to completion 
even when \Fpplus fails to perform at all. 
At two levels of nesting, the execution of \Shred begins 
with a localized join between the lowest-level 
Lineitem dictionary and \Part, then aggregates the result.
Unlike the flattening procedures, these operations 
avoid carrying redundant information from top-level tuples, 
thereby reducing the lowest-level dictionary as much as possible.
The evaluation continues by dropping all non-label attributes from 
the first-level Order dictionary, which reduces the 
first-level dictionary and results in a cheap join operation 
to reassociate the next two levels. This benefit 
persists even as the number of intermediate lookups increase for 
deeper levels of nesting, exhibiting resilience to the 
number of top-level tuples. 

Due to the nature of the non-skewed TPC-H data, 
\Fpplus does not 
benefit from local aggregation (before unnesting) 
and also does not benefit from pushing the aggregation 
past the nested join since the top-level information 
must be included in the aggregation key. These results 
show how the shredded representation can introduce 
more opportunities for optimization in comparison to 
traditional flattening methods, supporting 
more cases where pushed and localized operations 
can be beneficial.  


 \begin{figure}
 \center
 \includegraphics[width=0.995\linewidth]{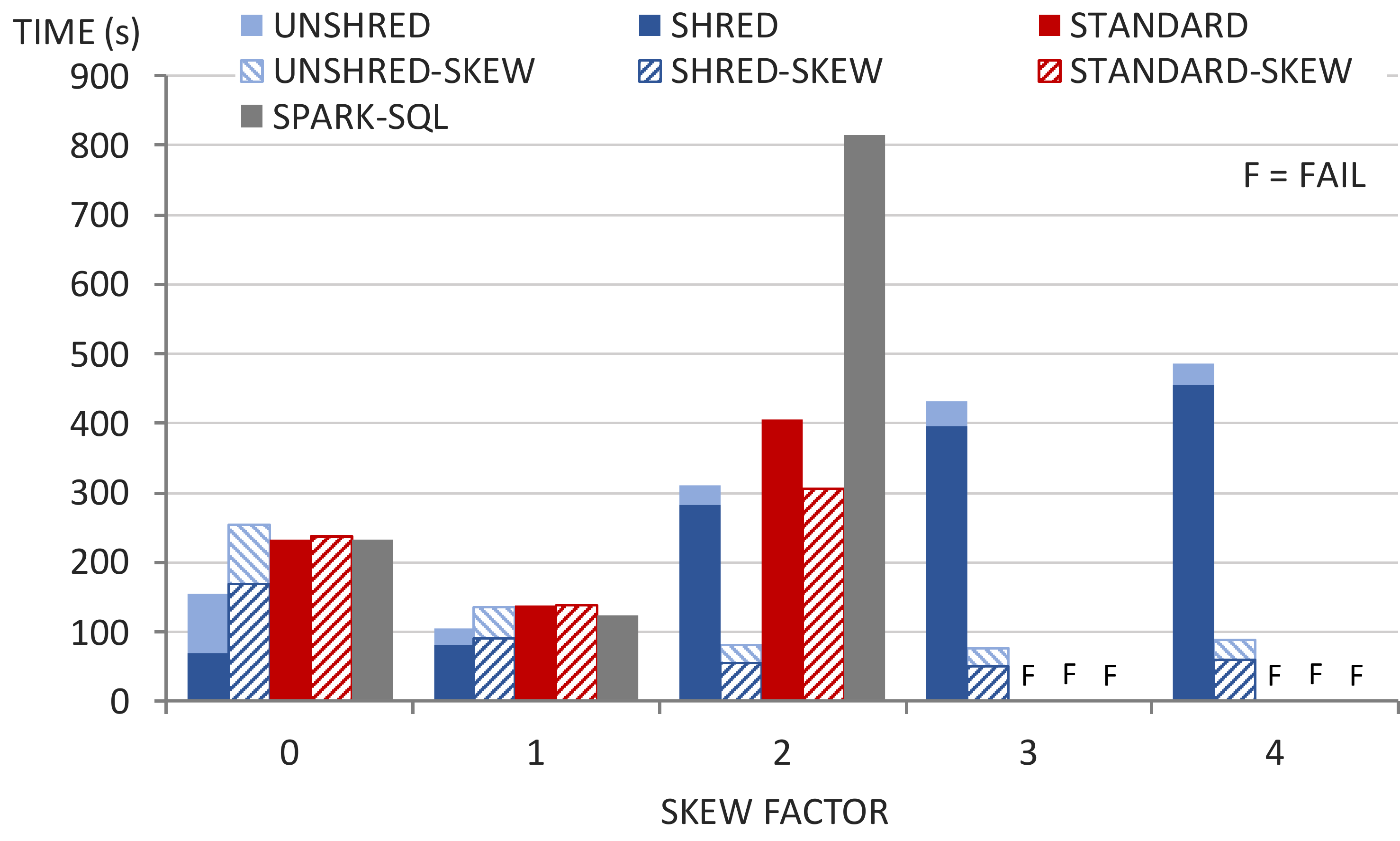}
 \vspace{-9pt}
 \caption{Nested-to-nested narrow TPC-H query with two levels of nesting on increasingly skewed datasets.}
 \vspace{-12pt}
 \label{fig:exp4b}
 \end{figure}

\myparagraph{Skew-handling}
We use the narrow variant of the nested-to-nested queries 
with two levels of nesting to evaluate our skew-handling procedure.
We use the materialized flat-to-nested narrow query ($\COP$) 
as input. 

The TPC-H data generator produces skew by duplicating 
values; thus, pushing aggregations will reduce 
the duplicated values associated to a heavy key 
thereby diminishing skew. 
When designing this experiment, we found that 
skew-unaware methods benefit from aggregation pushing, 
whereas skew-aware methods benefit more from maintaining 
the distribution of heavy keys (skew-resilience, Section \ref{sec:skew}). 
Figure~\ref{fig:exp4b} shows the runtimes for all methods 
for increasing amounts of skew. Skew-unaware methods are reported
with pushed aggregation and skew-aware methods without pushed aggregation. 
Appendix \ref{sec:skewlocalagg} 
provides results without aggregation 
pushing for all methods, which suggests
further benefits to skew-aware methods in the case where 
aggregation pushing does not reduce skew. 

The results in Figure~\ref{fig:exp4b} show that
\Sskew has up to a 15x performance gain over other 
methods for moderate
amounts of skew (skew-factor 2). 
The skew join of \Sskew shuffles 
up to 33x less for moderate skew and 
up to 74x less for high skew than the skew-unaware join of \Shred, 
which has also first performed aggregation pushing. 
At greater amounts of skew (skew-factor 3 and 4), 
SparkSQL, \Fpplus, and \Fskew crash due to memory saturation 
from flattening skewed inner collections. 

The performance gain of the skew-aware shredded compilation route over the 
skew-aware standard compilation route is attributed to the succinct representation 
of dictionaries, which provides better support for the skew-aware operators.
The shredded compilation route, regardless of 
skew-handling, runs to completion for all skew factors.
Beyond supporting the distribution of the large-skewed inner collections, 
the localized aggregation reduces the size of the lower-level dictionary 
thereby decreasing the skew. 
This highlights how the shredded representation is able to 
handle skew even before the skew-handling procedure is applied. 
Overall, we see that the skew-aware components ads additional 
benefits by maintaining skew-resilience and 
reducing the amount of data shuffle required to handle heavy keys. 

\myeat{The skew-aware shredded compilation route outperforms all 
strategies. \Sskew shows a 14x improvement over \Shred, and 
\Uskew shows a 10x improvement over \Unshred. 
Given the general performance gain of the localized aggregation, 
the difference between \Shred and \Sskew highlight the 
advantages of skew-resilience (Section~\ref{sec:skew}) --
the skew-aware join and \bagtomatdict 
maintaining the distribution of the values associated to heavy keys. 
Interestingly,
\Sskew and \Uskew show improved  performance as  skew increases. 
This suggests that as skew increases, the shredded compilation route is not affected by 
the size of the bags associated to the heavy keys, with
skew-resilience ensuring these bags will remain distributed. 
These results also validate the assumption that 
the set of heavy keys will be small, supporting the use of broadcast-based methods 
for the skew-aware join.
Overall, we see that the skew-aware, shredded compilation route outperforms all other methods, 
gracefully handling increasing skew, even when the flattening 
methods are unable to run at all.}


 \begin{figure}
     \center
     \includegraphics[width=0.95\linewidth]{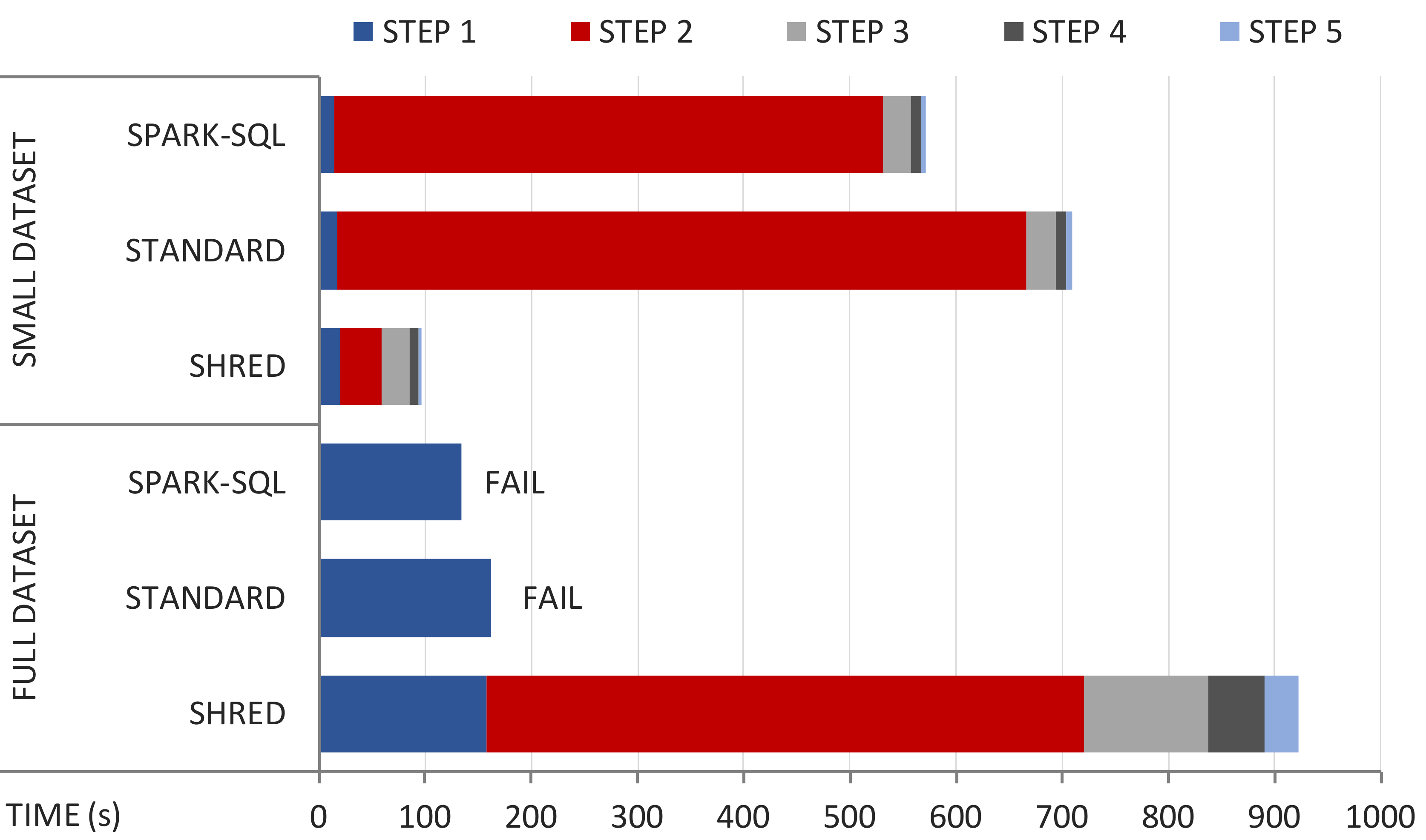}
     \vspace{3pt}
     \caption{End-to-end pipeline of the biomedical benchmark.}
     \vspace{-9pt}
     \label{fig:bioexp1}
\end{figure}

\myparagraph{Biomedical Benchmark} 
Figure~\ref{fig:bioexp1} shows the runtimes for SparkSQL,  \kw{Std}, 
and \Shred for the 
$\bioendtoend$ pipeline of the biomedical benchmark.
The final output of $\bioendtoend$ is flat, so no unshredding 
is needed. \Shred exhibits many advantages, overall surviving the 
whole pipeline.  Since \Fpplus and SparkSQL fail during \Stepii, 
we also provide results with a smaller  dataset
using only  6GB \OccurFull, 2GB \ExprFull, 
and 4GB \CNVFull. 
The results for \Stepi using the small and full dataset 
exhibit similar behavior to the 
nested-to-flat results for TPC-H; given that the whole 
\OccurFull input is flattened and there are many projections,
\Fpplus and SparkSQL are not too burdened by executing nested 
joins via flattening.

The methods diverge at \Stepii, where a nested join 
between \Network and the output of \Stepi leads to an 
explosion in the amount of data being shuffled. 
\Shred displays up to a 16x performance gain 
for \Stepii with the small dataset. 
Despite the 
small number of output attributes, the result of the join 
between flattened output of \Stepi and \Network produces over 
16 billion tuples and shuffles up to 2.1TB before 
crashing. The results for \Shred highlight that this 
is an expensive join, the longest running time for the 
whole pipeline; however, the succinct dictionary 
representation allows for a localized join at the 
first level of the output of \Stepi, reducing 
the join to 10 billion tuples and 
the shuffling to only 470GB. Thus we avoid
carrying around redundant information from the top-level 
and avoid the outer-join required to properly 
reconstruct the top-level tuples.

The $\bioendtoend$ queries  thus
exhibit how a succinct representation of dictionaries 
is vital for executing the whole of the pipeline, 
with a 7x advantage for \Shred on the small dataset 
and scaling to larger datasets when the  
flattening methods are unable to perform. 
The queries also highlight how an aggregation pipeline that eventually 
returns flat output can leverage the succinct representation of 
dictionaries without the need for reassociating the 
dictionaries. 

\section{Related Work}\label{sec:related}
Declarative querying against complex objects has been investigated for
decades in different contexts, from stand-alone implementations
on top of a functional language, as in Kleisli~\cite{kleisli}, language-integrated querying approaches such as
Links~\cite{links} and LINQ~\cite{linq}.  Alternative approaches such as Emma \cite{emma} give more fine-grained
programming abstractions, integrating into  a host language like Scala.

One of the first insights in the field was that
 queries returning flat output, intermediate subqueries
that produce   nesting absent in the input can be eliminated.
These \emph{conservativity} results \cite{conservativity,wongconservativity} were original
obtained in the presence of set  collections only, but were later extended
to multi-sets, where they are often phrased as  a \emph{normalization} step
in processing nested queries \cite{fegarasmaier}.

When extending to nested output, a natural approach is to consider  the simulation
of nested queries on flat representation.
This dates to early foundational studies of the nested relational model  \cite{limsoonthesis,vandenbussche},
where it was seen as an extension of conservativity results.

Implementation of this idea in the form of query shredding has been  less investigated, but
it  has been utilized by Cheney et al.~\cite{cheneyshred} in the 
Links system and Grust et al.~\cite{ferry, grust_aval} in the Ferry 
system. The thesis of Ultich \cite{ulrichphd} contains an in-depth overview of 
Both these systems focus on the generation of SQL queries. 
Koch et al. \cite{pods16} provide a shredding algorithm in order
to reduce incremental evaluation of nested queries to incremental evaluation
of flat queries.  They do not deal with scalable evaluation or target
any concrete processing platform. However, our symbolic shredding transformation
is closely related to the one provided by \cite{pods16}, differing primarily
in that  the language of \cite{pods16} supports only queries  returning bag type without
aggregation.


Query unnesting~\cite{fegarasmaier,nestedopt},
on the other hand,  deals with the programming model mismatch, both for flat queries
and nested ones. Fegaras and Maier's unnesting algorithm \cite{fegarasmaier} de-correlates standard $\nrc$ queries to support a
 more efficient bulk implementation, exploiting the richness
of the object data model. In the process they introduce an attractive calculus for manipulating
and optimizing nested queries. 
A number of recent applications of nested data models build on this calculus.
For example, CleanM~\cite{cleanm} uses it as the frontend language for data cleaning and generates Spark code. 
Similarly, DIQL~\cite{diql} and MRQL~\cite{mrql} provide embedded DSLs in Scala 
generating efficient Spark and Flink programs with query unnesting and normalization.
These works do not deal with limitations of standard nested representations, nor do they
provide support for skew handling.


Google's large scale analytics systems such as Spanner~\cite{spanner}, F1~\cite{f1google,f1query}, and Dremel~\cite{dremel}
support querying against complex objects. 
 Dremel performs evaluation over a ``semi-flattened''~\cite{afrati2014storing} format in order
to avoid the space inefficiencies caused by fully flattening data.
Skew-resilience and query processing performance are not discussed in \cite{afrati2014storing}, which focuses
on the impact  on storage, while details of the query-processing
techniques applied in the commercial systems are proprietary.
\revision{Skew-resilience in parallel processing \cite{skewparallel,suciusurvey}, and methods for efficient  identification  of heavy keys
\cite{flowjoin}
have been investigated for relational data. But for nested
data the only related work we know of targets parallel processing on  a low-level parallel language \cite{suciuoldvldb}, rather than
current frameworks like Spark.}

\section{Conclusion}\label{sec:conclusion}

Our work takes a step in exploring how components like shredded representations, query unnesting,  
and skew-resilient processing fit together to support scalable processing of nested collections.
Our results show that the platform has promise in 
automating the challenges that arise for large-scale, distributed 
processing of nested collections; showing scalable performance 
for deeper levels of nesting and skewed collections even 
when state-of-the-art flattening methods are unable to perform at all. 
In the process we have developed both a micro-benchmark
and a benchmark based on a real-world, biomedical use case.
A number of components of a full solution still remain to be explored. 
A crucial issue, and a target of our ongoing work, is cost estimation for these programs, and the application of such estimates to optimization decisions within the 
compilation framework.

~                                                                                                                                                
~         


\balance

\bibliographystyle{abbrv}
\bibliography{arxiv}

\appendix
The appendix, beginning on the next page, 
contains the implementation of the 
Spark operators, the details of the TPC-H and biomedical 
benchmark, as well as additional and extended experimental 
results.
\renewcommand{\Network}{\texttt{Network}\xspace}
\def\sojoin{\setbox0=\hbox{$\bowtie$}%
  \rule[-.02ex]{.25em}{.4pt}\llap{\rule[\ht0]{.25em}{.4pt}}}
\def\sleftouterjoin{\mathbin{\sojoin\mkern-5.8mu\bowtie}}

\onecolumn

\section{Spark Implementation of the Plan Operators}

Figure \ref{fig:plan_operators} details the Spark implementations of the 
plan operators (Section 2.2) 
applied during code generation (Section 3.2). All operations 
use Dataset[R], where \texttt{R} is an arbitrary Scala case class. 
For outer unnest operation, $\outerunnest^{a}(X)$, we use the function
\texttt{index} to create a unique index associated to the top-level 
input of the unnest operator.

As mentioned in Section 3.2, Spark Datasets are 
a special instance of RDD that use encoders to avoid the large memory footprint of 
using RDDs of Scala case classes. Though similar in their underlying data type, 
the Dataset and RDD APIs are different and thus have different implementations 
for the language operators. Figure \ref{fig:plan_operators_rdd} provides the implementation of the operators of the plan language using Spark RDDs. 
The experiment in Section \ref{sec:rddvsds} highlights the advantages of 
using Spark Datasets as the underlying type in the code generator. 

 \begin{figure}[t]
\begin{adjustbox}{max width=0.7\textwidth,center}
 \begin{small}
 \begin{tabular}{l|l@{}}
 \toprule
 Plan Operator & Definition for Dataset API \\[3pt]
 \midrule
 $\sigma_{p(x)}(X)$ &
 \begin{lstlisting}[language=Scala]
 X.filter(x => p(x))
 \end{lstlisting}  \\[6pt]
  \midrule
 $\pi_{\,a_1,\ldots,a_k}(X)$ &
 \begin{lstlisting}[language=Scala,mathescape] 
 X.select(a$_\texttt{1}$, $\ldots$, a$_\texttt{k}$)
 \end{lstlisting} \\[6pt]
  \midrule
 $\mu^{a_i}(X)$ &
 \begin{lstlisting}[language=Scala] 
 // R does not contain x.a$_\texttt{i}$
 X.flatMap(x => x.a$_\texttt{i}$.map(y =>                                             
    R(x.a$_\texttt{1}$, $\ldots$, x.a$_\texttt{k}$, y.b$_\texttt{1}$, $\ldots$, y.b$_\texttt{j}$))).as[R]
 \end{lstlisting} \\[6pt]
  \midrule
 $\outerunnest^{a}(X)$ &
 \begin{lstlisting}[language=Scala] 
 // R does not contain x.a$_\texttt{i}$ 
 X.withColumn(index, monotonically_increasing_id()).flatMap(x =>                     
   if (x.a$_\texttt{i}$.isEmpty) 
    R(x.index, x.a$_\texttt{1}$, $\ldots$, x.a$_\texttt{k}$, None, $\ldots$, None)
   else x.a$_\texttt{i}$.map(y => 
     R(x.index, x.a$_\texttt{1}$, $\ldots$, x.a$_\texttt{k}$, Some(y.b$_\texttt{1}$), $\ldots$, Some(y.b$_\texttt{j}$)))).as[R] 
 \end{lstlisting} \\[6pt]
   \midrule
 $X \join_{f(x) = g(y)} Y$ &
 \begin{lstlisting}[language=Scala] 
 X.join(Y, f === g)
 \end{lstlisting} \\[6pt]
   \midrule
 $X \sleftouterjoin_{f(x) = g(y)} Y$ &
 \begin{lstlisting}[language=Scala] 
 X.join(Y, f === g, "left_outer")
 \end{lstlisting} \\[6pt]
   \midrule
$\Gamma_{\,+,\, key(x)}^{value(x)}(X)$ &
 \begin{lstlisting}[language=Scala] 
 X.groupByKey(x => key(x)).agg(typed.sum(x => 
   value(x) match { case Some(v) => v; case _ => 0 }))
 \end{lstlisting} \\[6pt]
   \midrule
$\Gamma_{\,\bagunion,\, key(x)}^{value(x)}(X)$  &
 \begin{lstlisting}[language=Scala] 
 X.groupByKey(x => key(x)).mapGroups{ case (key, values) =>     
   val grp = values.flatMap{ case x => 
     value(x) match { case Some(t) => Seq(t); case _ => Seq() }}.toSeq
   (key, grp)
 }
 \end{lstlisting} \\[3pt]
 \bottomrule
 \end{tabular}
 \end{small}
 \end{adjustbox}
 \vspace{3pt}
  \caption{Plan language operators and their semantics for Spark Datasets.}\label{fig:plan_operators}
 \end{figure}

 \begin{figure}[t!]
   \begin{adjustbox}{max width=0.7\textwidth,center}
\begin{small}
\begin{tabular}{l|l@{}}
\toprule
Plan Operator & Definition for RDD API \\[3pt]
\midrule
$\sigma_{p(x)}(X)$ &
\begin{lstlisting}[language=Scala] 
X.filter(x => p(x))
\end{lstlisting}  \\[6pt]
   \midrule
$\pi_{\,a_1,\ldots,a_k}(X)$ &
\begin{lstlisting}[language=Scala,mathescape] 
X.map(x => R(x.a$_\texttt{1}$, $\ldots$, x.a$_\texttt{k}$))
\end{lstlisting} \\[6pt]
   \midrule
$\mu^{a_i}(X)$ &
\begin{lstlisting}[language=Scala] 
// $\texttt{x.drop(a)}$ excludes attribute $\texttt{a}$ from tuple $\texttt{x}$
X.flatMap(x => x.a.map(y => (x.drop(a), y)))    
\end{lstlisting} \\[6pt]
   \midrule
$\outerunnest^{a}(X)$ &
\begin{lstlisting}[language=Scala] 
// $\texttt{x.dropAddIndex(a)}$ excludes attribute $\texttt{a}$ from tuple $\texttt{x}$, 
// and adds a unique index for every x-tuple
X.flatMap(x => if (x.a.isEmpty) Vector((x, null)) 
  else x.a.map(y => (x.dropAddIndex(a), y)))
\end{lstlisting} \\[6pt]
   \midrule
$X \join_{f(x) = g(y)} Y$ &
\begin{lstlisting}[language=Scala] 
val keyX = X.map(x => (f(x), x))
val keyY = Y.map(y => (f(y), y))
keyX.join(keyY).values
\end{lstlisting} \\[6pt]
   \midrule
$X \sleftouterjoin_{f(x) = g(y)} Y$ &
\begin{lstlisting}[language=Scala] 
val keyX = X.map(x => (f(x), x))
val keyY = Y.map(y => (f(y), y))
keyX.leftOuterJoin(keyY).values
\end{lstlisting} \\[6pt]
   \midrule
$\Gamma_{\,+,\, key(x)}^{value(x)}(X)$ &
\begin{lstlisting}[language=Scala]
X.map(x => if (value(x) != null) (key(x), value(x)) 
  else (key(x), 0)).reduceByKey(_+_)
\end{lstlisting} \\[6pt]
   \midrule
$\Gamma_{\,\bagunion,\, key(x)}^{value(x)}(X)$  &
\begin{lstlisting}[language=Scala] 
X.map(x => if (value(x) != null) (key(x), Vector(value(x))) 
  else (key(x), Vector())).reduceByKey(_++_)
\end{lstlisting} \\[3pt]
\bottomrule
\end{tabular}
\end{small}
 \end{adjustbox}
  \vspace{3pt}
\caption{Plan language operators and their semantics for Spark RDDs.}\label{fig:plan_operators_rdd}
\end{figure}

\pagebreak

\section{Detailed Description of Nested TPC-H Benchmark}

We detail the nested TPC-H benchmark introduced in our experimental evaluation (Section 6).
The queries are designed for a systematic exploration of nested queries 
within a distributed environment, focusing on a small number of top-level tuples 
and large inner collections. The queries range from $0$ to $4$ levels of nesting, 
organized such that the number of top-level tuples decrease as the 
level of nesting increases. 
All queries start with the Lineitem table at level 0,
then group across Orders, Customer, Nation, 
then  Region, as the level of nesting increases.
Each query has a \emph{wide} variant where we keep all the attributes,
and a \emph{narrow} variant which follows the grouping with a projection at each level.

For each of the query categories below, 
we provided the input NRC, 
the optimal plan produced from the standard compilation route \Fpplus and the shredded variant \Shred. The shredded compilation route can produce a plan with 
or without unshredding (reconstructing the nested object). 
The shredded compilation route with unshredding will
materialized the output of the shredded query and pass it to the 
unshredding process. Where relevant, 
we describe the subplan corresponding to the unshredding process 
and discuss optimizations introduced by the code generator.

\subsection{Flat-to-nested}\label{sec:flattonest}

Here we detail the flat-to-nested queries of the benchmark, 
which build up nested objects from flat input.
For scale factor $100$, this organization gives query results with $600$ million, 
$150$ million, $15$ million, $25$, and $5$ top-level tuples. 
\emph{Flat-to-nested} queries perform the iterative grouping above to the relational inputs,
returning a nested output.  
This starts with the Lineitem table ($0$ levels), Lineitem grouped by Orders (\oparts), 
\oparts\ grouped by Customers (\corders),
\corders\ grouped by Nation (\ncusts), \ncusts\ grouped by Region (\rnations). 
At the lowest
level we keep the partkey and quantity of a Lineitem. 

Each of the queries have a narrow and wide variant. 
At the higher levels the narrow schema keeps only a single attribute,
e.g. orderdate for Orders, customername for Customer, etc. The 
wide variant returns all attributes except at the lowest level, 
which still just has the partkey and quantity of Lineitem. 
The ellipses represent the additional fields that may be present 
based on narrow and wide schemas. 

We provide the query with 4 levels, since 
queries with different levels are merely subsets of this query. 

\subsubsection{Input NRC Program}
\vspace{10pt}
\begin{lstlisting}[language=NRC]
$\cidfor ~ r ~ \cidin ~ \Region ~ \cidunion$
  $\{\<\, \rname \ateq r.\rname,$ $\ldots$, $\rnations \ateq$
    $\cidfor ~ n ~ \cidin ~ \Nation ~ \cidunion$
      $\cidif ~ r.\rid == n.\rid ~ \cidthen$
      $\{\<\, \nname \ateq n.\nname,$ $\ldots$, $\ncusts \ateq$
        $\cidfor ~ c ~ \cidin ~ \Customer ~ \cidunion$
          $\cidif ~ n.\nid == c.\nid ~ \cidthen$
          $\{\<\, \cname \ateq c.\cname,$ $\ldots$, $\corders \ateq$ 
            $\cidfor ~ o ~ \cidin ~ \Order ~ \cidunion$
              $\cidif ~ c.\cid == o.\cid ~ \cidthen$  
                $\{\<\, \odate \ateq o.\odate,$ $\ldots$, $\oparts \ateq$ 
                  $\cidfor ~ l ~ \cidin ~ \Lineitem ~ \cidunion$
                    $\cidif ~ o.\oid == l.\oid ~ \cidthen$
                      $\{\<\, \pid \ateq l.\pid,$ $\ldots$, $\,\lqty \ateq l.\lqty \,\>\}
                \,\>\}
            \,\>\}
        \,\>\}
    \,\>\}$
\end{lstlisting}

\subsubsection{Plan produced by the standard compilation route}

\begin{forest}
for tree={
  s sep=10mm,
  l sep = 0.75em, 
  l = 0,
},
[$\pi_{\rname, \ldots, \rnations}$,
[$\sleftouterjoin_{\rid}$
  [\Region]
  [${\mbox{\large$\Gamma^{\bagunion\,}$}}^{\rid}_{\texttt{\nname, \ldots, \ncusts}}$
    [$\sleftouterjoin_{\nid}$
        [\Nation]
        [${\mbox{\large$\Gamma^{\bagunion\,}$}}^{\nid}_{\texttt{\cname, \ldots, \corders}}$
          [$\sleftouterjoin_{\cid}$
              [\Customer]
              [${\mbox{\large$\Gamma^{\bagunion\,}$}}^{\cid}_{\texttt{\odate, \ldots, \oparts}}$
                [$\sleftouterjoin_{\oid}$
                  [\Order]
                  [${\mbox{\large$\Gamma^{\bagunion\,}$}}^{\oid}_{\texttt{\lqty, \pprice}}$
                    [\Lineitem]]]]]]]]]]
\end{forest}

The sequential join-nest operations in the above plan will be merged 
into cogroups during code generation (Section 3.3). We implement the 
cogroup in a left-outer fashion, persisting empty bags from the 
right relation for every matching tuple in the left relation. 
As an example, the join and nest over \Order and \Lineitem 
in the above plan will be translated to the following during 
code generation:

\medskip 

\begin{lstlisting}[language=Scala,basicstyle=\ttfamily] 
Orders.groupByKey(o => o.oid)
      $\,$.cogroup(Lineitem.groupByKey(l => l.oid))(
          case (key, orders, lineitems) => 
            val oparts = 
              lineitems.map(l => (l.pid, l.lqty)).toSeq
            orders.map(o => (o.odate, oparts)))
\end{lstlisting}

\subsubsection{Plan produced by the shredded compilation route}

Below we show the plan produced by the shredded variation of our 
compilation framework for the 
evaluation of the shredded query prior to unshredding. 
The shredded variant produces an additional plan 
that corresponds to the execution strategy of unshredding; this unshredding 
plan is identical to the plan produced by the standard compilation route, 
with each input relation represented as a top-level bag. 

\medskip 

$\RNCOP_{Top} := $
\begin{forest}
[$\pi_{\rname, \ldots, \rnations \ateq \rid }$($\Region$)]
\end{forest}

$\rnations_{Dict} := $
\begin{forest}
[$\pi_{\labelatt \ateq \rid, \nname, \ldots, \ncusts \ateq \nid}$($\Nation$)]
\end{forest}

$\ncusts_{Dict} := $
\begin{forest}
[$\pi_{\labelatt \ateq \nid, \cname, \ldots, \corders \ateq \cid}$($\Customer$)]
\end{forest}

$\corders_{Dict} := $
\begin{forest}
[$\pi_{\labelatt \ateq \cid, \odate, \ldots, \oparts \ateq \oid}$($\Order$)]
\end{forest}

$\oparts_{Dict} := $
\begin{forest}
[$\pi_{\labelatt \ateq \oid, \pid, \lqty}$($\Lineitem$)]
\end{forest}

\medskip 

\subsection{Nested-to-nested}

The \emph{nested-to-nested} queries operate on 
nested input; these queries
take the materialized 
result of the flat-to-nested queries as input, and perform a join with $\Part$ at
the lowest level, followed by $\sumby^{\lqty\times \pprice}_{\pname}$, 
as in Example 1. 
The nested-to-nested queries thus produce the same hierarchy 
and number of top-level tuples as
the flat-to-nested queries for all levels of nesting. 
The ellipses represent the additional fields that may be 
present in the narrow and wide version of the queries. 
We provide the query for 4 levels of nesting, 
all other queries can be derived from this query.

\subsubsection{Input NRC Program}

\medskip 

\begin{lstlisting}[language=NRC]
$\cidfor ~ r ~ \cidin ~ \RNCOP ~ \cidunion$
  $\{\<\, \rname \ateq r.\rname,$ $\ldots$, $\rnations \ateq$
    $\cidfor ~ n ~ \cidin ~ r.\rnations ~ \cidunion$
      $\{\<\, \nname \ateq n.\nname,$ $\ldots$, $\ncusts \ateq$
        $\cidfor ~ c ~ \cidin ~ n.\ncusts ~ \cidunion$
          $\{\<\, \cname \ateq c.\cname,$ $\ldots$, $\corders \ateq$ 
            $\cidfor ~ o ~ \cidin ~ c.\corders ~ \cidunion$
              $\{\<\, \odate \ateq o.\odate,$ $\ldots$, $\oparts \ateq$ 
                sumBy$_{\hspace{0.15mm}\pname}^{\hspace{0.15mm}\total}($
                  $\cidfor ~ l ~ \cidin ~ o.\oparts ~ \cidunion$
                    $\cidfor ~ p ~ \cidin ~ \Part ~ \cidunion$
                      $\cidif ~ l.\pid == p.\pid ~ \cidthen$
                        $\{\<\, \pname \ateq p.\pname,$
                          $\,\total \ateq l.\lqty * p.\pprice \,\>\})
                \,\>\}
            \,\>\}
        \,\>\}
    \,\>\}$
\end{lstlisting}

\pagebreak

\subsubsection{Plan produced by the standard compilation route}

\medskip

\begin{forest}
for tree={
  s sep=10mm,
  l sep = 0.75em, 
  l = 0,
},
[$\pi_{\rname, \ldots, \rnations}$,
  [${\mbox{\large$\Gamma^{\bagunion\,}$}}^{\nname, \ldots, \ncusts}_{\texttt{rncopID}, \rname, \ldots}$
    [${\mbox{\large$\Gamma^{\bagunion\,}$}}^{\cname, \ldots, \corders}_{\texttt{rncopID}, \texttt{ncopID},
    \rname, \nname, \ldots}$
      [${\mbox{\large$\Gamma^{\bagunion\,}$}}^{\odate, \ldots, \oparts}_{\texttt{rncopID}, \texttt{ncopID},\texttt{copID},\rname, \nname, \cname, \ldots}$
        [${\mbox{\large$\Gamma^{\bagunion\,}$}}^{\pname, \total}_{\texttt{rncopID}, \texttt{ncopID},\texttt{copID}, \texttt{coID}, \rname, \nname, \cname, \odate, \ldots}$
          [${\mbox{\large$\Gamma^{+\,}$}}^{\lqty*\pprice}_{\texttt{rncopID}, \texttt{ncopID},\texttt{copID}, \texttt{coID}, \rname, \nname, \cname, \odate, \pname, \ldots}$
            [ $\sleftouterjoin_{\pid}$
              [$\outerunnest^{\oparts}$
                [$\outerunnest^{\corders}$
                  [$\outerunnest^{\ncusts}$
                    [$\outerunnest^{\rnations}$
                      [$\RNCOP$]]]]]
              [$\Part$]
            ]]]]]]]
\end{forest}

\subsubsection{Plan produced by the shredded variant}

\medskip

$\RNCOP_{Top} := $
\begin{forest}
[$\pi_{\rname, \ldots, \rnations \ateq \rid }$($\RNCOP_{Top}$)]
\end{forest}

$\rnations_{Dict} := $
\begin{forest}
[$\pi_{\labelatt \ateq \rid, \nname, \ldots, \ncusts \ateq \nid}$($\rnations_{Dict}$)]
\end{forest}

$\ncusts_{Dict} := $
\begin{forest}
[$\pi_{\labelatt \ateq \nid, \cname, \ldots, \corders \ateq \cid}$($\ncusts_{Dict}$)]
\end{forest}

$\corders_{Dict} := $
\begin{forest}
[$\pi_{\labelatt \ateq \cid, \odate, \ldots, \oparts \ateq \oid}$($\corders_{Dict}$)]
\end{forest}

$\oparts_{Dict} := $

\begin{forest}
[$\bagtomatdict$,
  [$\pi_{\labelatt \ateq \oid, \pname, \total}$
    [$\Gamma^{+ / \oid, \pname / \lqty * \pprice}$,   
      [$\join_{\pid}$,
      [$\pi_{\pid, \ldots}$ [$\oparts_{Dict}$]]
      [$\pi_{\pname, \lqty}$ [$\Part$]]]
  ]]]
\end{forest}

\medskip

\subsection{Nested-to-flat}

The \emph{nested-to-flat} queries follow
the same construction as the nested-to-nested queries, but
apply $\sumby^{\lqty \times \pprice}_{name}$ at top-level, where $name$ is one of the top-level
attributes; this returns a flat collection persisting only 
attributes from the outermost level.

The ellipses represent the additional fields that may be 
present in the narrow and wide version of the queries. 
We provide the query for 4 levels of nesting, 
all other queries can be derived from this query.

\pagebreak

\subsubsection{Input NRC}

\medskip

\begin{lstlisting}[language=NRC]
sumBy$_{\hspace{0.15mm}\rname, \ldots}^{\hspace{0.15mm}\total}($
  $\cidfor ~ r ~ \cidin ~ \RNCOP ~ \cidunion$
    $\cidfor ~ n ~ \cidin ~ n.\rnations ~ \cidunion$
      $\cidfor ~ c ~ \cidin ~ n.\ncusts ~ \cidunion$
        $\cidfor ~ o ~ \cidin ~ c.\corders ~ \cidunion$
          $\cidfor ~ l ~ \cidin ~ o.\oparts ~ \cidunion$
            $\cidfor ~ p ~ \cidin ~ \Part ~ \cidunion$
              $\cidif ~ l.\pid == p.\pid ~ \cidthen$
                $\{\<\, \rname \ateq r.\rname, \dots$
                    $\,\total \ateq l.\lqty * p.\pprice \,\>\})$
\end{lstlisting}

\medskip

\subsubsection{Plan produced by the standard compilation route}

\medskip

\begin{forest}
for tree={
  s sep=10mm,
  l sep = 0.75em, 
  l = 0,
},
[${\mbox{\large$\Gamma^{+\,}$}}^{\lqty*\pprice}_{\rname, \ldots, \pname}$
  [ $\sleftouterjoin_{\pid}$
    [$\outerunnest^{\oparts}$
      [$\outerunnest^{\corders}$
        [$\outerunnest^{\ncusts}$
          [$\outerunnest^{\rnations}$
            [$\RNCOP$]]]]]
    [$\Part$]
    ]]
\end{forest}

\medskip

\myparagraph{Plan produced by the  shredded compilation route}

\medskip

\begin{forest}
[${\mbox{\large$\Gamma^{+\,}$}}^{\total}_{\rname, \ldots}$
  [$\sleftouterjoin_{\rnations = \rlabelatt}$,
    [$\pi_{\rname, \rnations, \ldots}$[$\RNCOP_{Top}$]]
    [$\sleftouterjoin_{\ncusts = \nlabelatt}$,
      [$\pi_{\rlabelatt, \ldots}$[$\rnations_{Dict}$]]
      [$\sleftouterjoin_{\corders = \clabelatt}$,
        [$\pi_{\nlabelatt, \corders}$[$\ncusts_{Dict}$]]
        [$\sleftouterjoin_{\oparts = \olabelatt}$, 
          [$\pi_{\clabelatt, \oparts}$[$\corders_{Dict}$]]
          [${\mbox{\large$\Gamma^{+\,}$}}^{\lqty*\pprice}_{\olabelatt, \pname, \ldots}$
            [$\join_{\pid}$,
              [$\pi_{\olabelatt, \pid, \lqty, \ldots}$ [$\oparts_{Dict}$]]
              [$\pi_{\pname, \pprice}$ [$\Part$]]
            ]
          ]
        ]
      ]
    ]
  ]
]
\end{forest}

\newpage

\section{Biomedical Query Benchmark}\label{sec:biobench}

This section overviews the biomedical benchmark, 
which was developed in collaboration with a precision 
medicine start-up.  
The biomedical benchmark includes 
an overview of each biomedical data source, queries of the
end-to-end ($\bioendtoend$) example from the experiments in Section 6, 
as well as additional clinical exploration queries and associated 
performance results.
The biomedical benchmark is a collection of $\nrcagg$ queries 
that perform multiomic analyses, including an 
end-to-end
pipeline $\bioendtoend$ that is based on an 
analysis that uses several genomic datasets
to identify driver genes in cancer\cite{weicancer}. 
Given that cancer progression is determined by the accumulation of mutations and other genomic aberrations within a sample \cite{chengcancer}, this analysis integrates 
somatic mutations, copy number information, protein-protein interactions and 
gene expression data. The benchmark also includes three queries reflecting web-based 
exploratory analysis that occurs through clinical user interfaces \cite{i2b2}. 
This section continues with details on these 
datasets and then describes the queries of the analysis and additional 
queries.

\subsection{Inputs}\label{subsec:inputs}

This section explains the inputs used within the biomedical benchmark. The 
majority of the datasets are provided from the Genomic Data Commons (GDC) \cite{gdcends}, which houses public datasets associated with the International Cancer Genome Consortium (ICGC). 
The types described below are often truncated for simplicity. 
The inputs include a two-level nested relation
\OccurFull (280GB) \cite{icgc,vep}, a one-level nested relation \NetworkFull (4GB) \cite{string}, 
and five relational inputs - the most notable of which are 
\ExprFull (23G), \CNVFull (34GB), and \SeqOnto (5KB) \cite{icgc, seqonto}. 
The implementation of the biomedical benchmark queries can be found in 
the code repository \cite{ourgit} at \url{https://github.com/jacmarjorie/trance/tree/master/compiler/src/main/scala/framework/examples/genomic}.

\subsubsection{\OccurFull: Occurrences}

An occurrence is a single, somatic mutation belonging to a single sample that 
has been annotated with candidate gene information.
\textit{Somatic mutations} are cancer-specific mutations that are identified within each sample, and are identified by comparing a sample's cancerous genome to a non-cancerous, reference genome. Note that the term mutation is often used interchangeably with 
variant. \textit{Candidate genes} are assigned to mutations based on proximity of a 
given mutation to a gene on the reference genome. 
In a naive assignment, a reference gene is a candidate if the mutation 
lies directly upstream, downstream, or on a gene; however, mutations have been shown to form 
long-range functional connections with genes \cite{longrange} and as such candidacy can best be assigned 
based on a larger flanking region of the genome. With this in mind, the same gene could be considered a 
candidate gene for multiple mutations within a sample.

\textit{Variant annotation} is the process that assigns 
candidate genes to every mutation within each sample. A popular annotation 
tool is the Variant Effect Predictor (VEP) \cite{vep}. 
The \Occur input is created by associating each simple somatic mutation file (MAF) 
with nested annotation information from VEP; this 
returns annotations in JSON format with mutation 
information at the top-level, 
a collection of candidate genes for that mutation and corresponding consequence information on the first level, and a collection of additional consequence information
on the second level. Each of the mutations within each sample 
will contain the corresponding nested annotation 
information. 

The tuples in the \cands collection contain attributes that correspond to 
the impact a mutation has on a gene. The \impact attribute is 
a value from 0 to 1 denoting 
any known detrimental consequence a mutation has to a candidate gene. \sift and \poly are 
additional impact scores that rely on prediction software \cite{sift, polyphen}. 
Given that genes code for proteins and proteins have functional consequences 
that are attributed disease, these scores reflect the predicted role 
a mutation has in functional changes to proteins based on alterations to a gene sequence.  
The \conseqs for each candidate gene contain qualitative descriptions of 
mutation impact to a gene sequence based on a standardized set of categorical descriptions 
from the sequence ontology (SO) \cite{seqonto}.

VEP also takes a distance flag to 
specify the upstream and downstream range from which to identify gene-based 
annotations. This flag is used to increase the flanking region of candidate genes 
associated to each somatic mutation for each sample. From a technical standpoint, 
increasing the distance will increase the size of \cands and potentially 
increase the amount of skew.  
The type of \Occur is:
\\[-2pt]
\begin{align*}
&\bagtype\,(\<\, \sample:\stringtype, \contig: \stringtype, \mstart: \inttype, \mend: \inttype, \\[-2pt]
& \hspace{1.2cm} \mref: \stringtype, \alt: \stringtype, \mutid: \stringtype, \\[-2pt] 
& \hspace{1.2cm} \cands:\bagtype\,(\<\, \gene: \stringtype, \impact: \stringtype, \\[-2pt]
& \hspace{2.4cm} \sift: \doubletype, \poly: \doubletype, \conseqs: \bagtype\,(\<\, \conseq: \stringtype \,\>)\, \>)\, \>).
\end{align*}\\[-2pt]

The shredded representation of $\Occur$ consists of a top-level flat bag ${\Occur}^{\flatc}$ of type:
\begin{align*}
&\bagtype\,(\<\, \sample:\stringtype, \contig: \stringtype, \mstart: \inttype, \mend: \inttype, \\[-2pt]
& \hspace{1.2cm} \mref: \stringtype, \alt: \stringtype, \mutid: \stringtype, \\[-2pt] 
& \hspace{1.2cm} \cands: \labeltype \>).
\end{align*}\\[-2pt]
  and a dictionary tree ${\Occur}^{\dictc}$ of tuple type
{\small
\begin{align*}
\<\,&\cands^\FUN: \\
&\tab\tab \labeltype \rightarrow \bagtype(\<\, \gene: \stringtype, \impact: \doubletype,
  \sift: \doubletype, \poly: \doubletype, \break \conseqs: \labeltype \,\>),\\
&\cands^\CHILD: \bagtype(\<\, \\
&\tab\tab \conseqs^\FUN: \labeltype \rightarrow \bagtype(\<\, \conseq: \stringtype \,\>),\\
&\tab\tab \conseqs^\CHILD: \bagtype(\<\, \,\>) \quad\>)\quad\>
\end{align*}
}

The shredded represented of $\Occur$ is representated as three datasets 
in the implementation. The top-level bag is 6GB, the 
first-level dictionary is 281GB, and the third-level dictionary is 35G.

\subsubsection{Somatic Mutations and VEP Annotations}
The \Occur data source described in the above section 
is based off an endpoint from the ICGC data portal; this resource 
is built from combining somatic mutation information 
and variant annotations. Somatic mutations are stored 
in the GDC as MAF files, which is a flat datadump file that 
includes a line for every mutation across all samples. The 
type of the somatic mutation information is:
\\[-2pt]
\begin{align*}
&\bagtype\,(\<\, \sample:\stringtype, \contig: \stringtype, \mstart: \inttype, \mend: \inttype, \\[-2pt]
& \hspace{1.2cm} \mref: \stringtype, \alt: \stringtype, \mutid: \stringtype \>) \\[-2pt] 
\end{align*}

The variant annotations are from the VEP software mentioned above, 
which returns top-level mutation 
information and two additional levels of gene and mutational 
impact information. The overall structure is similar to the occurrences
data source, except VEP is returning a unique set of variant annotations 
that are not associated to a specific sample. 
The type of the VEP annotation information is:
 
\begin{align*}
&\bagtype\,(\<\, \contig: \stringtype, \mstart: \inttype, \mend: \inttype, \\[-2pt]
& \hspace{1.2cm} \mref: \stringtype, \alt: \stringtype, \mutid: \stringtype, \\[-2pt] 
& \hspace{1.2cm} \cands:\bagtype\,(\<\, \gene: \stringtype, \impact: \stringtype, \\[-2pt]
& \hspace{2.4cm} \sift: \doubletype, \poly: \doubletype, \conseqs: \bagtype\,(\<\, \conseq: \stringtype \,\>)\, \>)\, \>).
\end{align*}\\[-2pt]


\subsubsection{\CNVFull: Copy Number}

\textit{Copy number} values correspond to the amplification or 
deamplification of a gene for each sample, and 
are also found by comparing against non-cancerous, reference copy number values.
The copy number information is provided per gene. The copy number information is 
reported for each physical sample taken from a patient; this is denoted \aliquot. 
The type of the copy number information is:
\\[-2pt]
\begin{align*}
&\bagtype\,(\<\, \aliquot: \stringtype, \gene: \stringtype, \cnum: \inttype \,\>).
\end{align*}\\[-2pt]

\subsubsection{\NetworkFull: Protein-protein Interactions}

Protein-protein interaction networks describe the relationship between proteins in a network. 
This \Network input is derived from the STRING \cite{string} database. The network is 
represented with a top-level node tuple and a nested bag of edges, where each edge tuple 
contains an edge protein and a set of node-edge relationship measurements. 
The type is:
\\[-2pt]
\begin{align*}
&\bagtype\,(\<\, \nodep: \stringtype, \edges: \bagtype\,(\<\, \edgep: \stringtype, \dist: \inttype \,\>)\,\>).
\end{align*}\\[-2pt]

\subsubsection{\ExprFull: Gene Expression}

Gene expression data is based on RNA sequencing data. Expression measurements are 
derived by counting the number of transcripts in an \aliquot and comparing it 
to a reference count. The expression measurement is represented as 
Fragments Per Kilobase of transcript per Million mapped read (FPKM), which 
is a normalized count. The type is:
\\[-2pt]
\begin{align*}
&\bagtype\,(\<\, \aliquot: \stringtype, \gene: \stringtype, \fpkm: \doubletype \,\>).
\end{align*}\\[-2pt]



\subsubsection{Mapping Files}

\myparagraph{Sample Metadata}
The \Biospec input maps samples to their aliquots; for the sake of this analysis \sample maps to a patient and \aliquot associates each biological sample taken from 
the patient. The type is:
\\[-2pt]
\begin{align*}
&\bagtype\,(\<\, \sample: \stringtype, \aliquot: \stringtype \,\>).
\end{align*}\\[-2pt]

\myparagraph{\SeqOnto: Sequence Ontology}
The \SOImpact input is a table derived from the sequence ontology \cite{seqonto} 
that maps a qualitative consequence to a quantitative consequence score (\conseq). 
This is a continuous measurement from 0 to 1, with larger values representing 
more detrimental consequences. The type is:
\\[-2pt]
\begin{align*}
&\bagtype\,(\<\, \conseq: \stringtype, \valueatt: \doubletype \,\>).
\end{align*}\\[10pt]

\myparagraph{Biomart Gene Map}
The \Biomart input is 
exported from \cite{biomart}. It is a  map from  gene identifiers to 
protein identifiers. This map is required to associate genes from \Occur and \CopyNum 
to proteins that make up \Network.
The type is:
\\[-2pt]
\begin{align*}
&\bagtype\,(\<\, \gene: \stringtype, \protein: \stringtype \,\>).
\end{align*}\\[-2pt]

The inputs described in this section are used in the queries described 
in the next section.

\subsection{$\bioendtoend$: Pipeline Queries}
The queries of the cancer driver gene analysis are 
an adaptation of the methods from \cite{weicancer}. 
They  work in pipeline fashion to integrate annotated somatic mutation information (\Occur), 
copy number variation (\CopyNum), protein-protein network (\Network), 
and gene expression (\GeneExpr) data. 
The idea is to provide an integrated look at the impact cancer has 
on the underlying biological system. 
The analysis takes into account the effects a mutation has on 
a gene, the accumulation of genes with respect to both copy number 
and expression, and the interaction of genes within the system.

Mutations that play a driving role in cancer often occur at low frequency \cite{somatic}, 
making cohort analysis across many samples important in their identification. 
Further, cancer is not just the consequence of a single mutation on a single gene. 
The interaction between genes in a network, the number of such genes, 
and their expression levels can provide a more thorough look 
at cancer progression \cite{genoabb}. The
queries below define
the analysis. The queries work in pipeline fashion where the 
materialized output from one query is used as input to a query later on in 
the pipeline. 

The 
pipeline starts with the integration of mutation and copy number variation 
to produce a set of hybrid-scores for each sample. The hybrid-scores 
are then combined with network interactions to determine effect-scores. 
The effect-scores are further combined with gene expression information 
to determine the connection scores for each sample. The queries 
conclude by combining the connection scores across all samples, 
returning connectivity scores for each gene. 
The genes with the highest connectivity scores are considered drivers.

\pagebreak

\subsubsection{\Stepi: Hybrid scores}

The hybrid score query is the first step in the pipeline. 
A \textit{hybrid-score} 
is calculated for each candidate gene within a sample by combining mutation impact and copy number 
information for that sample, thus providing a score that corresponds to the likelihood that 
the gene is a driver. 
The output of this step remains grouped by sample in order to continue 
integrating sample-specific genomic datasets that 
can further contribute to our understanding of driver genes in cancer in 
the steps below.

The query below describes the process of creating hybrid scores 
based on the \Occur input. The main difference between this version and the 
running example is that \Biospec provides a map between \sample and \aliquot 
used to join \CopyNum, and the hybrid score is determined for every \aliquot. 
In addition, conditionals are used to assign qualitative scores based on the 
human-interpretable level of impact (\impact). 

\medskip 

\begin{lstlisting}[language=NRC]
$\Hybrid \assigneq$
$\cidfor ~  s ~ \cidin ~ \Biospec ~ \cidunion$
  $\{\<\, \sample \ateq s.\sample,$ $\aliquot \ateq s.\aliquot,$ $\hyscores \ateq$
    sumBy$_{\hspace{0.15mm}\gene}^{\hspace{0.15mm}\hscore}($
      $\cidfor ~ o ~ \cidin ~ \Occur ~ \cidunion$
       $\cidif ~ o.\sample == b.\sample ~ \cidthen$
        $\cidfor ~ t ~ \cidin ~ o.\trans ~ \cidunion$
            $\cidfor ~ n ~ \cidin ~ \CopyNum ~ \cidunion$
              $\cidif ~ s.\aliquot == n.\aliquot ~ \&\& ~ n.\gene == t.\gene ~ \cidthen$
                $\cidfor ~ c ~ \cidin ~ t.\conseqs ~ \cidunion$
                  $\cidfor ~ v ~ \cidin ~ \SOImpact ~ \cidunion$
                    $\cidif ~ c.\conseq == v.\conseq ~ \cidthen$
                      $\{\<\, \gene \ateq t.\gene,$ $\ldots$ $\hscore \ateq$ $\ldots$
                        $\cidlet ~ \impact \ateq$
                          $\cidif ~ t.\impact == "HIGH" ~ \cidthen ~ 0.8$
                          $\cidelse ~ \cidif ~ t.\impact == "MODERATE" ~ \cidthen ~ 0.5$
                          $\cidelse ~ \cidif ~ t.\impact == "LOW" ~ \cidthen ~ 0.3$
                          $\cidelse ~ \cidif ~ t.\impact == "MODIFIER" ~ \cidthen ~ 0.15$ 
                          $\cidelse ~ 0.01$
                        $\cidin ~ \impact * v.\valueatt * (n.\cnum + 0.01) * \sift * \poly
                      \,\>\})
    \,\>\}$
\end{lstlisting}

\medskip 

The output type of this query is:
\\[-2pt]
\begin{align*}
&\bagtype\,(\<\, \sample: \stringtype, \aliquot: \stringtype, \hyscores: \bagtype\,(\<\, \gene: \stringtype, \hscore: \doubletype \,\>)\, \>).
\end{align*}\\[-2pt]

The plan produced by our standard compilation route is:

\medskip 

\begin{forest}
for tree={
  s sep=10mm,
  l sep = 0.75em, 
  l = 0,
},
[$\pi_{\cid, \aid, \hyscores}$,
  [${\mbox{\large$\Gamma^{\bagunion,}$}}^{\gid, \hscore}_{\texttt{cid}, \texttt{oID},\texttt{aid}, \texttt{aID}}$
    [${\mbox{\large$\Gamma^{+\,}$}}^{\impact*\valueatt*(\cnum+0.01)}_{\texttt{cid}, \texttt{oID},\texttt{aid}, \texttt{aID}, \gid, \impact, \conseq, \cnum}$
      [$\sleftouterjoin^{\conseq}$
        [$\SOImpact$]
          [$\outerunnest^{\conseqs}$
            [$\sleftouterjoin_{\aid \&\& \gid}$
              [$\CopyNum$]
                [$\sleftouterjoin_{\cid}$
                  [$\Biospec$]
                  [$\outerunnest^{\trans}$
                    [$\Occur$]
                  ]
                ]
              ]
            ]
        ]
      ]
    ]
  ]
\end{forest}

\pagebreak




\subsubsection{\Stepii: By Sample Network}
The second step in the pipeline 
aggregates on an individual\textsc{\char13}s
sample network, based 
on the hybrid scores. The goal is to associate 
each gene in the nested \edges collection of \Network
with the corresponding hybrid scores for a sample. For each 
gene in this collection, the product of the hybrid score and 
the relationship measurement for that edge (\dist) are summed 
for each node in the network for each sample. 

\vspace{10pt}
\begin{lstlisting}[language=NRC]
$\SNetwork \assigneq$
$\cidfor ~  h ~ \cidin ~ \Hybrid ~ \cidunion$
  $\{\<\, \sample \ateq h.\sample,$ $\aliquot \ateq h.\aliquot,$ $\nodes \ateq$
    sumBy$_{\hspace{0.15mm}\gene}^{\hspace{0.15mm}\score}($
      $\cidfor ~  n ~ \cidin ~ \Network ~ \cidunion$
        $\cidfor ~  e ~ \cidin ~ n.\edges ~ \cidunion$
          $\cidfor ~  b ~ \cidin ~ \Biomart ~ \cidunion$
            $\cidif ~ e.\edgep == b.\protein ~ \cidthen$
              $\{\<\, \nodep \ateq n.\nodep, \score \ateq e.\dist * h.hscore \,\>\})
      \,\>\}$
\end{lstlisting}
\vspace{10pt}
The output type of this query is:
\\[-2pt]
\begin{align*}
&\bagtype\,(\<\, \sample: \stringtype, \aliquot: \stringtype, \nodes: \bagtype\,(\<\, \nodep: \stringtype, \score: \doubletype \,\>)\, \>).
\end{align*}\\[-2pt]

\subsubsection{\Stepiii: Effect scores} The effect scores are calculated using the materialized 
output of the previous two steps, denoted \SNetwork and \Hybrid. 
The nested \nodes collection of \SNetwork contains the sum 
of the combined gene interaction and hybrid score across the \edges collection 
for each node gene in the top-level tuples of \Network. 
These values are then combined with 
the hybrid scores for each node gene to produce the effect matrix.

\vspace{10pt}
\begin{lstlisting}[language=NRC]
$\Effect \assigneq$
$\cidfor ~  h ~ \cidin ~ \Hybrid ~ \cidunion$
  $\{\<\, \sample \ateq h.\sample,$ $\aliquot \ateq h.\aliquot,$ $\hyscores \ateq$
    $\cidfor ~  s ~ \cidin ~ \SNetwork ~ \cidunion$
      $\cidif ~ h.\sample == s.\sample \&\& h.\aliquot == s.\aliquot ~ \cidthen$
        $\cidfor ~  n ~ \cidin ~ s.\nodes ~ \cidunion$
          $\cidfor ~  b ~ \cidin ~ \Biomart ~ \cidunion$
            $\cidif ~ n.\nodep == b.\protein ~ \cidthen$
              $\cidfor ~  y ~ \cidin ~ h.\hyscores ~ \cidunion$
                $\cidif ~ y.\gene == b.\gene ~ \cidthen$
                  $\{\<\, \gene \ateq y.\gene, \score \ateq n.\score * h.\hscore \,\>\}
      \,\>\}$
\end{lstlisting}
\vspace{10pt}
The output type is:
\\[-2pt]
\begin{align*}
&\bagtype\,(\<\, \sample: \stringtype, \aliquot: \stringtype, \hyscores: \bagtype\,(\<\, \gene: \stringtype, \score: \doubletype \,\>)\, \>).
\end{align*}\\[-2pt]

\myparagraph{\Stepiv: Connection scores} Connection scores are determined by combining the 
effect scores and gene expression data. Gene expression data uses the normalized 
count (FPKM) measurement discussed in the input section above.

\vspace{10pt}
\begin{lstlisting}[language=NRC]
$\Connect \assigneq$
$\cidfor ~  s ~ \cidin ~ \Effect ~ \cidunion$
  $\{\<\, \sample \ateq e.\sample,$ $\aliquot \ateq e.\aliquot,$ $\hyscores \ateq$
    sumBy$_{\hspace{0.15mm}\gene}^{\hspace{0.15mm}\score}($
      $\cidfor ~ e  ~ \cidin ~ s.\hyscores ~ \cidunion$
        $\cidfor ~  g ~ \cidin ~ \GeneExpr ~ \cidunion$
          $\cidif ~ e.\gene == g.\gene ~ \cidthen$
            $\{\<\, \gene \ateq e.\gene, \score \ateq e.\score * g.fpkm \,\>\})
      \,\>\}$
\end{lstlisting}
\vspace{10pt}
The output type is:
\\[-2pt]
\begin{align*}
&\bagtype\,(\<\, \sample: \stringtype, \aliquot: \stringtype, \hyscores: \bagtype\,(\<\, \gene: \stringtype, \score: \doubletype \,\>)\, \>).
\end{align*}\\[-2pt]

\pagebreak

\subsubsection{\Stepv: Gene connectivity} The gene connectivity sums up connection scores 
for each gene across all samples. The genes with the highest connection scores are 
taken to be drivers.

\vspace{10pt}
\begin{lstlisting}[language=NRC]
$\Connects \assigneq$
sumBy$_{\hspace{0.15mm}\gene}^{\hspace{0.15mm}\score}($
$\cidfor ~  s ~ \cidin ~ \Connect ~ \cidunion$
  $\cidfor ~  c ~ \cidin ~ s.\hyscores ~ \cidunion$
    $\{\<\, \gene \ateq c.\gene, \score \ateq s.\score\,\>\})$
\end{lstlisting}
\vspace{10pt}
The output type is:
\\[-2pt]
\begin{align*}
&\bagtype\,(\<\, \gene: \stringtype, \score: \doubletype \,\>).
\end{align*}\\[-2pt]

\subsection{Clinical exploration queries}\label{sec:clinexplore}

This section provides an overview of the clinical 
exploration queries in the biomedical benchmark. 
The clinical exploration queries reflect requests a clinician may make from 
a user-interface; for example, an electronic health record system 
that provides access to \Occur and \CopyNum. The 
queries contain a combination of restructuring, nested joins, 
and aggregation. 
The queries are nested-to-nested and
each query applys an additional operation on the 
next. \Ci groups \OccurFull to return a three-level 
nested output. \Cii  
joins \CNVFull at level 1 of \OccurFull then groups 
as in \Ci. \Ciii proceeds the same way and aggregates the result 
of the join prior to grouping. 

\subsubsection{\Ci: Group occurrences by sample}
This query groups occurrences by sample producing a bag of nested 
mutation information for each sample. The query also associates 
a quantitative value to the consequences at the lowest level in the 
process. The output has four levels of nesting. 
\vspace{10pt}
\begin{lstlisting}[language=NRC]
$\OccurGroup \assigneq$
$\cidfor ~  s ~ \cidin ~ \Biospec ~ \cidunion$
  $\{\<\, \sample \ateq s.\sample,$ $\mutations \ateq$ 
    $\cidfor ~  o ~ \cidin ~ \Occur ~ \cidunion$
      $\cidif ~ s.\sample == o.\sample ~ \cidthen$
      $\{\<\, \mutid \ateq o.\mutid,$ $\ldots$, $\cands \ateq$
        $\cidfor ~ t  ~ \cidin ~ o.\cands ~ \cidunion$
          $\{\<\, \gene \ateq t.\gene$, $\ldots$, $\conseqs \ateq$
            $\cidfor ~ c ~ \cidin ~ t.\conseqs ~ \cidunion$
              $\cidfor ~ i ~ \cidin ~ \SOImpact ~ \cidunion$
              $\cidif ~ c.\conseq == i.\conseq ~ \cidthen$
              $\{\<\, \conseq \ateq i.\conseq, \score \ateq i.\valueatt \,\>\}
            \,\>\}
          \,\>\}
        \,\>\}$
\end{lstlisting}
\vspace{10pt}

\subsubsection{\Cii: Integrate copy number and occurrences, group by sample}
This query is similar to above, but joins copy number data on the second 
level - per each gene - while constructing mutation groups per sample. 

\vspace{10pt}
\begin{lstlisting}[language=NRC]
$\OccurJoin \assigneq$
$\cidfor ~  s ~ \cidin ~ \Biospec ~ \cidunion$
  $\{\<\, \sample \ateq s.\sample,$ $\mutations \ateq$ 
    $\cidfor ~  o ~ \cidin ~ \Occur ~ \cidunion$
      $\cidif ~ s.\sample == o.\sample ~ \cidthen$
      $\{\<\, \mutid \ateq o.\mutid,$ $\ldots$, $\cands \ateq$
        $\cidfor ~ t  ~ \cidin ~ o.\cands ~ \cidunion$
          $\cidfor ~ g ~ \cidin ~ \SOImpact ~ \cidunion$
            $\cidif ~ g.\gene == t.\gene ~ \cidthen$
            $\{\<\, \gene \ateq t.\gene$, $\cnum \ateq g.\cnum$, $\ldots$, $\conseqs \ateq$
              $\cidfor ~ c ~ \cidin ~ t.\conseqs ~ \cidunion$
                $\cidfor ~ i ~ \cidin ~ \SOImpact ~ \cidunion$
                $\cidif ~ c.\conseq == i.\conseq ~ \cidthen$
                $\{\<\, \conseq \ateq i.\conseq, \score \ateq i.\valueatt \,\>\}
              \,\>\}
            \,\>\}
          \,\>\}$
\end{lstlisting}

\pagebreak

\subsubsection{\Ciii: Aggregate copy number and occurrences, group by sample}
This query groups by sample, joins copy number at the second level, 
joins quantitative consequence values at the third level,
and aggregates the product of copy number and consequence score 
for each gene. 
\vspace{20pt}
\begin{lstlisting}[language=NRC]
$\OccurAgg \assigneq$
$\cidfor ~  s ~ \cidin ~ \Biospec ~ \cidunion$
  $\{\<\, \sample \ateq s.\sample,$ $\mutations \ateq$ 
    $\cidfor ~  o ~ \cidin ~ \Occur ~ \cidunion$
      $\cidif ~ s.\sample == o.\sample ~ \cidthen$
      $\{\<\, \mutid \ateq o.\mutid,$ $\ldots$, $\cands \ateq$
        sumBy$_{\hspace{0.15mm}\gene}^{\hspace{0.15mm}\score}($
        $\cidfor ~ t  ~ \cidin ~ o.\cands ~ \cidunion$
          $\cidfor ~ g ~ \cidin ~ \SOImpact ~ \cidunion$
            $\cidif ~ g.\gene == t.\gene ~ \cidthen$
              $\cidfor ~ c ~ \cidin ~ t.\conseqs ~ \cidunion$
                $\cidfor ~ i ~ \cidin ~ \SOImpact ~ \cidunion$
                $\cidif ~ c.\conseq == i.\conseq ~ \cidthen$
                $\{\<\, \gene \ateq t.\gene, \score \ateq c.\cnum * i.\valueatt \,\>\})
              \,\>\}
          \,\>\}$
\end{lstlisting}

\subsubsection{Performance}

The clinical queries extend the biomedical benchmark to explore how
queries perform that are not necessarily part 
of a whole analysis pipeline. These queries are 
indicative of a process that would return all data to display 
in a user-interface, which is synonymous to fewer projections 
in the output.

\begin{figure}
 \center
 \includegraphics[width=0.95\linewidth]{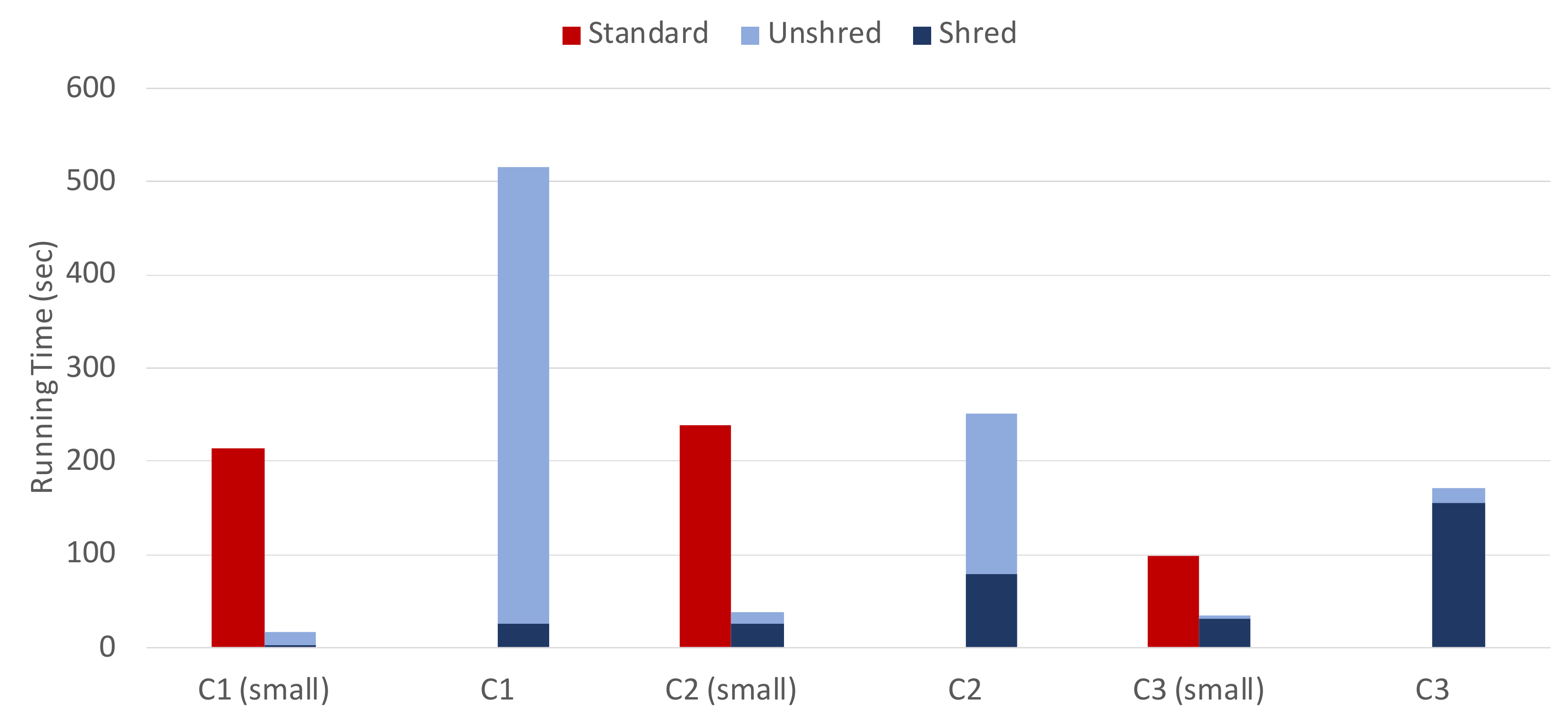}
 \vspace{3pt}
 \caption{Results for the clinical exploration queries.}
 \vspace{-6pt}
 \label{fig:bioexp2}
 \end{figure}

The clinical exploration queries were evaluated using varying 
sizes of the \Occur input; one small collection based on a 168M 
mutation file (identified by small) and a 
large collection of 42G annotated mutations. \textit{Note this 
is a smaller dataset that is used in the $\bioendtoend$}.
Figure \ref{fig:bioexp2} displays these results. 
\Fpplus was unable to run to completion for the larger input 
for all queries,
overloading the available memory on the system each time. 
The shredded variant of our compilation was able to complete for all queries 
exhibiting resilience by distributing the large inner collections.


\newpage

\section{Note about succinct representation and sharing}\label{sec:succintexp}
We use the somatic mutations 
and VEP annotations described in Section \ref{subsec:inputs} 
to explore the benefits of sharing. 
This microexperiment uses one
MAF file from the breast cancer dataset containing 120988 tuples, 
and the associated VEP annotations for the unique set of all mutations 
(58121 tuples). 
When the mutations are joined with the annotation table 
in the standard compilation route,
the result contains 5170132 nested transcript consequence tuples. 
In the shredded compilation route, 
the mutations are joined 
with the top-level annotation dictionary, and first-level dictionary 
of the output is just the first-level transcript consequences 
dictionary from the input (3777092 tuples). 
In comparison to the standard compilation route, 
the dictionary representation 
has reduced the total size of the transcript tuples by over 1 million tuples. 
These results are based off a small subset of the data. Since many of 
the samples will share mutations specific to cancer, 
the benefits of sharing will increase as the number of samples 
increases. 

\section{Additional and Augmented Experimental Results}

This section contains additional experimental results, along with 
extended experimental results and some of the experimental
plots in the paper augmented with memory information.

 All results are for 
100GB of non-skewed TPC-H data (scale factor 100, skew factor 0). 
We use the flat-to-nested and 
nested-to-nested queries to 
compare the performance of the standard and shredded variants of our framework using both RDDs and Datasets.
As in the body of the paper, we use \Fpplus (standard compilation route), \Shred 
(shredded variant without unshredding), and \Unshred (shredded variant with unshredding of the final output) 
to represent runs from our framework.
We also explore the effects of introducing database-style optimizations on the 
plans of the standard compilation route.
The results highlight advantages of using Datasets over RDDs in code generation, particularly for 
nested collections. The results also show how introducing database-style optimizations can 
significantly improve performance of the standard compilation route, generating programs similar to 
programs that have been optimized by hand. 



\begin{figure*}
  \begin{subfigure}{0.95\textwidth} 
    \centering   
    \includegraphics[width=\linewidth]{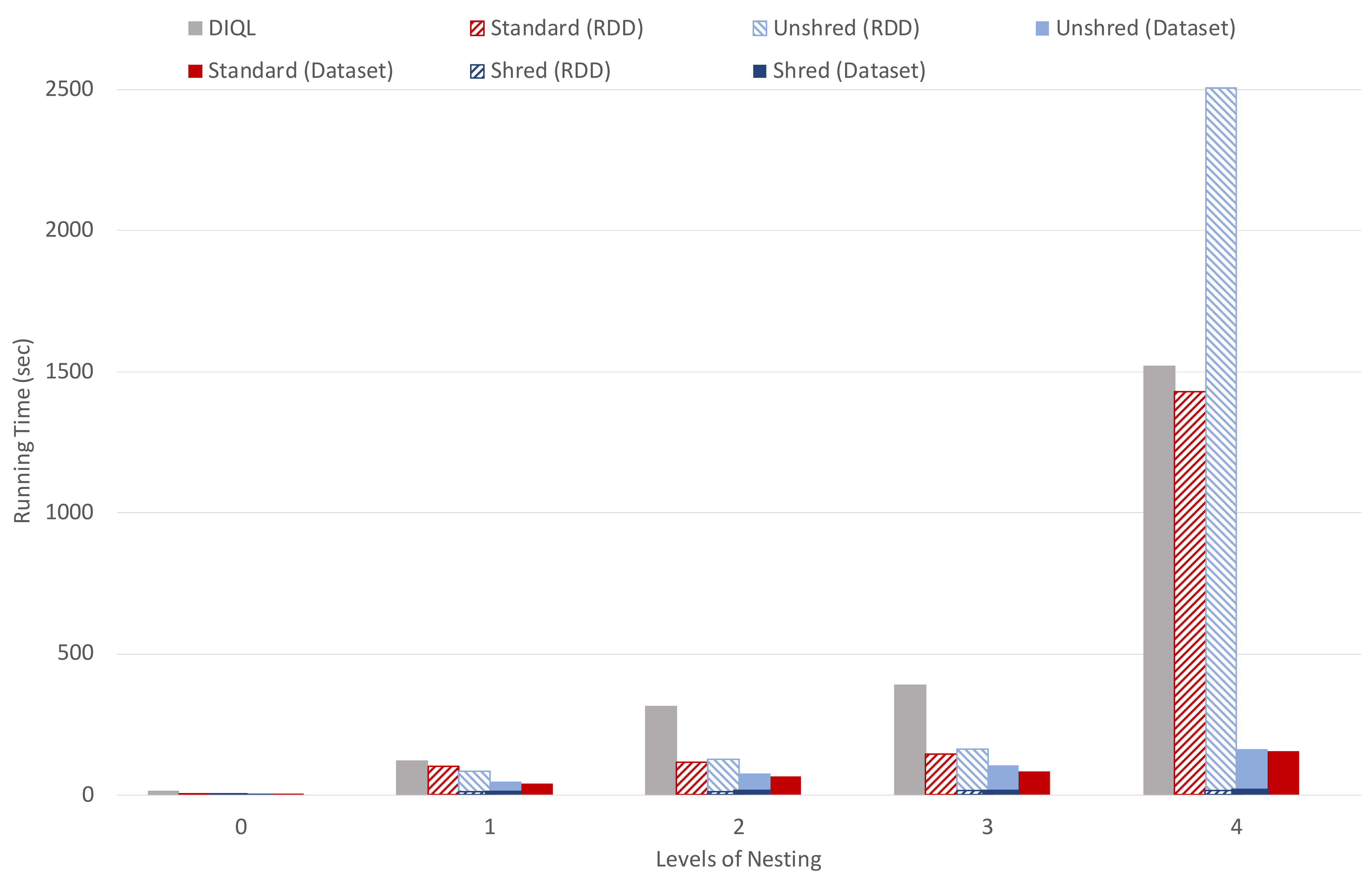}
    \caption{Narrow schema}
    \label{fig:rddvsdfs}
  \end{subfigure}
  \begin{subfigure}{0.95\textwidth} 
      \centering   
    \includegraphics[width=\linewidth]{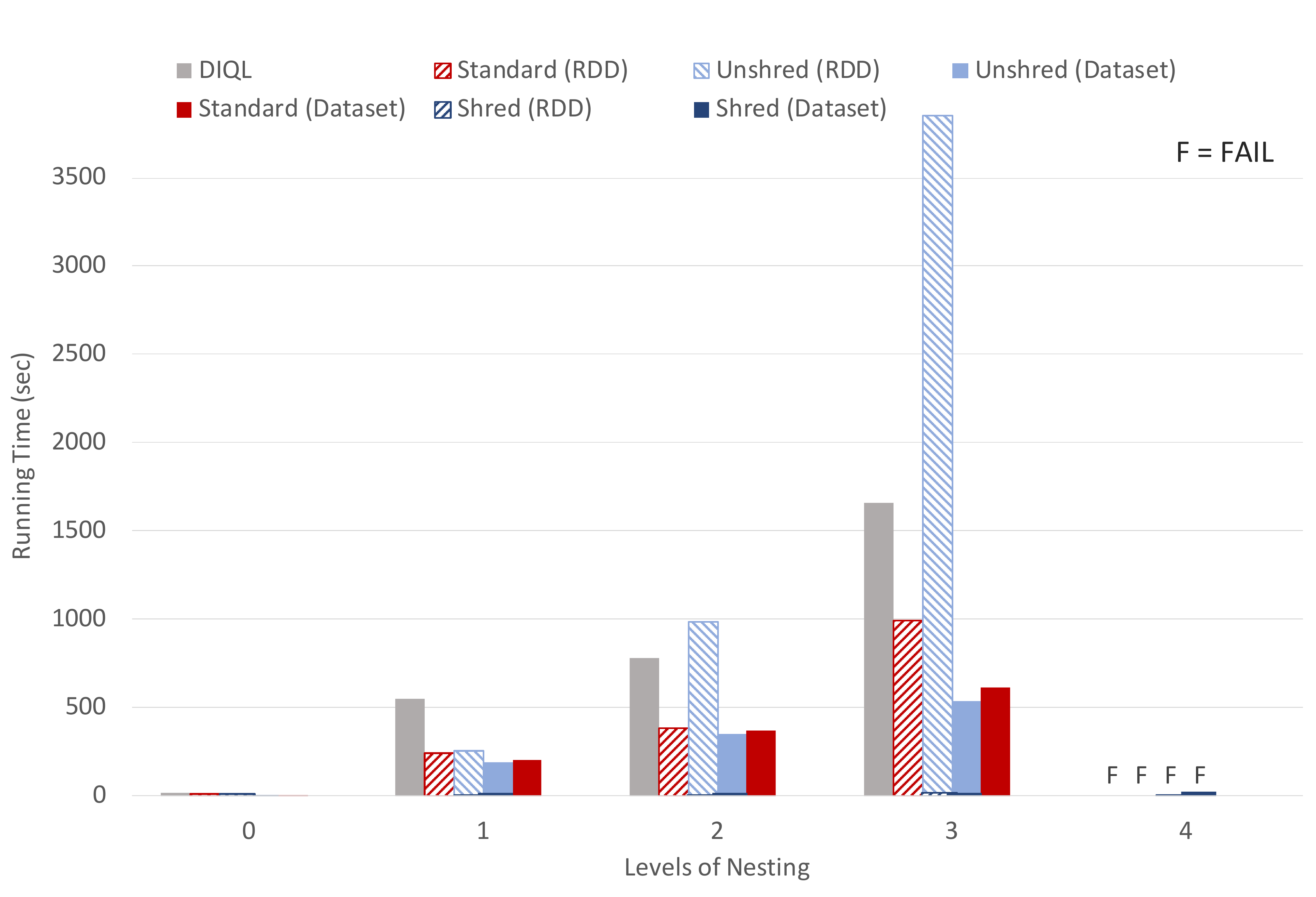}
    \caption{Wide schema}
    \label{fig:rddvsdfswide}
  \end{subfigure}
    \vspace{5pt}
    \caption{Performance comparison of RDDs and Datasets for the flat-to-nested benchmarked queries, 
    including additional competitor DIQL that uses an RDD implementation.}
    \label{fig:rdddfsfn}
\end{figure*}

\subsection{Spark RDDs vs Spark Datasets}\label{sec:rddvsds}

The flat-to-nested and nested-to-nested queries are  
used to compare the performance of \Fpplus (standard compilation route), \Shred 
(shredded compilation route without unshredding of output), and \Unshred (shredded compilation with unshredding)
using RDDs and Datasets. 
Figure \ref{fig:rdddfsfn} displays the results for the 
flat-to-nested queries with (\ref{fig:rddvsdfs}) 
and without projections (\ref{fig:rddvsdfswide}). 
With projections, the results show that all strategies exhibit similar performance 
up to three levels of nesting. The strategies diverge at four levels of nesting where
\Unshred and \Fpplus with RDDs have a spike in total run time. The 
results without projections follow a similar trend, with strategies diverging 
earlier at lower levels of nesting. Without projections, \Unshred 
with RDDs shows increasingly worse performance starting at two levels of nesting. 
\Shred with RDDs and \Shred with Datasets have 
similar performance. 

As stated in  Section \ref{sec:flattonest}, the plan produced with unshredding 
in the shredded compilation and the plan for the standard compilation are identical; thus, 
the difference in performance is attributed to code generation for the unshredding procedure. Both methods use a series 
of cogroups to build up a nested set; however, unshredding requires 
additional map operations and intermediate object creation that are required 
for reconstructing nested objects from dictionaries. Case classes that lack binary 
encoders require a 
significant amount of time and space to create and store, which is a cost that 
only increases with the levels of nesting. \Unshred with RDDs grows 
exponentially as the number of nesting increases, bringing along 
the previous level with each level of nesting. 
This is also a problem for \Fpplus, 
which sees worse performance with 
increasing levels of nesting but grows at a slower rate due to less object creation.

To consider what this means from a code generation perspective, compare the 
project operator of Figure \ref{fig:plan_operators} to the project 
operator in Figure \ref{fig:plan_operators_rdd}. The RDD API maps 
over a relation and creates a new case class (\texttt{R}), whereas 
the Dataset API avoids this map and uses a select 
operator that accepts a series of attribute names. By explicitly stating the 
attributes, the Dataset API has delayed an explicit map operation allowing 
for further performance benefits from the Spark optimizer. Further, when the time comes 
to construct \texttt{R} 
objects, the Dataset API leverages the binary format for a much smaller memory 
footprint.

\begin{figure*} 
  \begin{subfigure}{0.95\textwidth} 
    \centering   
    \includegraphics[width=\linewidth]{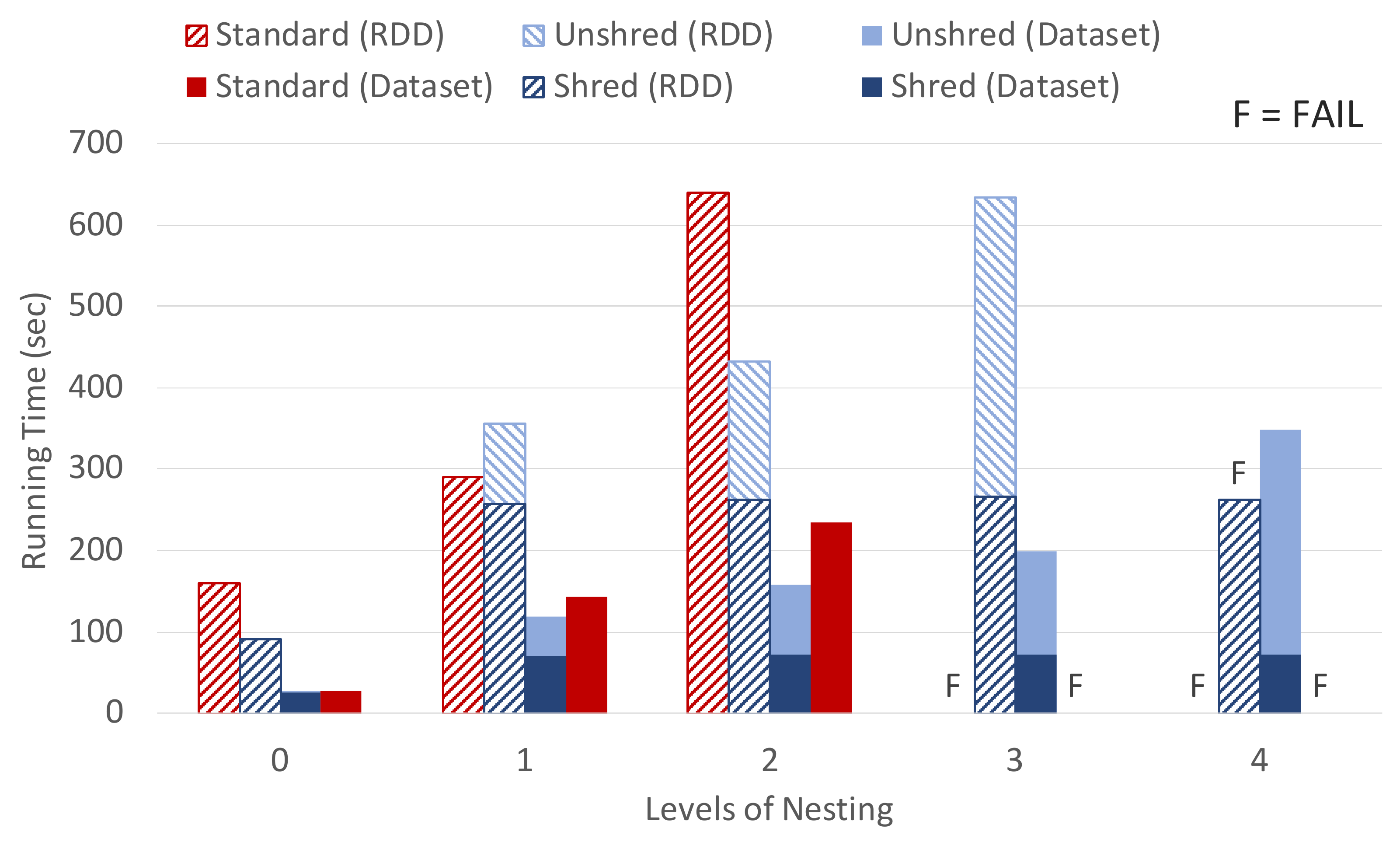}
    \caption{Narrow schema}
    \label{fig:rddvsdfs2}
  \end{subfigure}
  \begin{subfigure}{0.95\textwidth} 
      \centering   
    \includegraphics[width=\linewidth]{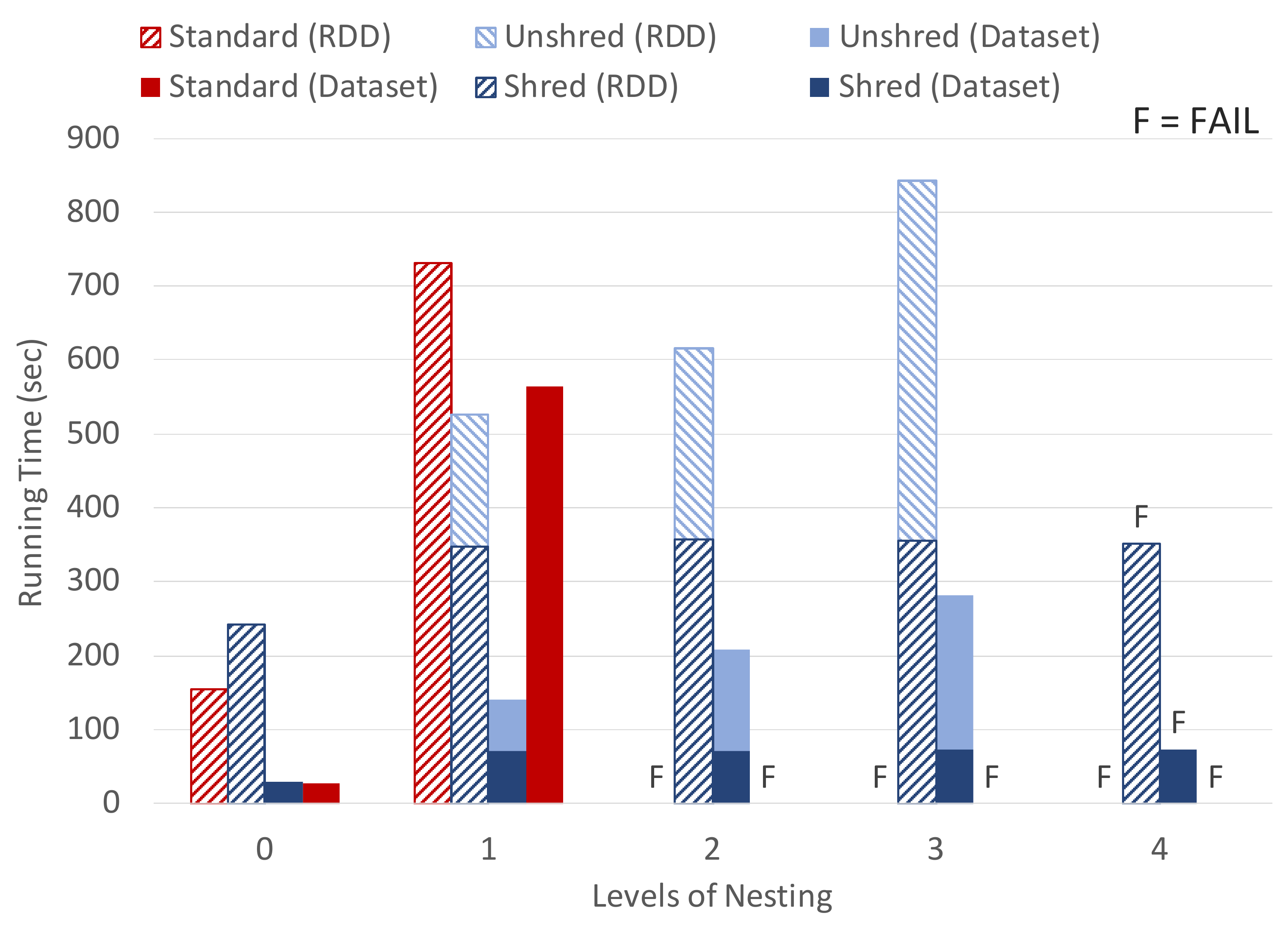}
    \caption{Wide schema}
    \label{fig:rddvsdfswide2}
  \end{subfigure}
    \vspace{5pt}
    \caption{Performance comparison of RDDs verse Datasets for the nested-to-nested benchmarked queries.}
    \label{fig:rdddfsnn}
\end{figure*}

Figure \ref{fig:rdddfsnn} further highlights the benefits of Datasets with 
nested-to-nested queries. Both with and without 
projections, the difference between 
\Shred with RDDs and \Shred with Datasets shows a 2$\times$ performance 
improvement of the explicit statement of attributes within the Dataset API, 
and \Fpplus has decreased 
performance with RDDs. While the unnest operators used in the code generators 
both use flat-map operations, the Dataset API maintains a low-memory footprint 
for the newly created objects (\texttt{.as[R]} in the code generator).

These results show that code generation with Datasets has minimized overhead 
in object creation and gains further improvements from passing meta-information 
to the Spark optimizer. Beyond the application to Spark, these results should be 
useful for further implementations of automated nested query processing on 
distributed systems. 

\pagebreak



\subsection{Experiments Comparing With Additional Competitors}

We used the TPC-H benchmark to compare to a wide array of external competitors:
an implementation via encoding in SparkSQL \cite{sparksql};
Citus, a distributed version of Postgres \cite{citus}; MongoDB \cite{mongodb}, 
and the recently-developed nested relational
engine DIQL \cite{diql}. We include the results for SparkSQL since it outperformed 
all the other competitors. This section shows the extended 
results, including all competitors,
for the flat-to-nested, nested-to-nested, and nested-to-flat queries. 
The source code for each of the queries is available in the github. 
\\[20pt]
\myparagraph{Evaluation strategies and competitors} 
We explored several potential competitors for use in the comparison. 
The following competitors were able to perform at least one of the TPC-H 
benchmark queries, and thus are represented in the subsequent results. 
\begin{itemize}
\item \textit{SparkSQL}: \break
The SparkSQL queries were manually written based on two restrictions.
First, SparkSQL does not support explode (i.e., UNNEST) operations in the SELECT clause, requiring 
the operator to be kept with the source relation which forces flattening 
for queries that take nested input. Second, an (outer) join cannot 
follow an explode statement; this means the query must be written in the following 
form:
\begin{lstlisting}[language=SQL,basicstyle=\ttfamily] 
...
FROM (
  SELECT 
  FROM Q
  LATERAL VIEW explode(Q.A) AS B
  -- cannot have here another join
) t1 
LEFT OUTER JOIN Parts P ...  
\end{lstlisting} 
\item \textit{DIQL}:\break
The syntax of DIQL fully supports all the queries in the TPC-H 
benchmark; however, this is an experimental system and we uncovered bugs during this 
process. With this in mind, we provide the results for the flat-to-nested queries only. 
The DIQL Spark API has slightly different system requirements and we were only able to 
compile and run the queries with Spark 2.4.3 and Scala 2.11.
\item \textit{Postgres+Citus}: \break
We use a distributed version of Postgres (Citus) as the representative relational database engine. 
We use a coordinator Postgres instance with five workers, exactly like the Spark. We 
cached inputs and parallelized processes as much as possible with the following:

\begin{lstlisting}[language=Scala,basicstyle=\ttfamily] 
shared_buffers = 80GB
effective_cache_size = 200GB
work_mem = 64MB
max_worker_processes = 20
max_parallel_workers_per_gather = 20
max_parallel_workers = 20
\end{lstlisting} 

However, the results using default values had better performance. 
The Citus queries were manually written, using both arrays and JSON with and without 
caching inputs. We report the runtimes
of the array based queries, without caching inputs, and default worker configurations 
since this continuously outperformed the others. 

All Citus queries are based on several caveats.
First, Citus does not support nested subqueries in the target of a SELECT, failing with 
\textit{could not run distributed query with subquery outside the FROM, WHERE and HAVING clauses}.
Second, queries can be rewritten using GROUP BY and ARRAY\_AGG, but joins between relations partitioned on different columns - known as complex joins in Citus terminology - 
are not supported; this fails with \textit{complex joins are only supported when 
all distributed tables are co-located and joined on their distribution columns}. 
Outer joins can be done in a binary fashion with one table being a common table expression (CTE). 
For instance, the following query where t1 and t2 are partitioned on the join key 
but not on the join key for t3:

\begin{lstlisting}[language=SQL,basicstyle=\ttfamily] 
SELECT 
  FROM (
    OUTER JOIN t1 and t2 
  ) 
  OUTER JOIN t3
\end{lstlisting} 

The result of the subquery (t1 join t2) will be collected entirely at the master and 
then partitioned to workers according to the next join key. 
This is obviously inefficient and has restrictions logged in Citus as:

\vspace{5pt}
\textit{DETAIL: Citus restricts the size of intermediate results of complex subqueries 
and CTEs to avoid accidentally pulling large result sets into once place}.
\vspace{5pt}

Third, left outer joins between tables partitioned on different keys are not yet supported \url{https://github.com/citusdata/citus/issues/2321}.
Finally, to avoid pulling data back to master and enable outer joins between relations partitioned 
on different keys, we manually created execution plans where at each step we (outer) join two relations 
partitioned on the same key and write the result back into a distributed materialized view partitioned 
on the next join key in sequence. That means we had to materialize the entire flattened nested object to 
get it repartitioned by partkey before joining with Part. Each nested-to-flat requires 2 queries, 
while each nested-to-nested has one extra query for the final regrouping.
\item \textit{MongoDB}: \break
We use MongoDB with one master and five workers, as in the Spark and Citus setup. 
The queries were hand-written based on the following 
restrictions. Only one collection can be sharded when performing lookups (joins), 
the inner one must be local. The only join strategy is to iterate (in parallel) over the outer collection and do lookups on the inner collection, which is located on one machine; thus, 
this is a bottleneck. We find that MongoDB has good performance with selective filters over a single collection, not designed for queries over multiple collections or even single-collection queries that return many documents. Nested collections formed using the \texttt{\$push} accumulator are 
currently capped at 100MB; pipelines using more than 100MB will fail.
\end{itemize}

\myparagraph{Additional competitors explored} 
The following systems were also explored, but were unable 
to support the queries of the benchmark.
\begin{itemize}
\item \textit{Rumble}: \break
Rumble transforms JSONiq to Spark and supports local and distributed execution. 
We discovered problems running even toy examples doing data denormalization. 
Initially, outer joins were not supported: 
\url{https://github.com/RumbleDB/rumble/issues/760}. 
We reported this and it was fixed, but now outer joins with distributed collections 
are transformed into Cartesian products\footnote{Accessed 10 September 2020.}. In general, the Rumble JSONiq language 
is not providing several of the operations available in Spark 
(e.g., caching, schema handling) are not available through their JSONiq language. 
\item \textit{Zorba}: \break
Zorba (JSONiq) has no support for distributed execution, and 
has not been maintained in the past 4 years \cite{zobra}.
\item \textit{MonetDB}: \break 
MonetDB \cite{monet} has no support for array type and array operations. 
The system does support JSON operations over strings, but there is no easy way 
to transform tables to JSON objects; manually creating JSON strings throws errors. 
\item \textit{Cockroach}: \break
CockroachDB does not support nested arrays or ordering by arrays \cite{croach}.
\item \textit{VoltDB}: \break 
VoltDB does not have 
support for arrays. There is support for JSON, but
is is up to the application to perform the conversion from an in-memory structure to the textual representation. In addition, there is a size limit for JSON values. The VARCHAR columns used to store JSON values are limited to one megabyte (1048576 bytes). JSON support allows for augmentation 
of the existing relational model with VoltDB; however, it is not intended or appropriate as a replacement for pure blob-oriented document stores.
\item \textit{YugabyteDB}: \break
YugabyteDB seemed like a good candidate 
as it supports distributed execution and much of SQL, but the performance 
was too poor to explore further. For example, the following query took four minutes 
with 18760 orders tuples and 2500 user tuples:
\begin{lstlisting}[language=SQL,basicstyle=\ttfamily] 
SELECT users.id, 
  (SELECT ARRAY_AGG(orders.id) 
    FROM orders 
    WHERE orders.user_id=users.id) 
FROM users
\end{lstlisting} 
\end{itemize}

\begin{figure*} 
  \begin{subfigure}{0.95\textwidth} 
    \centering   
    \includegraphics[width=\linewidth]{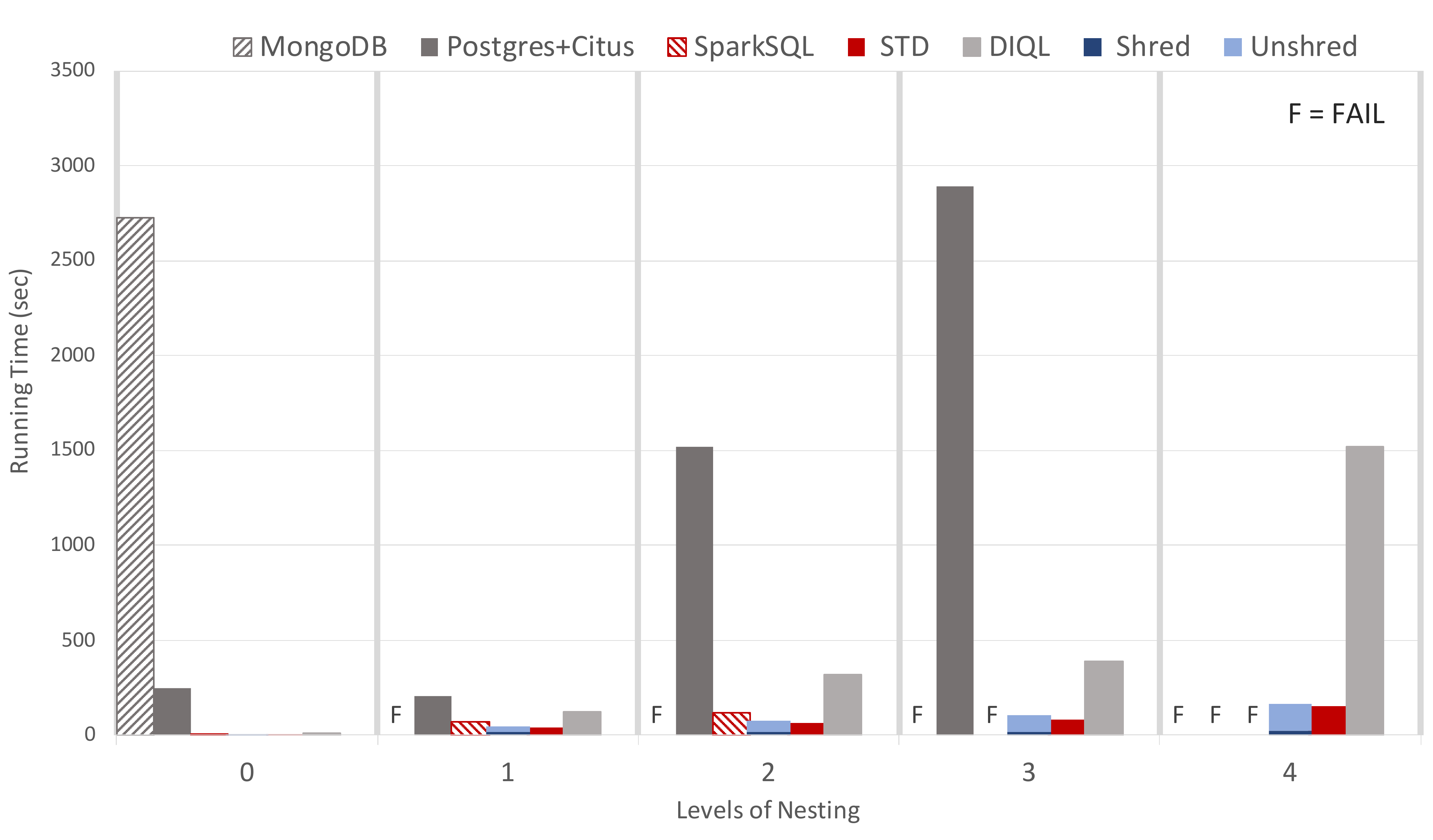}
    \caption{Narrow schema}
    \label{fig:fnthin}
  \end{subfigure}
  \begin{subfigure}{0.95\textwidth} 
      \centering   
    \includegraphics[width=\linewidth]{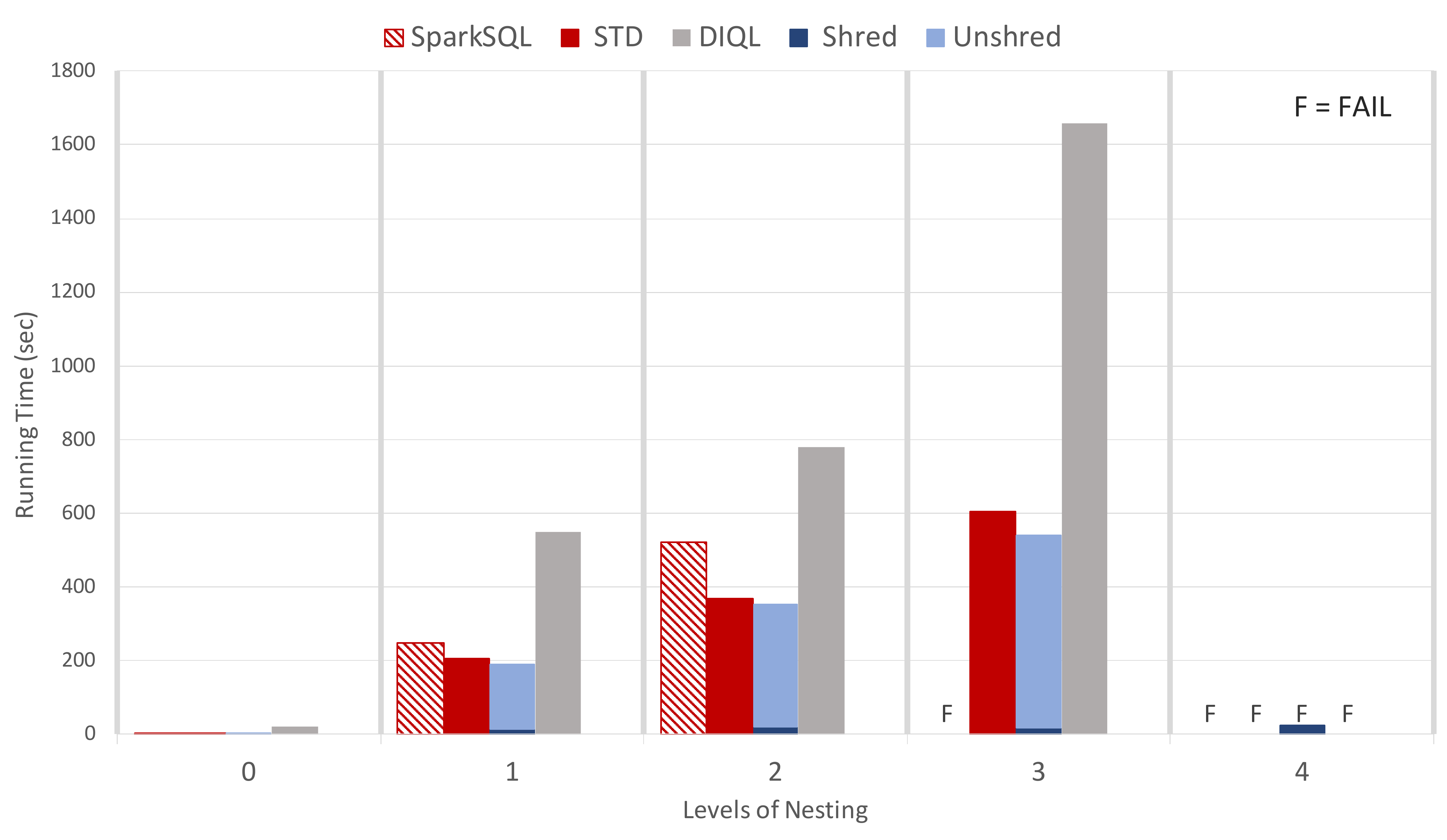}
    \caption{Wide schema}
    \label{fig:fnwide}
  \end{subfigure}
    \vspace{5pt}
    \caption{Performance comparison of flat-to-nested queries including all competitors.}
    \label{fig:flatnested}
\end{figure*}

\myparagraph{Flat-to-nested for non-skewed data}
Figure \ref{fig:fnthin} displays the results for MongoDB, Postgres Citus,
DIQL, SparkSQL, \Fpplus, \Shred, and \Unshred for the narrow 
flat-to-nested queries of the TPC-H benchmark. 
Given the performance for all systems is worse for 
wide tuples, we did not explore the performance of MongoDB and 
Citus for the wide variants.  Figure \ref{fig:fnwide} displays the 
results for DIQL, SparkSQL, \Fpplus, \Shred, and \Unshred 
for the wide flat-to-nested queries.

\begin{figure*} 
  \includegraphics[width=\linewidth]{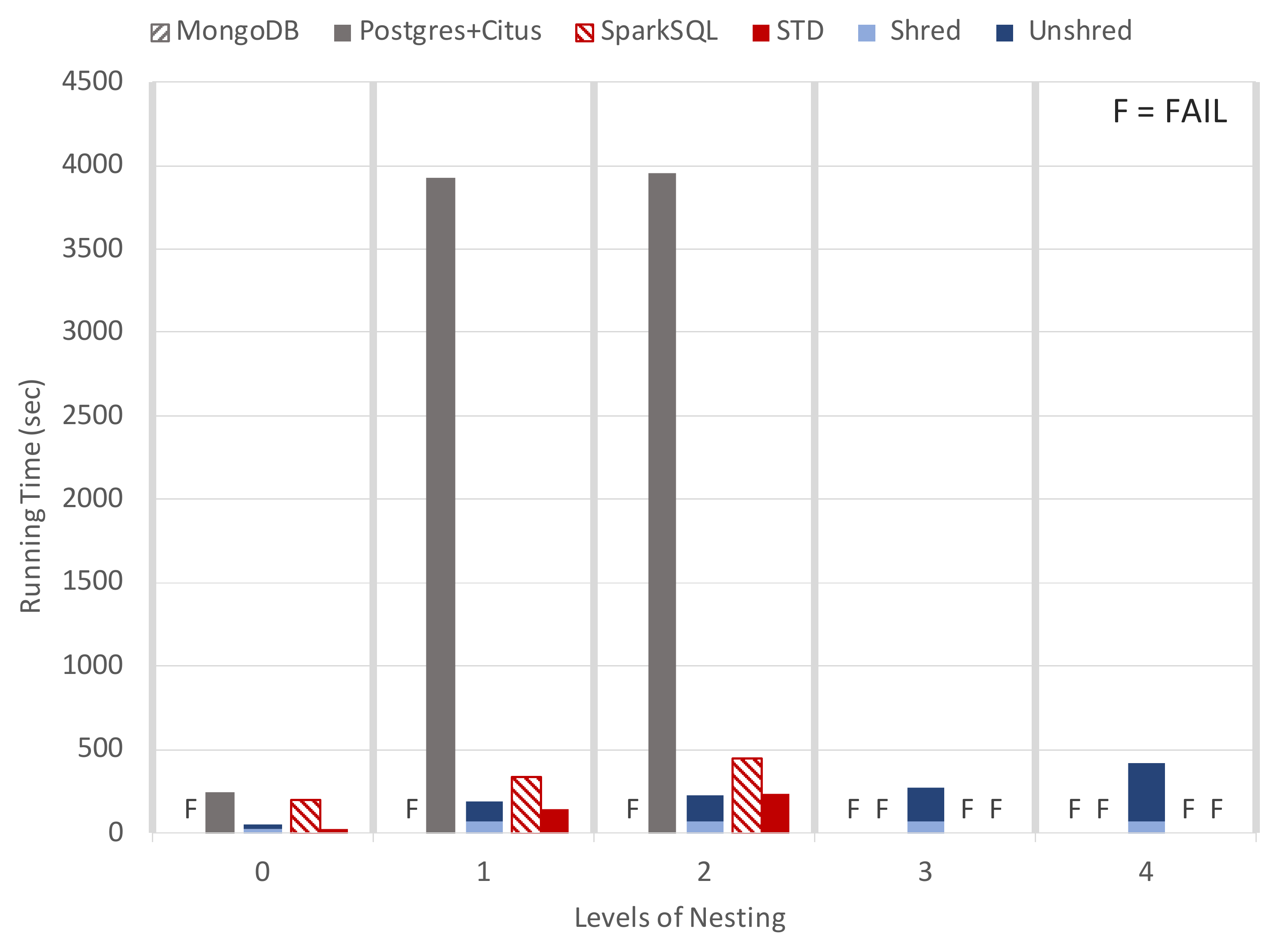}
  \caption{Performance comparison of narrow nested-to-nested queries including all competitors.}
  \label{fig:nnthin}
\end{figure*}

\myparagraph{Nested-to-nested for non-skewed data}
Figure \ref{fig:nnthin} displays the results for MongoDB, Postgres Citus,
SparkSQL, \Fpplus, \Shred, and \Unshred for the narrow 
nested-to-nested queries of the TPC-H benchmark. 
Given the performance for all systems is worse for 
wide tuples, we did not explore the performance of MongoDB and 
Citus for the wide variants; thus, the wide variants for 
SparkSQL, \Fpplus, \Shred, and \Unshred can be found in the 
main body of the paper. 

\begin{figure*} 
  \includegraphics[width=\linewidth]{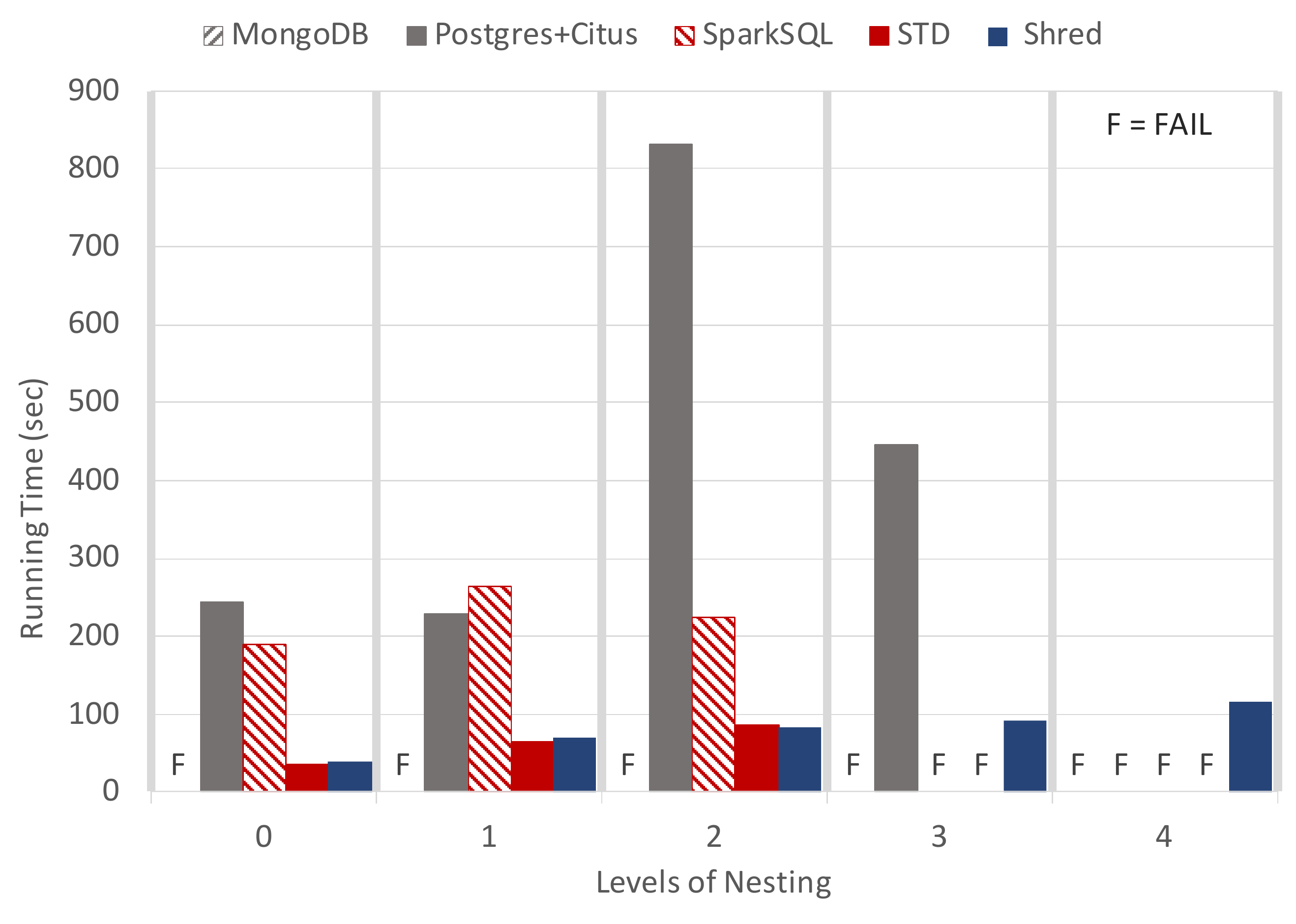}
  \caption{Performance comparison of narrow nested-to-flat queries including all competitors.}
  \label{fig:nfthin}
\end{figure*}

\pagebreak 

\myparagraph{Nested-to-flat for non-skewed data}
Figure \ref{fig:nfthin} displays the results for MongoDB, Postgres Citus,
SparkSQL, \Shred, \Unshred and \Fpplus for the narrow 
nested-to-flat queries of the TPC-H benchmark. 
As with the previous two query categories, we provide 
only the narrow variants for MongoDB and Postgres Citus. 
The wide variants for SparkSQL, \Shred, 
\Unshred, and \Fpplus can be found in the main body of the paper. Due to 
the poor performance of these systems, the  
nested inputs for MongoDB and Postgres Citus were preprocessed with 
projections pushed; thus, MongoDB and Postgres Citus take 
the materialized narrow flat-to-nested query as input. The other systems take the 
materialzied wide flat-to-nested query as input to better explore the effects 
of projection. 

\subsection{Total shuffle for non-skewed TPC-H benchmark}
\begin{figure*} 
  \includegraphics[width=\linewidth]{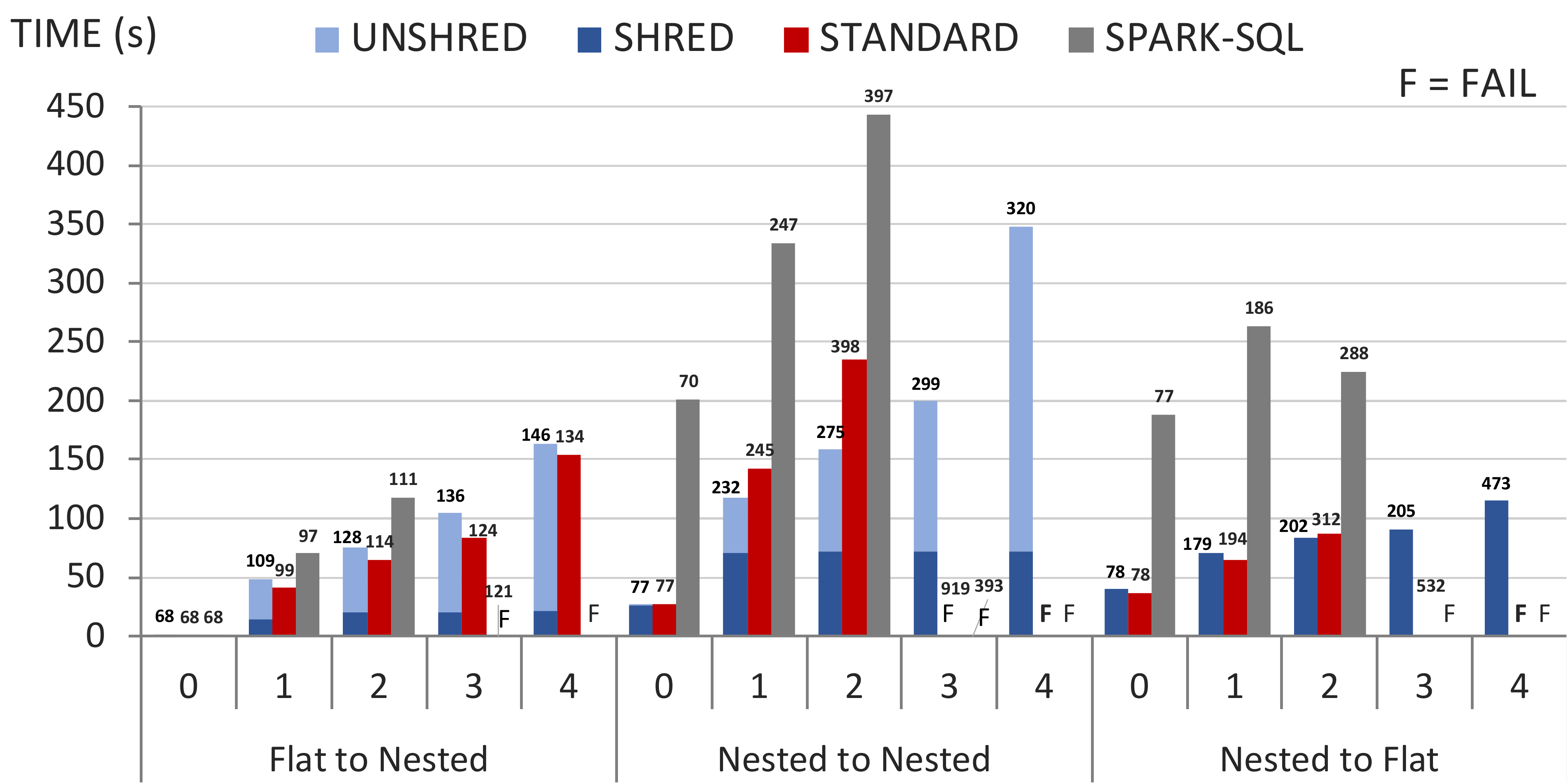}
  \caption{Performance comparison of narrow TPC-H benchmark queries with total shuffle memory (GB).}
  \label{fig:tpchmemn}
\end{figure*}

\begin{figure*} 
  \includegraphics[width=\linewidth]{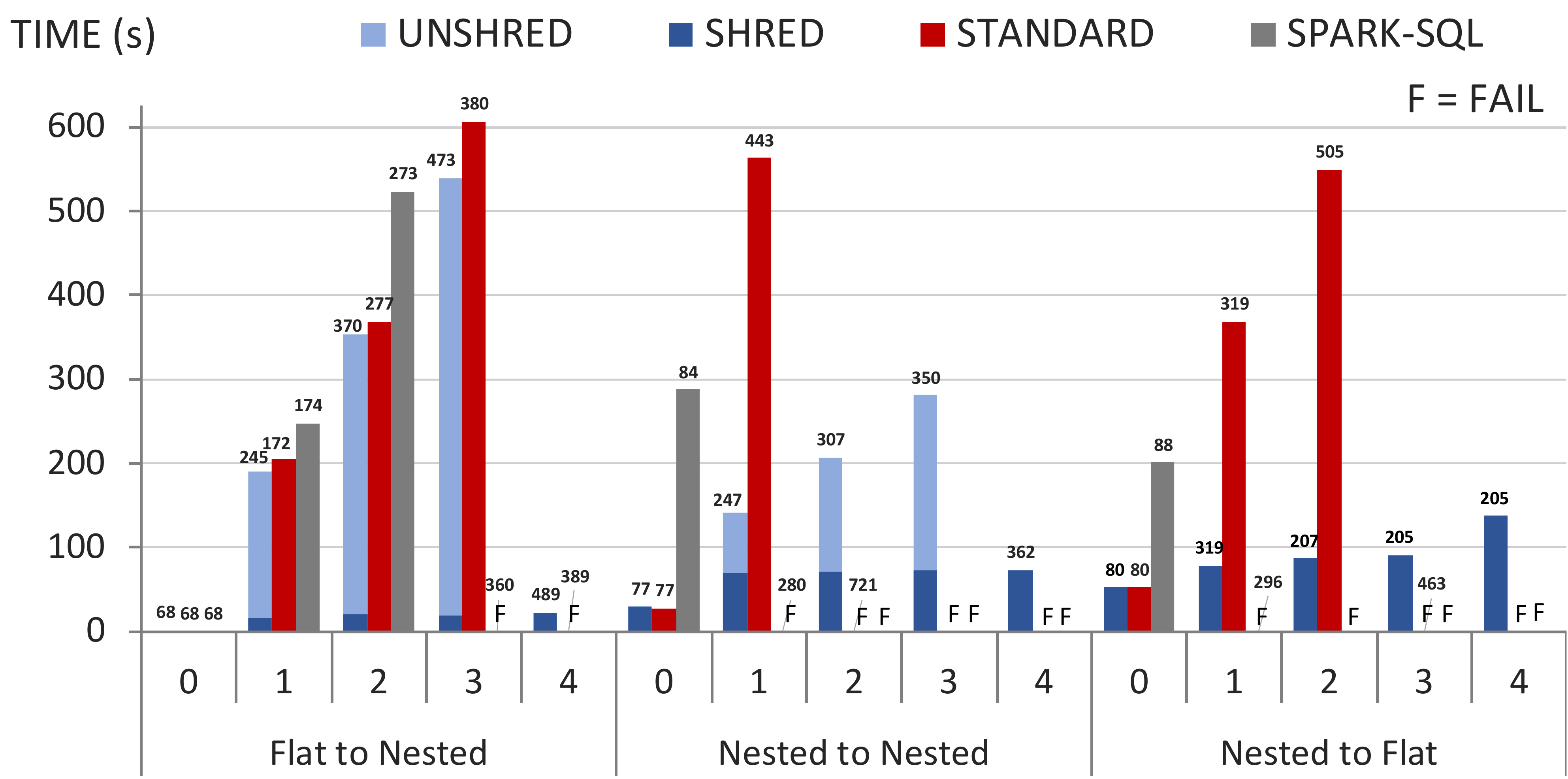}
  \caption{Performance comparison of wide TPC-H benchmark queries with total shuffle memory (GB).}
  \label{fig:tpchmemw}
\end{figure*}

Figure \ref{fig:tpchmemn} and Figure \ref{fig:tpchmemw} 
provides annotated results for Figure 7 in Section 6 of paper, which includes 
the total shuffled memory (GB) for each run. If a job crashes at a particular 
nesting level, we do not report any further total shuffle memory. 
As above, MongoDB and Postgres Citus take the materialized narrow flat-to-nested 
query as input; whereas, the other methods take the materialized wide flat-to-nested 
query as input. 

\subsection{Standard complation framework optimizations}

\begin{figure*}
  \begin{subfigure}{0.95\textwidth} 
    \centering   
    \includegraphics[width=\linewidth]{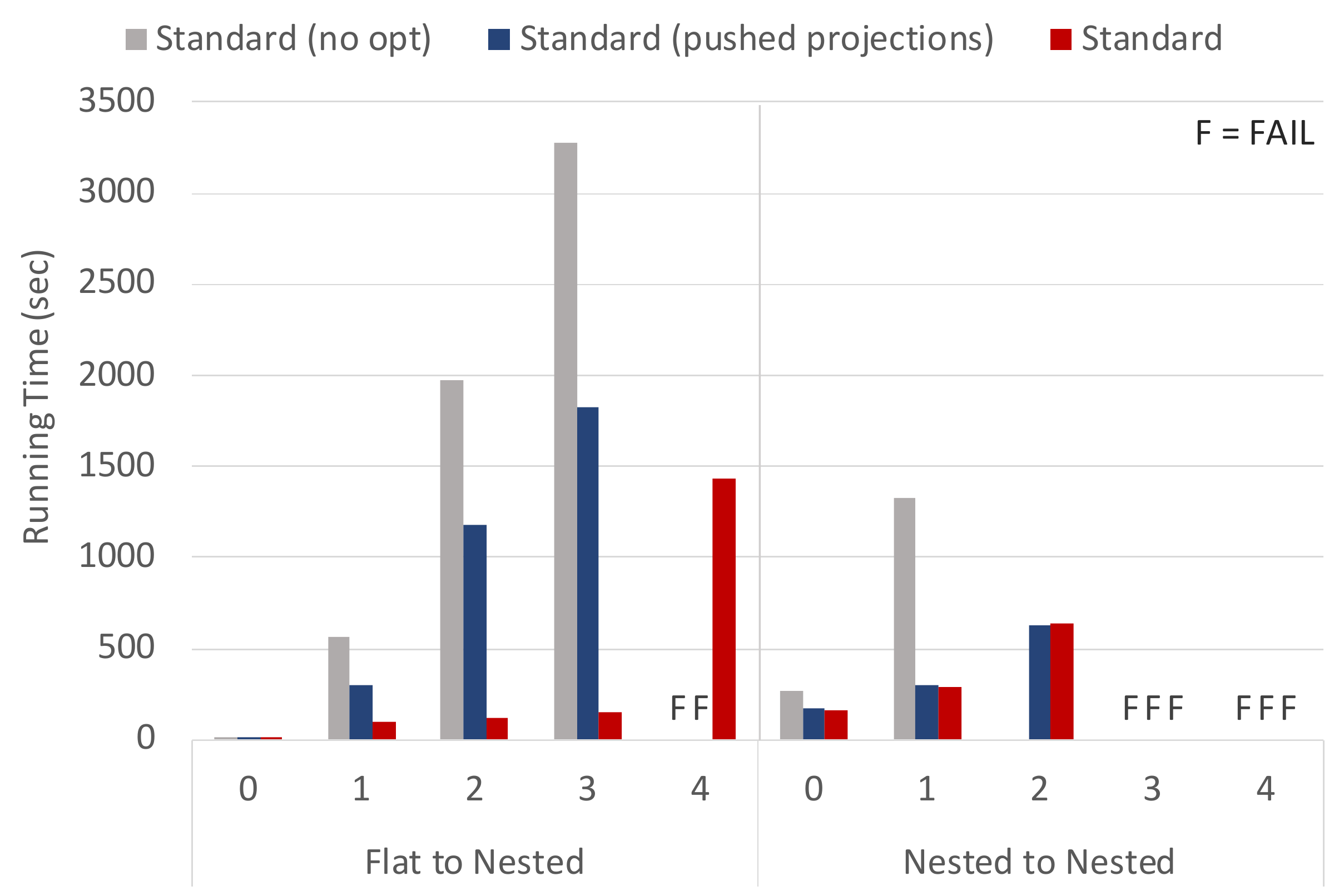}
    \caption{Narrow schema}
    \label{fig:flatsfn}
  \end{subfigure}
  \vspace{5pt}
  \begin{subfigure}{0.95\textwidth} 
      \centering   
    \includegraphics[width=\linewidth]{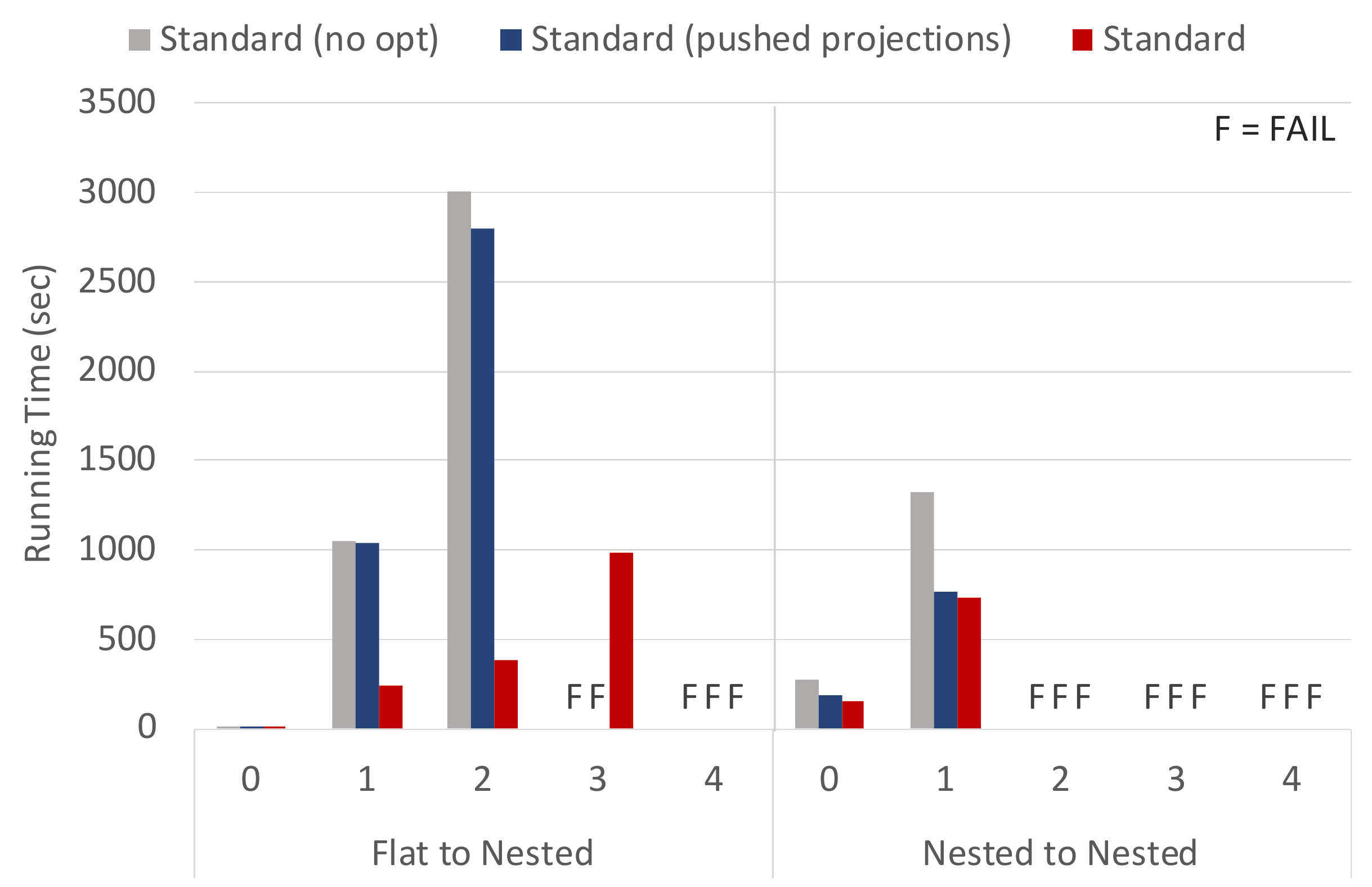}
    \caption{Wide schema}
    \label{fig:flatsnn}
  \end{subfigure}
    \vspace{5pt}
    \caption{Performance comparison of benchmarked queries for increasing optimization levels of the standard 
compilation route.}
    \label{fig:flats}
\end{figure*}

This experiment highlights how the framework can leverage database-style optimizations to 
automatically generate programs that are comparable to hand-optimized programs. 
Plans are generated using the standard compilation route with an increasing level 
of optimization
applied to both the flat-to-nested and nested-to-nested queries. 
Figure \ref{fig:flats} shows the 
results of this experiment. \Fpplus (no opt) 
is the standard compilation route with no optimizations; this corresponds to the 
plan that comes directly from unnesting. Standard (pushed projections) 
is the standard compilation route with projections pushed. Standard 
applies all optimizations that produce the optimal plan, which is the 
plan in all experiments in the body of the paper; this includes 
pushed projections, nesting operators pushed past join operators, and 
merged into cogroups where relevant. 

The results show that even simple optimizations 
like pushing projections can provide major performance benefits for flattening methods. For example, 
Figure \ref{fig:flatsfn} shows that projections have not only increased performance of the 
standard compilation route, but have allowed the strategy to survive to deeper levels of nesting. This is 
expected since the experiments in the previous sections have shown that the 
performance of \Fpplus is heavily impacted by the 
presence of projections (ie. the number of attributes an output tuple). 
For nested-to-nested queries, \Fpplus is the only strategy to survive past one level of 
nesting. These results show that database-style optimizations are not only beneficial to improve performance, 
but are necessary when using flattening methods even with shallow-nested objects.

\newpage

\subsection{\COP shuffle in skew-handling results}

\begin{figure*} 
  \includegraphics[width=\linewidth]{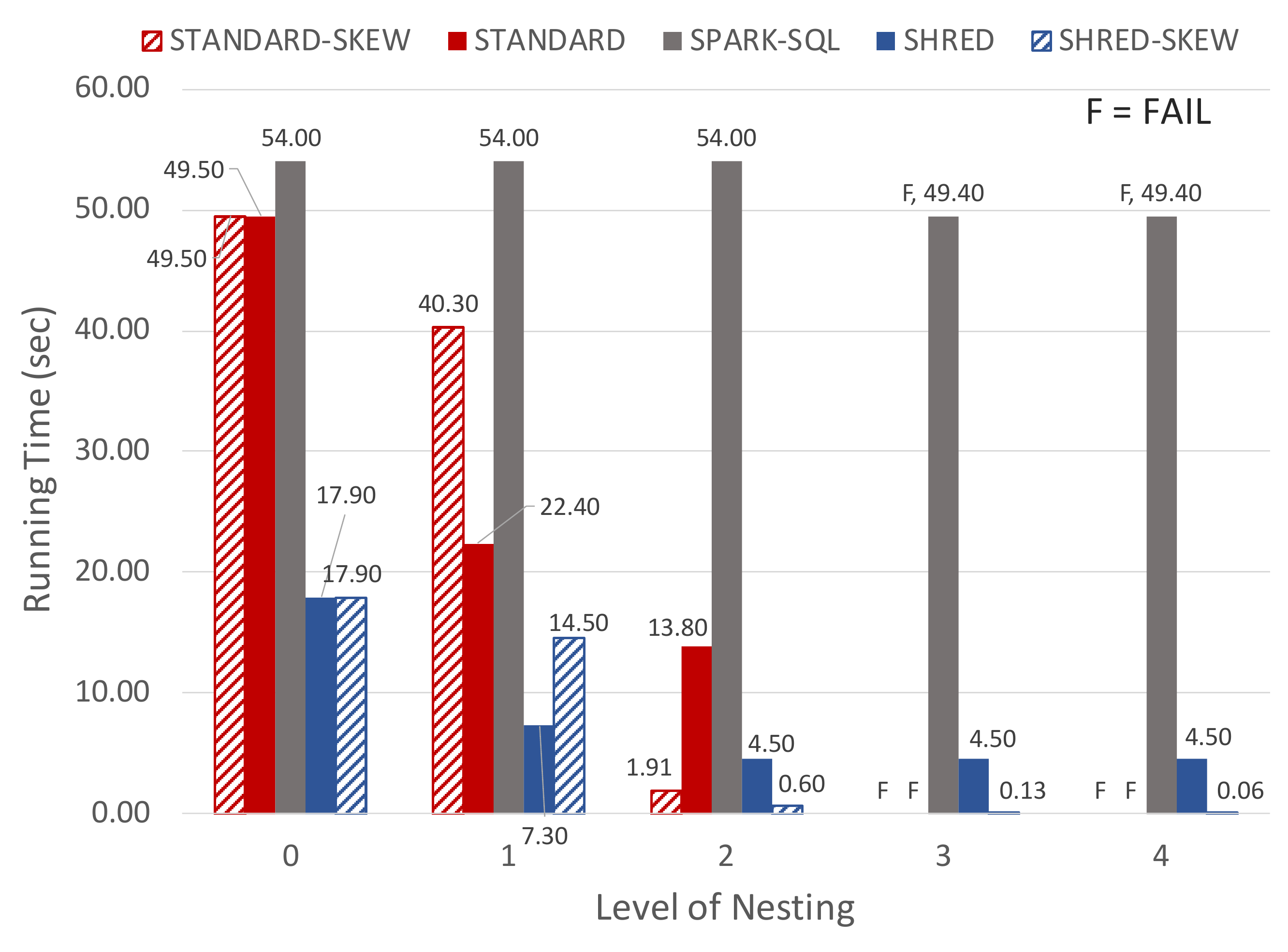}
  \caption{Amount of shuffled data from \COP prior to joining with \Part in level 2, narrow, nested-to-nested TPC-H query for skew-aware and skew-unaware variants, as well as SparkSQL.}
  \label{fig:skewshuff}
\end{figure*}

Figure \ref{fig:skewshuff} shows the amount of data shuffled from \COP prior to the 
nested join with \Part for the nested-to-nested TPC-H query used in the skew-handling 
results in the paper. The results highlight how the skew-aware shredded compilation route
leads to less than a gigabyte of shuffling for moderate and high levels of skew. 
There are no \COP shuffle results for the standard compilation route since 
query fails during execution while attempting to flatten \COP. 
SparkSQL survives flattening for high levels of skew, 
but fails while performing the join with \Part. The shuffling of the standard 
compilation route shows that at lower levels of skew the local aggregation is beneficial. 
At higher levels of skew the local aggregation reduces the data in the 
skew-handling variation of the compilation route 
to about 4.5G; however, the skew-aware 
compilation route has 
reduced this to only megabytes of data leading to 74x less shuffle than the 
skew-unaware compilation route.

\subsection{Skew-handling results without aggregations pushed}\label{sec:skewlocalagg}
\begin{figure*} 
  \includegraphics[width=\linewidth]{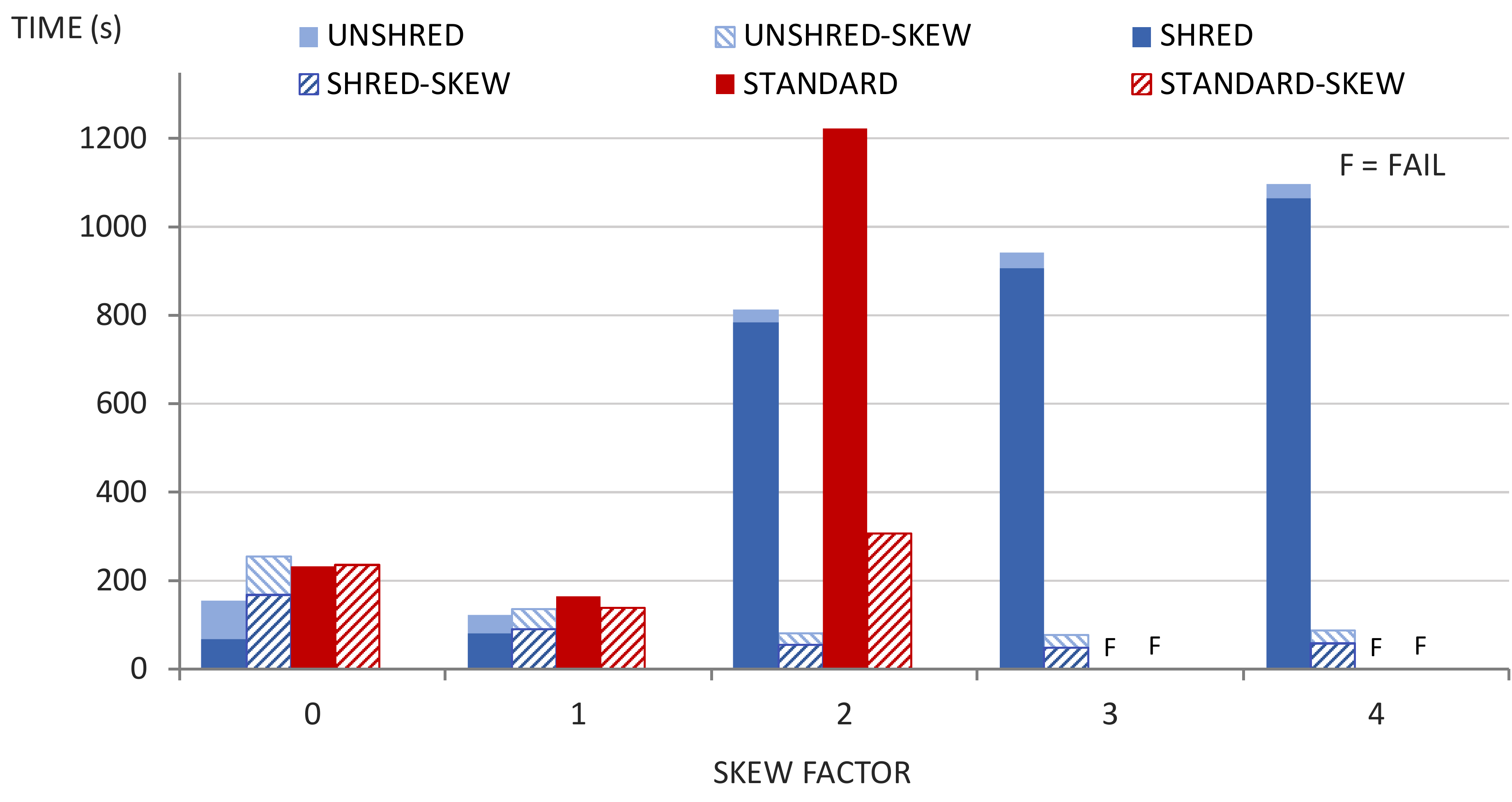}
  \caption{Performance comparison of skew-aware and non-skew aware standard and shredded  compilation
  without aggregation pushing.}
  \label{fig:skewnoagg}
\end{figure*}

Figure \ref{fig:skewnoagg} shows the runtimes for the standard and shredded 
compilation routes when aggregations are not pushed, 
for both the skew-aware and skew-unaware variants. The results show that 
without pushing aggregation, the skew-unaware methods decline 
in performance in comparison to the skew-unaware runs in the 
body of the paper (with aggregation pushed). 
These results highlight 
how pushing aggregation plays a key factor in the TPC-H skewed dataset 
to help with skewed data; the benefit in this 
case is a consequence of the TPC-H data generator, which duplicates 
values to create skew. We include these results to provide insight into 
how the skew-aware method could perform if aggregation pushing was 
not a key factor in the performance of the skew-unaware methods. 

\subsection{Overhead of skew-handling}

\begin{figure}
 \center
 \includegraphics[width=0.8\linewidth]{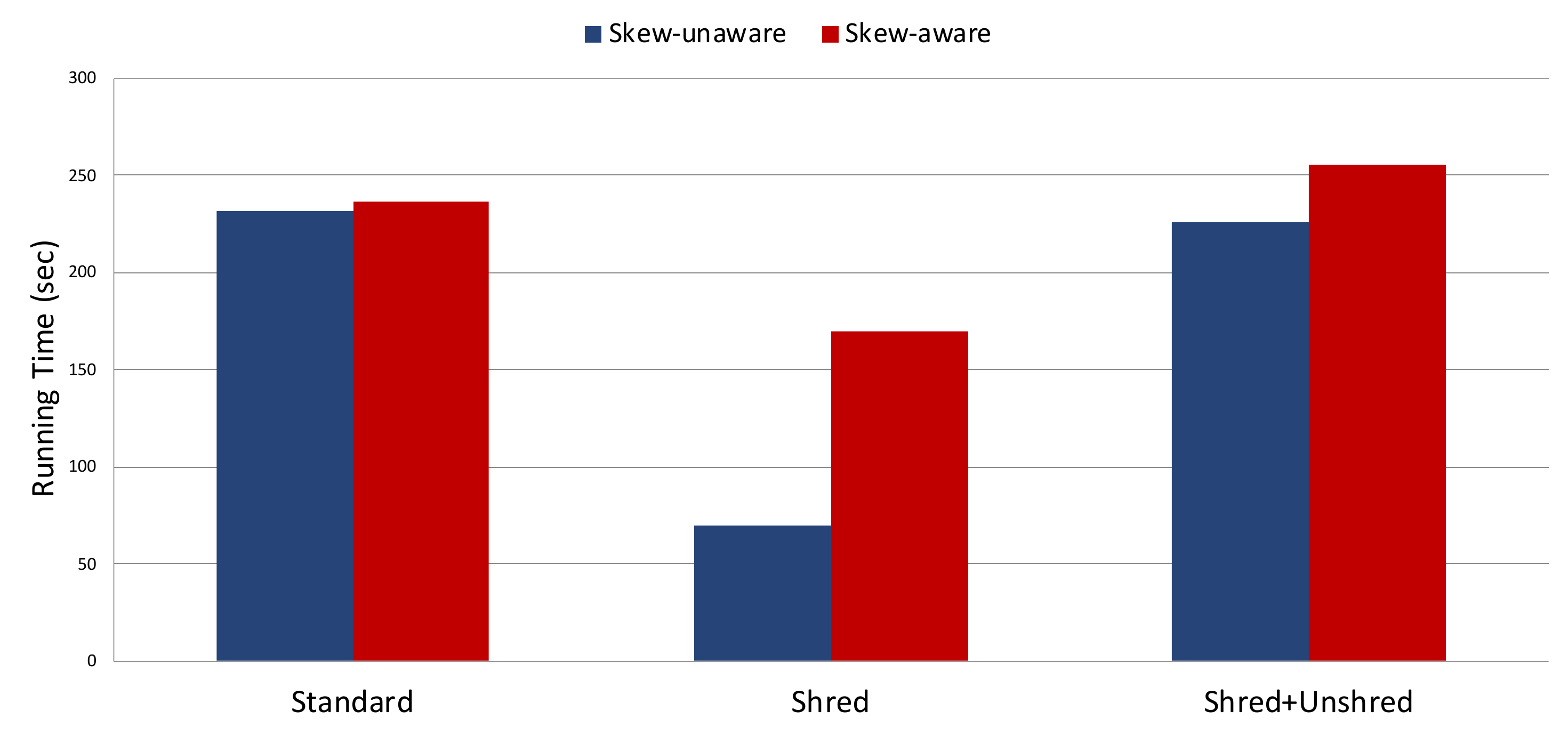}
 \vspace{3pt}
 \caption{Skew-handling overhead for skew-unaware and skew-aware variants of the 
 standard and shredded compilation routes.}
 \vspace{-6pt}
 \label{fig:exp4a}
 \end{figure}

Figure~\ref{fig:exp4a} shows the overhead of skew-handling for \emph{non-skewed} data, 
comparing the results for the skew-unaware version of the compilation framework 
to the skew-aware version of the compilation framework for both the shredded and 
standard compilation routes. The shredded compilation route with unshredding is 
also shown.
\Sskew exhibits the largest overhead for heavy key collection. Both \Fskew 
and \Sskew calculate heavy keys for \Lineitem $\join$ \Part, which takes about 12s.
The main overhead of \Sskew is the heavy key calculation within the final casting of 
the lowest-level dictionary with \bagtomatdict. 
Unshredding does no heavy key calculations, thus there is no additional overhead 
for \Uskew.

\end{document}


\lstdefinelanguage{NRC}{
  morekeywords={for, in, union, if, then, else, match, let},%
  sensitive,%
  morecomment=[l]//,%
  morecomment=[s]{/*}{*/},%
  morestring=[b]",%
  morestring=[b]',%
  showstringspaces=false,%
  breaklines=true,%
  mathescape=true,%
  showspaces=false,
  showtabs=false, 
  showstringspaces=false,
  breakatwhitespace=true,
  aboveskip=1pt,
  belowskip=1pt,
  lineskip=0pt,
  basicstyle=\small\ttfamily\color{white!15!black},
  keywordstyle=\small\ttfamily\bfseries\color{blue!80!black},%
  columns=fullflexible,
  escapeinside={(*@}{@*)}
}[keywords,comments,strings]%

\lstdefinelanguage{Scala}%
{morekeywords={abstract,%
  case,catch,char,class,%
  def,else,extends,final,finally,for,%
  if,import,implicit,%
  match,module,%
  new,null,%
  object,override,%
  package,private,protected,public,%
  for,public,return,super,%
  this,throw,trait,try,type,%
  val,var,%
  with,while,%
  yield%
  },%
  sensitive,%
  morecomment=[l]//,%
  morecomment=[s]{/*}{*/},%
  morestring=[b]",%
  morestring=[b]',%
  showstringspaces=false,%
    breaklines=true,%
  mathescape=true,%
  showspaces=false,
  showtabs=false,
  showstringspaces=false,
  breakatwhitespace=true,
  xleftmargin=2em,
  aboveskip=1pt,
  belowskip=1pt,
  lineskip=-0.2pt,
   basicstyle=\ttfamily,
  keywordstyle=\ttfamily\color{blue!80!black},
  columns=fullflexible,
}[keywords,comments,strings]%

\newcommand{\myeat}[1]{}
\newcommand{\punto}{$\hspace*{\fill}\Box$}
\newcommand{\NULL}{\texttt{NULL}}
\newcommand{\tab}{\;\;\;}
\newcommand\mydots{\makebox[1em][c]{.\hfil.\hfil.}}

\newcounter{magicrownumbers}
\newcommand\rownumber{\stepcounter{magicrownumbers}\arabic{magicrownumbers}}
\newcommand{\linenumber}{\makebox[2ex][r]{$\scriptscriptstyle\rownumber\;\;$}}

\newcommand{\michael}[1]{{\bf\color{green} Michael: #1}}
\newcommand{\jac}[1]{{\bf \color{violet} Jaclyn: #1}}
\newcommand{\milos}[1]{{\bf \color{red} Milos: #1}}

\newcommand{\bagtype}{\textit{Bag}\hspace{0.5mm}}
\newcommand{\labeltype}{\textit{Label}\xspace}
\newcommand{\stringtype}{\textit{string}}
\newcommand{\datetype}{\textit{date}}
\newcommand{\doubletype}{\textit{real}}
\newcommand{\inttype}{\textit{int}}
\newcommand{\booltype}{\textit{bool}}

\newcommand{\Part}{\texttt{\textup{Part}}\xspace}
\newcommand{\Lineitem}{\texttt{Lineitem}\xspace}
\newcommand{\COP}{\texttt{\textup{COP}}\xspace}
\newcommand{\cid}{\texttt{\textup{cid}}\xspace}
\newcommand{\cname}{\texttt{\textup{cname}}\xspace}
\newcommand{\cop}{\texttt{\textup{cop}}\xspace}
\newcommand{\corders}{\texttt{\textup{corders}}\xspace}
\newcommand{\odate}{\texttt{\textup{odate}}\xspace}
\newcommand{\oparts}{\texttt{\textup{oparts}}\xspace}
\newcommand{\pid}{\texttt{\textup{pid}}\xspace}
\newcommand{\pname}{\texttt{\textup{pname}}\xspace}
\newcommand{\pprice}{\texttt{\textup{price}}\xspace}
\newcommand{\lqty}{\texttt{\textup{qty}}\xspace}
\newcommand{\total}{\texttt{\textup{total}}\xspace}
\newcommand{\pkey}{\texttt{\textup{id}}\xspace}
\newcommand{\sumBy}{\texttt{\textup{sumBy}}}

\newcommand{\Order}{\texttt{Order}\xspace}
\newcommand{\Customer}{\texttt{Customer}\xspace}
\newcommand{\Region}{\texttt{Region}\xspace}
\newcommand{\rid}{\texttt{\textup{rid}}\xspace}
\newcommand{\rname}{\texttt{\textup{rname}}\xspace}
\newcommand{\rnations}{\texttt{\textup{rnations}}\xspace}
\newcommand{\Nation}{\texttt{Nation}\xspace}
\newcommand{\nid}{\texttt{\textup{nid}}\xspace}
\newcommand{\nname}{\texttt{\textup{nname}}\xspace}
\newcommand{\ncusts}{\texttt{\textup{ncusts}}\xspace}
\newcommand{\oid}{\texttt{\textup{oid}}\xspace}
\newcommand{\RNCOP}{\texttt{\textup{RNCOP}}\xspace}
\newcommand{\cogroup}{\texttt{\kw{COGROUP}}\xspace}

\newcommand{\nrcagg}{\kw{NRC}}
\newcommand{\nrcaggshr}{\nrcagg^{Lbl+\lambda}}
\newcommand{\nrcaggshrmat}{\nrcagg^{Lbl}}

\newcommand{\kwd}[1]{\texttt{\textup{#1}}\xspace}
\newcommand{\idfor}{\kwd{for}}
\newcommand{\idin}{\kwd{in}}
\newcommand{\idunion}{\kwd{union}}
\newcommand{\idif}{\kwd{if}}
\newcommand{\idthen}{\kwd{then}}
\newcommand{\idelse}{\kwd{else}}
\newcommand{\idlet}{\kwd{let}}
\newcommand{\idmatch}{\kwd{match}}

\newcommand{\ckwd}[1]{\kwd{\color{blue!80!black} #1}}
\newcommand{\cidfor}{\ckwd{\idfor}}
\newcommand{\cidin}{\ckwd{\idin}}
\newcommand{\cidunion}{\ckwd{\idunion}}
\newcommand{\cidif}{\ckwd{\idif}}
\newcommand{\cidthen}{\ckwd{\idthen}}
\newcommand{\cidelse}{\ckwd{\idelse}}
\newcommand{\cidlet}{\ckwd{\idlet}}
\newcommand{\cidmatch}{\ckwd{\idmatch}}

\newcommand{\getkw}{\kwd{get}}
\newcommand{\dedup}{\kwd{dedup}}
\newcommand{\groupby}{\kwd{groupBy}}
\newcommand{\sumby}{\kwd{sumBy}}
\newcommand{\lookup}{\kwd{Lookup}}
\newcommand{\dicttreeunion}{\kwd{DictTreeUnion}}
\newcommand{\newlabelof}{\kwd{NewLabel}}
\newcommand{\matlookup}{\kwd{MatLookup}}
\newcommand{\flatc}{\textup{F}}
\newcommand{\dictc}{\textup{D}}
\newcommand{\shredf}[1]{\mathcal{F}(#1)}
\newcommand{\shredd}[1]{\mathcal{D}(#1)}
\newcommand{\FUN}{\kwd{fun}}
\newcommand{\CHILD}{\kwd{child}}

\newcommand{\groupatt}{\kwd{group}}
\newcommand{\labelatt}{\kwd{label}}
\newcommand{\valueatt}{\kwd{value}}

\newcommand{\NetworkFull}{$\kw{BN}_1$\xspace}
\newcommand{\SeqOnto}{$\kw{BF}_3$\xspace}

\newcommand{\olabelatt}{\kwd{olabel}}
\newcommand{\clabelatt}{\kwd{clabel}}
\newcommand{\nlabelatt}{\kwd{nlabel}}
\newcommand{\rlabelatt}{\kwd{rlabel}}

\newcommand{\topexpr}{\kw{TopExpr}}
\newcommand{\topbag}{\kw{TopBag}}
\newcommand{\labeldomain}{\kw{LabDomain}}

\newcommand{\bagunion}{\uplus}

\newcommand{\vshred}{\kw{vshred}}
\newcommand{\vunshred}{\kw{vunshred}}
\newcommand{\labelcode}{\kw{idgen}}

\newcommand{\matdict}{\kw{MatDict}}
\newcommand{\shredalg}{\kw{ShredQ}}
\newcommand{\sysname}{\kw{BINDS}}
\newcommand{\abinterp}{\kw{AbInterp}}
\newcommand{\langeq}{==}
\newcommand{\join}{\bowtie}
\newcommand{\coloneq}{~:=~}
\newcommand{\ateq}{:=}
\newcommand{\leteq}{:=}
\newcommand{\assigneq}{\Leftarrow}
\newcommand{\sng}[1]{\{ #1 \}}

\newcommand{\linearize}{\kw{Linearize}\xspace}
\newcommand{\materialize}{\kw{Materialize}\xspace}
\newcommand{\translate}{\kw{Translate}\xspace}
\newcommand{\bagtomatdict}{\texttt{BagToDict}\xspace}
\newcommand{\agg}{\kw{AGG}\xspace}
\newcommand{\heavykeys}{\texttt{heavyKeys}\xspace}

\newcommand{\oplang}{\kw{OPL}}
\newcommand{\totalmult}{\kw{TotalMult}}
\newcommand{\topout}{\kw{OutTopSubq}}
\newtheorem{proposition}{Proposition}
\newcommand{\renaming}{\kw{Relabeling}}
\newcommand{\relabeling}{\renaming}
\newcommand{\mapunion}{\kw{MapUnion}}
\newcommand{\topdict}{\kw{TopDict}}
\newcommand{\fulldict}{\dict}
\newcommand{\inputlabels}{\kw{InputLabels}}
\newcommand{\nrcplus}{NRC^+}
\newcommand{\nrc}{\kw{NRC}}
\newcommand{\freevars}{\kw{FreeVars}}
\newcommand{\transformbag}{\kw{TransformQueryBag}}
\newcommand{\transformtuple}{\kw{TransformQueryTuple}}
\newcommand{\transformtupleelement}{\kw{TransformQueryTupleElement}}
\newcommand{\pretransformbag}{\kw{TransformQueryBagAux}}
\newcommand{\outputlabels}{\kw{OutputLabels}}
\newcommand{\inputdict}{\kw{InputDict}}
\newcommand{\extractfromlabel}{\kw{Extrctval}}
\newcommand{\extracttag}{\kw{Extrcttag}}
\newcommand{\extract}{\extractfromlabel}
\newcommand{\extractfromlabelall}{\kw{ExtractFromLabelAll}}
\newcommand{\simplenrc}{\kw{SimpleNRC}}
\newcommand{\simplenrclet}{\kw{SimpleNRC}^{\kw{Let}}}
\newcommand{\nextmaximal}{\kw{NextMaximal}}
\newcommand{\mult}{\kw{Mult}}
\newcommand{\maximalsubexpressions}{\kw{MaxSub}}
\newcommand{\parent}{\kw{Parent}}
\newcommand{\typeof}{\kw{TypeOf}}

\newcommand{\variants}{\kw{Variants}}
\newcommand{\gensingleton}{\kw{WeightedSing}}
\newcommand{\flatten}{\kw{Flatten}}
\newcommand{\indextype}{\kw{IndexType}}
\newcommand{\indicesof}{\kw{IndicesOf}}
\newcommand{\evalexpr}[2]{{\llbracket #1\rrbracket}_{#2}}
\newcommand{\shredobj}{\kw{Shredded}}
\newcommand{\shreddedvars}{\kw{ShreddedVars}}
\newcommand{\outcome}[3][]{\ensuremath{\llbracket#2\ifstrempty{#1}{}{\mid#1}\rrbracket_{#3}}}
\newcommand{\interpret}[2]{\outcome{#1}{#2}}
\newcommand{\IdFor}{\kw{For}}
\newcommand{\emptytype}{()}
\newcommand{\IdUnion}{\kw{Union}}
\newcommand{\IdIn}{\kw{In}}
\newcommand{\queryunion}{\uplus}
\newcommand{\tuple}[1]{{\langle#1\rangle}}
\newcommand{\pred}{pred}
\newcommand{\emptybag}{\{\}_{\bagtype(T)}}
\newcommand{\emptybagq}{\emptyset_{\bagtype(F)}}
\newcommand{\emptybaggen}[1]{\emptyset_{\bagtype(#1)}}
\newcommand{\emptysetq}{\emptyset_{\settype(T)}}
\def\lBrack{\lbrack\!\lbrack}
\def\rBrack{\rbrack\!\rbrack}
\newcommand{\Bracks}[1]{\lBrack#1\rBrack}
\newcommand{\<}{\langle}
\renewcommand{\>}{\rangle}

\newcommand{\datalogarrow}{ :=}
\newcommand{\aschema}{\mathcal{S}}
\newcommand{\powerset}[1]{\mathcal{P}(#1)}
\newcommand{\LFPP}[3]{\LFP_{#1,#2}^{#3}}
\newcommand{\LFPA}[2]{\LFP_{#1,#2}}
\newcommand{\GFPA}[2]{\GFP_{#1,#2}}
\newcommand{\nf}{\text{strict normal form}\xspace}
\newcommand{\logic}[1]{\textup{\small #1}\xspace}
\renewcommand{\vec}[1]{\boldsymbol{#1}}
\newcommand{\V}{\mathcal{V}}
\newcommand{\cA}{\mathcal{A}}
\newcommand{\cB}{\mathcal{B}}
\newcommand{\cC}{\mathcal{C}}
\newcommand{\cD}{\mathcal{D}}
\newcommand{\cE}{\mathcal{E}}
\newcommand{\cF}{\mathcal{F}}
\newcommand{\cG}{\mathcal{G}}
\newcommand{\cM}{\mathcal{M}}
\newcommand{\cN}{\mathcal{N}}
\newcommand{\cO}{\mathcal{O}}
\newcommand{\cS}{\mathcal{S}}
\newcommand{\cT}{\mathcal{T}}
\newcommand{\cU}{\mathcal{U}}
\newcommand{\cV}{\mathcal{V}}
\newcommand{\cW}{\mathcal{W}}
\newcommand{\cH}{\mathcal{H}}
\newcommand{\cI}{\mathcal{I}}
\newcommand{\cP}{\mathcal{P}}
\newcommand{\cQ}{\mathcal{Q}}
\newcommand{\A}{\mathbb{A}}
\newcommand{\B}{\mathbb{B}}
\newcommand{\C}{\mathbb{C}}
\newcommand{\D}{\mathbb{D}}
\newcommand{\U}{\mathbb{U}}
\newcommand{\N}{\mathbb{N}}
\newcommand{\Ninf}{\mathbb{N_\infty}}
\newcommand{\fA}{\mathfrak{A}}
\newcommand{\fB}{\mathfrak{B}}
\newcommand{\fU}{\mathfrak{U}}
\newcommand{\fM}{\mathfrak{M}}
\newcommand{\set}[1]{\left\{{#1}\right\}}
\newcommand{\sset}[1]{\{{#1}\}}
\newcommand{\size}[1]{\lvert #1 \rvert}
\newcommand{\width}[1]{\operatorname{width}( #1 )}
\newcommand{\restrict}[2]{ #1\restriction_{#2} }
\let\Lslash\L
\renewcommand{\L}{\mathrm{L}}
\newcommand{\R}{\mathrm{R}}
\newcommand{\interpolant}[1]{\theta_{#1}}
\newcommand{\free}[1]{\operatorname{free}(#1)}
\newcommand{\dom}[1]{\operatorname{dom}(#1)}
\newcommand{\gn}{\textup{GN}\xspace}
\newcommand{\gnfo}{\logic{GNF}}
\newcommand{\gso}{\logic{GSO}}
\newcommand{\mso}{\logic{MSO}}
\newcommand{\gnf}{\gnfo}
\newcommand{\gf}{\logic{GF}}
\newcommand{\lfp}{\logic{LFP}}
\newcommand{\unfo}{\logic{UNF}}
\newcommand{\fo}{\logic{FO}}
\newcommand{\unf}{\unfo}
\newcommand{\Lmu}{\textup{L}_{\mu}}
\newcommand{\gnfpup}{\logic{GNFP-UP}}
\newcommand{\unfpup}{\logic{UNFP-UP}}
\newcommand{\gnfpplus}{\gnfpup}
\newcommand{\unfpplus}{\unfpup}
\newcommand{\GFP}{\operatorname{\bf gfp}}
\newcommand{\gfp}{\logic{GFP}}
\newcommand{\LFP}{\operatorname{\bf lfp}}
\newcommand{\unravel}[1]{\cT(#1)}
\newcommand{\gnfp}{\logic{GNFP}}
\newcommand{\unfp}{\logic{UNFP}}
\newcommand{\unfpk}{\unfp^k}
\newcommand{\gnfk}{\gnf^k}
\newcommand{\unfk}{\unf^k}
\newcommand{\lfpl}{\logic{LFP}}
\newcommand{\complexity}[1]{\textsc{#1}}
\newcommand\ptime{\complexity{PTime}\xspace}
\newcommand\exptime{\complexity{ExpTime}\xspace}
\newcommand\nexptime{\complexity{NExpTime}\xspace}
\newcommand\twoexptime{\complexity{2-ExpTime}\xspace}
\newcommand{\twoexp}{\twoexptime}
\newcommand\np{\complexity{NP}\xspace}
\newcommand\conp{\complexity{co-NP}\xspace}
\newcommand{\dmodality}[1]{\langle #1 \rangle}
\newcommand{\bmodality}[1]{[ #1 ]}
\newcommand{\valuation}{\ell}
\newcommand{\codom}[1]{\operatorname{rng}(#1)}
\newcommand{\decode}[1]{\mathfrak{D}(#1)}
\newcommand{\tree}{\cT}
\newcommand{\cL}{\mathcal{L}}
\newcommand{\gnfpk}{\text{$\gnfp^k$}\xspace}
\renewcommand{\varphi}{\phi}
\newcommand{\kw}[1]{\textsc{#1}}
\newcommand{\atypesig}[3]{\kw{AT}_{#2,#3}(#1)}
\newcommand{\ExactLabel}[1]{\kw{ExactLabel}(#1)}
\newcommand{\GNLabel}[1]{\kw{GNLabel}(#1)}
\newcommand{\sigmag}{\sigma_g}
\newcommand{\sigmaprimeg}{\sigma'_g}
\newcommand{\gnl}{\gn^{l}}
\newcommand{\sigk}{\tilde{\sigma}_k}
\newcommand{\gnkinvar}{\logic{GN}^k}
\newcommand{\sgnkinvar}{\logic{BGN}^k}
\newcommand{\sunkinvar}{\logic{BUN}^k}
\newcommand{\gnlinvar}{\logic{GN}^l}
\newcommand{\gnninvar}{\logic{GN}^n}
\newcommand{\gnminvar}{\logic{GN}^m}
\newcommand{\ginvar}{\logic{G}}
\newcommand{\bunravelk}[2]{\cU_{\textup{BGN}^k[#2]}( #1 )}
\newcommand{\unravelk}[2]{\cU_{\textup{GN}^k[#2]}( #1 )}
\newcommand{\unravelm}[2]{\cU_{\textup{GN}^m[#2]}( #1 )}
\newcommand{\unravell}[2]{\cU_{\textup{GN}^l[#2]}( #1 )}
\newcommand{\sunravelk}[2]{\cU^{\textup{plump}}_{\textup{BGN}^k[#2]}( #1 )}
\newcommand{\gunravel}[2]{\cU_{\textup{G}[#2]}( #1 )}
\newcommand{\sigcode}[2]{\Sigma^{\text{code}}_{#1,#2}}
\newcommand{\dloglite}{\logic{DATALOG-LITE}}
\newcommand{\dlog}{\logic{DATALOG}}
\newcommand{\unravellingcode}{\cU_{\logictarget}(\fB)}
\newcommand{\lemmafwd}{\hyperref[lemma:forward-gso]{Lemma~Fwd}\xspace}
\newcommand{\lemmagfpbwd}[1]{\hyperref[lemma:backwards-gfp]{#1}\xspace}
\newcommand{\lemmagnfpbwd}[1]{\hyperref[lemma:backwards-gnfp]{#1}\xspace}
\newcommand{\logictarget}{\cL_1}
\newcommand{\sigtarget}{\sigma'}
\newcommand{\sigoriginal}{\sigma}
\newcommand{\Bekic}{Beki\v{c}\xspace}
\newcommand{\convertnf}[1]{\operatorname{convert}(#1)}
\newcommand{\indices}[1]{\textup{names}(#1)}
\newcommand{\dup}{\uparrow}
\newcommand{\dstay}{0}
\newcommand{\ddown}{\downarrow}
\newcommand{\Pri}{\mathrm{Pri}}
\newcommand{\Dir}{\mathrm{Dir}}
\newcommand{\gameonstart}[3]{\cG(#1, #2, #3)}
\newcommand{\gameon}[2]{\cG(#1, #2)}
\newcommand{\Sigmap}{\Sigma_p}
\newcommand{\Sigmaa}{\Sigma_a}
\newcommand{\QE}{Q_{E}}
\newcommand{\QA}{Q_{A}}
\newcommand{\simgn}[1]{\sim_{#1}}
\newcommand{\subsetgn}[1]{\subseteq_{#1}}
\newcommand{\remove}[2]{\text{remove}_{#2}(#1)}
\newcommand{\cJ}{\mathcal{J}}
\newcommand{\namesk}{U_k}
\newcommand{\atype}[2]{\kw{AT}_{#2}(#1)}
\newcommand{\propsguardedneg}[1]{\kw{GNProps}(#1)}
\newcommand{\propsneg}[1]{\kw{NProps}(#1)}
\newcommand{\bagextend}[1]{\kw{BagLabels}(#1)}
\newcommand{\GuardedLabels}{\kw{GLabels}}
\newcommand{\GuardedLabelsNeighbor}[2]{\kw{GLabels}(#1,#2)}
\newcommand{\IntNodeLabels}{\kw{InterfaceLabels}}
\newcommand{\BagNodeLabels}{\kw{BagLabels}}
\newcommand{\EdgeLabels}{\kw{Edges}}
\newcommand{\EdgeLabelsRestrict}[1]{\kw{Edges}(#1)}
\newcommand{\NodeLabels}{\kw{NodeLabels}}
\newcommand{\tforward}[1]{ {#1}^{\rightarrow} }
\newcommand{\tbackward}[1]{ {#1}^{\leftarrow} }
\newcommand{\struct}[1]{\mathfrak{D}(#1)}
\newcommand{\readymode}{\text{select}}
\newcommand{\waitmode}{\text{move}}
\newcommand{\views}{\kw{Views}}
\newcommand{\view}{I}
\newcommand{\dneighbor}{\updownarrow}
\newcommand{\strategy}{\zeta}
\newcommand{\correct}[1]{\textup{correct}(#1)}
\newcommand{\correctr}[1]{\textup{correct}_{r}(#1)}
\newcommand{\Fpart}[1]{\textup{match}(#1)}
\newcommand{\id}{\textup{id}}
\newcommand{\consistencyform}{\phi_{\textup{consistent}}}
\newcommand{\elem}[1]{\operatorname{elem}(#1)}
\newcommand{\guardg}[2]{\operatorname{guard}^{\sigmag}_{#1}(#2)}
\newcommand{\Guarded}[1]{\kw{Gdd}(#1)}
\newcommand{\GuardedDom}[1]{\guardedg_{\dom{#1}}} 
\newcommand{\GuardedRng}[1]{\guardedg_{\codom{#1}}} 
\newcommand{\GuardDom}[2]{\operatorname{guard-dom}(#1,#2)} 
\newcommand{\GuardRng}[2]{\operatorname{guard-rng}(#1,#2)}
\newcommand{\phiL}{\phi_{\L}}
\newcommand{\phiR}{\phi_{\R}}
\newcommand{\cform}[2]{\text{consistent}_{#1,#2}}
\newcommand{\sigR}{\sigma_{\R}}
\newcommand{\fG}{\mathfrak{G}}
\newcommand{\gnnf}{\text{\textup{GN}-normal form}\xspace}
\newcommand{\myparagraph}[1]{{\bf #1.}}
\newcommand{\true}{\kw{True}}
\newcommand{\false}{\kw{False}}

\newcommand{\mattranstwo}[2]{\mathcal{M}\lBrack#1\rBrack_{#2}}
\newcommand{\mattransbasic}[2]{\mathcal{M}\lBrack#1\rBrack_{#2}}
\newcommand{\mattrans}[3]{\mattranstwo{#1}{#2,#3}}

\newcommand{\Flat}{\kw{Flat}\xspace}
\newcommand{\Fplus}{$\kw{Flat}^+$\xspace}
\newcommand{\Fpplus}{\kw{Standard} } 
\newcommand{\bioendtoend}{\kw{E2E}}
\newcommand{\Stepi}{$\kw{Step}_1$\xspace}
\newcommand{\Stepii}{$\kw{Step}_2$\xspace}
\newcommand{\Stepiii}{$\kw{Step}_3$\xspace}
\newcommand{\Stepiv}{$\kw{Step}_4$\xspace}
\newcommand{\Stepv}{$\kw{Step}_5$\xspace}
\newcommand{\Ci}{$\kw{C}_1$\xspace}
\newcommand{\Cii}{$\kw{C}_2$\xspace}
\newcommand{\Ciii}{$\kw{C}_3$\xspace}
\newcommand{\OccurFull}{$\kw{BN}_2$\xspace}
\newcommand{\CNVFull}{$\kw{BF}_2$\xspace}
\newcommand{\ExprFull}{$\kw{BF}_1$\xspace}
\newcommand{\Network}{$\kw{BN}_1$\xspace}

\newcommand{\superunshred}{\kw{U}}
\newcommand{\Shred}{\kw{Shred}\xspace}
\newcommand{\Unshred}{\kw{Unshred}\xspace}
\newcommand{\Shrdom}{$\Shred^{\kw{Dom}}$\xspace}
\newcommand{\Fskew}{$\Fpplus_{\kw{skew}}$\xspace}
\newcommand{\Sskew}{$\Shred_{\kw{skew}}$\xspace}
\newcommand{\Uskew}{$\Shred_{\kw{skew}}^{+ \superunshred}$\xspace}
\newcommand{\Sdskew}{$\Shred^{\kw{Dom}}_{\kw{skew}}$\xspace}

\DeclareRobustCommand{\ojoin}{\rule[0.10ex]{.3em}{.4pt}\llap{\rule[1.40ex]{.3em}{.4pt}}}
\newcommand{\leftouterjoin}{\mathrel{\ojoin\mkern-6.5mu\Join}}
\newcommand{\rightouterjoin}{\mathrel{\Join\mkern-6.5mu\ojoin}}
\DeclareRobustCommand{\ounnest}{\rule[2.0ex]{.3em}{.4pt}\llap{\rule[1.0ex]{.3em}{.4pt}}}
\newcommand\textequal{%
 \rule[.3ex]{3pt}{0.4pt}\llap{\rule[.7ex]{3pt}{0.4pt}}}
\newcommand{\outerunnest}{\mathrel{\textequal\!\mu}}

\newtheorem{example}{Example}

\newcommand{\revcolor}{black}
\newcommand{\revision}[1]{{\color{\revcolor} #1}}
\newcommand{\revisionomit}[1]{} 

\newcommand{\Occur}{\texttt{Occurrences}\xspace}
\newcommand{\OccurGroup}{\texttt{OccurGrouped}\xspace}
\newcommand{\OccurJoin}{\texttt{OccurCNVJoin}\xspace}
\newcommand{\OccurAgg}{\texttt{OccurCNVAgg}\xspace}
\newcommand{\CopyNum}{\texttt{CopyNumber}\xspace}
\newcommand{\Pathway}{\texttt{Pathway}\xspace}
\newcommand{\GeneExpr}{\texttt{GeneExpression}\xspace}
\newcommand{\Biospec}{\texttt{Samples}\xspace}
\newcommand{\SOImpact}{\texttt{SOImpact}\xspace}
\newcommand{\MNetwork}{\texttt{MappedNetwork}\xspace}
\newcommand{\FNetwork}{\texttt{FNetwork}\xspace}
\newcommand{\SNetwork}{\texttt{SampleNetwork}\xspace}
\newcommand{\Biomart}{\texttt{Biomart}\xspace}
\newcommand{\Hybrid}{\texttt{HybridMatrix}\xspace}
\newcommand{\Effect}{\texttt{EffectMatrix}\xspace}
\newcommand{\Connect}{\texttt{ConnectMatrix}\xspace}
\newcommand{\Connects}{\texttt{Connectivity}\xspace}
\newcommand{\aid}{\texttt{\textup{aid}}\xspace}
\newcommand{\aliquot}{\texttt{\textup{aliquot}}\xspace}
\newcommand{\sample}{\texttt{\textup{sample}}\xspace}
\newcommand{\gid}{\texttt{\textup{gid}}\xspace}
\newcommand{\gene}{\texttt{\textup{gene}}\xspace}

\newcommand{\protein}{\texttt{\textup{protein}}\xspace}
\newcommand{\impact}{\texttt{\textup{impact}}\xspace}
\newcommand{\conseq}{\texttt{\textup{conseq}}\xspace}
\newcommand{\cnum}{\texttt{\textup{cnum}}\xspace}

\newcommand{\trans}{\texttt{\textup{transcripts}}\xspace}
\newcommand{\cands}{\texttt{\textup{candidates}}\xspace}
\newcommand{\nodes}{\texttt{\textup{nodes}}\xspace}
\newcommand{\conseqs}{\texttt{\textup{consequences}}\xspace}
\newcommand{\hyscores}{\texttt{\textup{scores}}\xspace}
\newcommand{\hscore}{\texttt{\textup{score}}\xspace}
\newcommand{\sift}{\texttt{\textup{sift}}\xspace}
\newcommand{\siftpred}{\texttt{\textup{siftpredict}}\xspace}
\newcommand{\poly}{\texttt{\textup{poly}}\xspace}
\newcommand{\polypred}{\texttt{\textup{polypredict}}\xspace}
\newcommand{\proj}{\texttt{\textup{project}}\xspace}
\newcommand{\nodep}{\texttt{\textup{nodeProtein}}\xspace}
\newcommand{\edges}{\texttt{\textup{edges}}\xspace}
\newcommand{\edgep}{\texttt{\textup{edgeProtein}}\xspace}
\newcommand{\score}{\texttt{\textup{score}}\xspace}
\newcommand{\genes}{\texttt{\textup{genes}}\xspace}
\newcommand{\mutations}{\texttt{\textup{mutations}}\xspace}
\newcommand{\dist}{\texttt{\textup{distance}}\xspace}
\newcommand{\case}{\texttt{\textup{case}}\xspace}
\newcommand{\contig}{\texttt{\textup{contig}}\xspace}
\newcommand{\mstart}{\texttt{\textup{start}}\xspace}
\newcommand{\mend}{\texttt{\textup{end}}\xspace}
\newcommand{\mref}{\texttt{\textup{reference}}\xspace}
\newcommand{\alt}{\texttt{\textup{alternate}}\xspace}
\newcommand{\mutid}{\texttt{\textup{mutationId}}\xspace}
\newcommand{\uid}{\texttt{\textup{uid}}\xspace}
\newcommand{\gibbs}{\texttt{\textup{gibbs}}\xspace}
\newcommand{\extern}{\texttt{\textup{external}}\xspace}
\newcommand{\interm}{\texttt{\textup{E}}\xspace}
\newcommand{\fpkm}{\texttt{\textup{fpkm}}\xspace}
\newcommand{\pathway}{\texttt{\textup{pathway}}\xspace}
\newcommand{\pathways}{\texttt{\textup{pathways}}\xspace}
\newcommand{\burden}{\texttt{\textup{burden}}\xspace}

\title{Supplementary}

\section{Operator Mappings: Code Generation}

\begin{small}
\begin{tabular}{ll@{}}
\toprule
Plan Operator & Definition \\[3pt]
\midrule
$\sigma_{p(x)}(X)$ &
\begin{lstlisting}[language=Scala] 
X.filter(x => p(x))
\end{lstlisting}  \\[3pt]
$\pi_{\,a_1,\ldots,a_k}(X)$ &
\begin{lstlisting}[language=Scala,mathescape] 
X.select(a$_\texttt{1}$, $\ldots$, a$_\texttt{k}$)
\end{lstlisting} \\[3pt]
$\mu^{a_i}(X)$ &
\begin{lstlisting}[language=Scala] 
X.flatMap(x => x.a$_\texttt{i}$.map(y => R(x.a$_\texttt{1}$, $\ldots$, x.a$_\texttt{k}$, y.b$_\texttt{1}$, $\ldots$, y.b$_\texttt{j}$))).as[R]  $\qquad$// R does not contain x.a$_\texttt{i}$
\end{lstlisting} \\[3pt]
$\outerunnest^{a}(X)$ &
\begin{lstlisting}[language=Scala] 
X.withColumn("index", monotonically_increasing_id()).flatMap(x => 
  if (x.a$_\texttt{i}$.isEmpty) R(x.index, x.a$_\texttt{1}$, $\ldots$, x.a$_\texttt{k}$, None, $\ldots$, None))).as[R]
  else x.a$_\texttt{i}$.map(y => R(x.index, x.a$_\texttt{1}$, $\ldots$, x.a$_\texttt{k}$, Some(y.b$_\texttt{1}$), $\ldots$, Some(y.b$_\texttt{j}$))).as[R] $\qquad$// R does not contain x.a$_\texttt{i}$
\end{lstlisting} \\[4pt]
$X \join_{X(f) = Y(g)} Y$ &
\begin{lstlisting}[language=Scala] 
X.join(Y, X(f) === Y(g))
\end{lstlisting} \\[6pt]
$X \leftouterjoin_{X(f) = Y(g)} Y$ &
\begin{lstlisting}[language=Scala] 
X.join(Y, X(f) === Y(g), "left_outer")
\end{lstlisting} \\[6pt]
$\Gamma^{\,+,\, key(x),\, value(x)}(X)$ &
\begin{lstlisting}[language=Scala] 
X.groupByKey(x => key(x)).agg(typed.sum[R](x => 
  value(x) match { case Some(v) => v; case _ => 0 }))
\end{lstlisting} \\[6pt]
$\Gamma^{\,\bagunion,\, key(x),\, value(x)}(X)$  &
\begin{lstlisting}[language=Scala] 
// key persists the index, value never includes an index
X.groupByKey(x => key(x)).mapGroups{ case (key, values) =>     
  val grp = values.flatMap{ case x => 
    value(x) match { case Some(t) => Seq(t); case _ => Seq() }}.toSeq
  (key, grp)
}
\end{lstlisting} \\[3pt]
\bottomrule
\end{tabular}
\end{small}

\section{TPCH Schema: Standard}

The TPCH benchmark builds up nested queries using the tables represented 
in the TPCH schema. To explore effects of large inner collections, the 
queries focus on the number of top level tuples. Each level of nesting 
decreases the amount of top level tuples: Region (5) < Nation (25) < Customer (150000) < Order (1500000) < Lineitem (6001215). The hierarchy is as follows:

\begin{itemize}
\item 0: Lineitem 
\item 1: Order - Lineitem
\item 2: Customer - Order - Lineitem
\item 3: Nation - Customer - Order - Lineitem 
\item 4: Region - Nation - Customer - Order - Lineitem
\end{itemize}

\section{Flat-to-nested}

This builds up a nested object from flat input. 
The ellipses represent the additional fields that may be 
present, such as when output tuples are wide (ie. no 
projections).

\subsection{NRC}

\begin{lstlisting}[language=NRC]
$\cidfor ~ r ~ \cidin ~ \Region ~ \cidunion$
  $\{\<\, \rname \ateq r.\rname,$ $\ldots$, $\rnations \ateq$
    $\cidfor ~ n ~ \cidin ~ \Nation ~ \cidunion$
      $\{\<\, \nname \ateq n.\nname,$ $\ldots$, $\ncusts \ateq$
        $\cidfor ~ c ~ \cidin ~ \Customer ~ \cidunion$
          $\{\<\, \cname \ateq c.\cname,$ $\ldots$, $\corders \ateq$ 
            $\cidif ~ c.\cid == o.\cid ~ \cidthen$  
              $\cidfor ~ o ~ \cidin ~ \Order ~ \cidunion$
                $\{\<\, \odate \ateq o.\odate,$ $\ldots$, $\oparts \ateq$ 
                  $\cidfor ~ l ~ \cidin ~ \Lineitem ~ \cidunion$
                    $\cidif ~ o.\oid == l.\oid ~ \cidthen$
                      $\{\<\, \pid \ateq l.\pid,$ $\ldots$, $\,\lqty \ateq l.\lqty \,\>\}
                \,\>\}
            \,\>\}
        \,\>\}
    \,\>\}$
\end{lstlisting}

\subsection{Optimized Plan}

\begin{forest}
[$\pi_{(r.\rname, \ldots, \rnations)}$,
[ $COGROUP_{r.\rid = n.\rid}$
  [$\Region$]
  [ $\pi_{(n.\nname, \ldots, \ncusts)}$,
    [ $COGROUP_{n.\nid = c.\nid}$
      [$\Nation$]
      [ $\pi_{(c.\cname, \ldots, \corders)}$,
        [ $COGROUP_{c.\cid = o.\cid}$
          [$\Customer$]
          [ $\pi_{(o.\odate, \ldots, \oparts)}$
            [ $COGROUP_{o.\oid = l.\oid}$
              [$\Order$]
              [$\Lineitem$]
            ]]]]]]]]
\end{forest}

\subsection{Shredded Plan}

Each dictionary is partitioned by the attribute 
wrapped in \texttt{Label}. 

$\RNCOP_{Top} := $
\begin{forest}
[$\pi_{(r.\rname, \ldots)}$($\Region$)]
\end{forest}

$\rnations_{Dict} := $
\begin{forest}
[$\pi_{(n.\nname, \ldots, Label(n.\rid))}$($\Nation$)]
\end{forest}

$\ncusts_{Dict} := $
\begin{forest}
[$\pi_{(c.\cname, \ldots, Label(c.\nid))}$($\Customer$)]
\end{forest}

$\corders_{Dict} := $
\begin{forest}
[$\pi_{(o.\odate, \ldots, Label(o.\cid))}$($\Order$)]
\end{forest}

$\oparts_{Dict} := $
\begin{forest}
[$\pi_{(l.\pid, \ldots, Label(l.\oid))}$($\Lineitem$)]
\end{forest}

\subsection{Unshredded Plan} 

\begin{forest}
[ $\pi_{(o.\odate, \ldots, \oparts)}$
  [ $LOOKUP_{o.\oparts = \labeltype}$
    [$\corders_{Dict}$]
    [$\oparts_{Dict}$]
  ]]
\end{forest}

\begin{forest}
[$\pi_{(r.\rname, \ldots, \rnations)}$,
[ $LOOKUP_{r.\rid = n.\labeltype}$
  [$\RNCOP_{Top}$]
  [ $\pi_{(n.\nname, \ldots, \ncusts)}$,
    [ $LOOKUP_{n.\nid = \labeltype}$
      [$\rnations_{Dict}$]
      [ $\pi_{(c.\cname, \ldots, \corders)}$,
        [ $LOOKUP_{c.\cid = \labeltype}$
          [$\ncusts_{Dict}$]
          [ $\pi_{(o.\odate, \ldots, \oparts)}$
            [ $LOOKUP_{o.\oparts = \labeltype}$
              [$\corders_{Dict}$]
              [$\oparts_{Dict}$]
            ]]]]]]]]
\end{forest}

\section{Nested-to-nested}

This is a query that operates on nested input. 
The ellipses represent the additional fields that may be 
present, such as when output tuples are wide (ie. no 
projections).

\subsection{NRC}

\begin{lstlisting}[language=NRC]
$\cidfor ~ r ~ \cidin ~ \RNCOP ~ \cidunion$
  $\{\<\, \rname \ateq r.\rname,$ $\ldots$, $\rnations \ateq$
    $\cidfor ~ n ~ \cidin ~ r.\rnations ~ \cidunion$
      $\{\<\, \nname \ateq n.\nname,$ $\ldots$, $\ncusts \ateq$
        $\cidfor ~ c ~ \cidin ~ n.\ncusts ~ \cidunion$
          $\{\<\, \cname \ateq c.\cname,$ $\ldots$, $\corders \ateq$ 
            $\cidif ~ c.\cid == o.\cid ~ \cidthen$  
              $\cidfor ~ o ~ \cidin ~ c.\corders ~ \cidunion$
                $\{\<\, \odate \ateq o.\odate,$ $\ldots$, $\oparts \ateq$ 
                  sumBy$_{\hspace{0.15mm}\pname}^{\hspace{0.15mm}\total}($
                    $\cidfor ~ l ~ \cidin ~ o.\oparts ~ \cidunion$
                      $\cidif ~ o.\oid == l.\oid ~ \cidthen$
                        $\cidfor ~ p ~ \cidin ~ \Part ~ \cidunion$
                          $\cidif ~ l.\pid == p.\pid ~ \cidthen$
                            $\{\<\, \pname \ateq p.\pname,$
                              $\,\total \ateq l.\lqty * p.\pprice \,\>\})
                \,\>\}
            \,\>\}
        \,\>\}
    \,\>\}$
\end{lstlisting}

\subsection{Optimized Plan}

\begin{forest}
[$\pi_{(r.\rname, \ldots)}$,
  [$\Gamma^{\cup / r / (n.\nname, \ldots)}$,
    [$\Gamma^{\cup / (r, n) / (c.\cname, \ldots)}$,
      [$\Gamma^{\cup / (r, n, c) / (o.\odate, \ldots)}$,
        [$\pi_{o.\odate, t.\pname, t.\total}$,
          [$\Gamma^{+ / (r, n, c, o, p.\pname) / l.\lqty * p.\pprice}$,        
            [$\leftouterjoin_{l.\pid = p.\pid}$,
              [$\outerunnest_{o.\oparts}$,
                [$\outerunnest_{c.\corders}$,
                  [$\outerunnest_{n.\ncusts}$,
                    [$\outerunnest_{r.\rnations}$($\RNCOP$)]]]]
              [$\pi_{\pname, \lqty}$ [$\Part$]]]]]]]]]
\end{forest}

\subsection{Shredded Plan}

Each dictionary is partitioned by the attribute 
wrapped in \texttt{Label}. 

$\RNCOP_{Top} := $
\begin{forest}
[$\pi_{(r.\rname, \ldots)}$($\RNCOP_{Top}$)]
\end{forest}

$\rnations_{Dict} := $
\begin{forest}
[$\pi_{(n.\nname, \ldots, Label(n.\rid))}$($\rnations_{Dict}$)]
\end{forest}

$\ncusts_{Dict} := $
\begin{forest}
[$\pi_{(c.\cname, \ldots, Label(c.\nid))}$($\ncusts_{Dict}$)]
\end{forest}

$\corders_{Dict} := $
\begin{forest}
[$\pi_{(o.\odate, \ldots, Label(o.\cid))}$($\corders_{Dict}$)]
\end{forest}

$\oparts_{Dict} := $

\begin{forest}
[$\pi_{(Label(l.\oid), \ldots)}$, 
  [$\Gamma^{+ / (o, p.\pname) / l.\lqty * p.\pprice}$,   
    [$\leftouterjoin_{l.\pid = p.\pid}$,
    [$\pi_{(l.\pid, \ldots)}$ [$\oparts_{Dict}$]]
    [$\pi_{\pname, \lqty}$ [$\Part$]]]
  ]]
\end{forest}

\subsection{Unshredded Plan} 

\begin{forest}
[$\pi_{(r.\rname, \ldots, \rnations)}$,
[ $LOOKUP_{r.\rid = n.\labeltype}$
  [$\RNCOP_{Top}$]
  [ $\pi_{(n.\nname, \ldots, \ncusts)}$,
    [ $LOOKUP_{n.\nid = \labeltype}$
      [$\rnations$]
      [ $\pi_{(c.\cname, \ldots, \corders)}$,
        [ $LOOKUP_{c.\cid = \labeltype}$
          [$\ncusts_{Dict}$]
          [ $\pi_{(o.\odate, \ldots, \oparts)}$
            [ $LOOKUP_{o.\oparts = \labeltype}$
              [$\corders_{Dict}$]
              [$\oparts_{Dict}$]
            ]]]]]]]]
\end{forest}

\section{Nested-to-flat}

\subsection{NRC}

\begin{lstlisting}[language=NRC]
sumBy$_{\hspace{0.15mm}\cname, \ldots}^{\hspace{0.15mm}\total}($
  $\cidfor ~ c ~ \cidin ~ \COP ~ \cidunion$
    $\cidfor ~ o ~ \cidin ~ c.\corders ~ \cidunion$
      $\cidfor ~ l ~ \cidin ~ o.\oparts ~ \cidunion$
        $\cidfor ~ p ~ \cidin ~ \Part ~ \cidunion$
          $\cidif ~ l.\pid == p.\pid ~ \cidthen$
            $\{\<\, \cname \ateq c.\cname, \dots$
                $\,\total \ateq l.\lqty * p.\pprice \,\>\})$
\end{lstlisting}

\subsection{Optimized plan}

\begin{forest}
[$\Gamma^{+ / (c.\cname, \ldots) / l.\lqty * p.\pprice}$,        
  [$\leftouterjoin_{l.\pid = p.\pid}$,
    [$\outerunnest_{o.\oparts}^{\pi_{\pid, \lqty}}$
      [$\outerunnest_{c.\corders}^{\pi_{\odate, \ldots}}$,
        [$\pi_{\cname, \ldots}$ [$\COP$]]]]
    [$\pi_{\pname, \pprice}$ [$\Part$]]]]
\end{forest}

\subsection{Shredded plan}

\begin{forest}
[$\Gamma^{+ / c.\cname / t.\total}$,
[$\leftouterjoin_{o.\oparts = t.\labelatt}$,
  [$\pi_{c.\cname, \ldots}$[$\COP_{Top}$]]
  [$\leftouterjoin_{o.\oparts = t.\labelatt}$, 
    [$\pi_{o.\odate, \ldots}$[$\corders_{Dict}$]]
    [$\Gamma^{+ / Label(\l.oid) / l.\lqty * p.\pprice}$,   
      [$\join_{l.\pid = p.\pid}$,
        [$\pi_{(l.\pid, \ldots)}$ [$\oparts_{Dict}$]]
        [$\pi_{\pname, \pprice}$ [$\Part$]]
      ]]]]]
\end{forest}

\section{Skew-handling}

\subsection{Experiment 3.0: Flat-to-Nested}

\begin{tabular}{lllll}
\bottomrule
Input & Key & Heavy Keys \\
\toprule
$\rnations$ & NA & NA \\
\midrule
$\ncusts$ & $\nid$ & 20 \\
\midrule
$\corders$ & $\cid$ & 5 \\
\midrule
$\oparts$ & $\oid$ & 0 \\
\bottomrule
\end{tabular}

\subsection{Experiment 3.1: Flat-to-Nested}

Input produced from Experiment 2.1 (nested wide tuples).
Focus on three levels of nesting here, since four levels 
of nesting crashes for flat and unshredding runs.

\subsection{Optimized Plan}

TODO


\subsection{Shredded Plan}

Each input dictionary is partitioned by Label, the 
heavy keys are known before running the query.

\begin{tabular}{lllll}
\bottomrule
Input & Light & Heavy & Key & Count \\
\toprule
$\rnations_{Top}$ & $\rnations_{Top}^{Light}$ & $\emptyset$ & NA & NA \\
\midrule
$\ncusts_{Dict}$ & $\ncusts_{Dict}^{Light}$ & $\ncusts_{Dict}^{Heavy}$ & $Label(\nid)$ & 20 \\
\midrule
$\corders_{Dict}$ & $\corders_{Dict}^{Light}$ & $\corders_{Dict}^{Heavy}$ & $Label(\cid)$ & 5 \\
\midrule
$\oparts_{Dict}$ & $\oparts_{Dict}^{Light}$ & $\emptyset$ & $Label(\oid)$ & 0 \\
\bottomrule
\end{tabular}


Each dictionary is partitioned by the attribute 
wrapped in \texttt{Label}. After executing the query plan below, 
the light, heavy, and heavy key sets are the same as the input. 


$\rnations_{Top} := $
\begin{forest}
[$\pi_{(n.\nname, \ldots, Label(n.\nid))}$($\rnations_{Dict}$)]
\end{forest}

$\ncusts_{Dict} := $
\begin{forest}
[$\pi_{(c.\cname, \ldots, Label(c.\nid))}$($\ncusts_{Dict}$)]
\end{forest}

$\corders_{Dict} := $
\begin{forest}
[$\pi_{(o.\odate, \ldots, Label(o.\cid))}$($\corders_{Dict}$)]
\end{forest}

$\oparts_{Dict} := $

\begin{forest}
[$\pi_{(Label(l.\oid), \ldots)}$, 
  [$\Gamma^{+ / (o, p.\pname) / l.\lqty * p.\pprice}$,   
    [$\leftouterjoin_{l.\pid = p.\pid}$,
    [$\pi_{(l.\pid, \ldots)}$ [$\oparts_{Dict}$]]
    [$\pi_{\pname, \lqty}$ [$\Part$]]]
  ]]
\end{forest}

\subsection{Unshredded Plan}

The unshredding plan recognizes that there are no heavy keys for the 
oparts relation, the light and heavy components of $\corders$ are unioned 
to $\corders_{Dict}$ and the lookup is performed. \break

$\corders_{Partial} :=$

\begin{forest}
[ $\pi_{(o.\odate, \ldots, \oparts)}$
  [ $LOOKUP_{o.\oparts = \labeltype}$
    [$\corders_{Dict}$]
    [$\oparts_{Light}$]
  ]]
\end{forest} 

At this point, all partition information is lost: \break

\begin{tabular}{lllll}
\toprule
$\corders_{Partial}$ & $\corders_{Partial}^{Light}$ & $\emptyset$ & NA & NA \\
\bottomrule
\end{tabular} \break

\textbf{POSSIBLE IMPROVEMENT}: The above assumes that we know nothing about the 
heavy key information of $\corders$, when we do know that there are five heavy 
keys associated to this dictionary. The plan continues as suggested below, 
recalculating the heavy keys for the $\corders_{Partial}$ dictionary, when we 
already knew what these were before. \break

$\ncusts_{Partial} :=$

\begin{forest}
[ $\pi_{(c.\cname, \ldots, \corders)}$,
  [ $LOOKUP_{c.\cid = \labeltype}$
    [$\ncusts_{Dict}$]
    [$\corders_{Partial}$]]]
\end{forest}

This happens for the next dictionary as well. \break

$Result :=$

\begin{forest}
[$\pi_{(r.\rname, \ldots, \rnations)}$,
[ $LOOKUP_{r.\rid = n.\labeltype}$
  [$\RNCOP_{Top}$]
  [$\ncusts_{Partial}$]]]
\end{forest}



